\documentclass[a4paper,12pt]{article}
\usepackage{amsmath,amsthm,amssymb}
\usepackage{graphicx}
\usepackage{color}
\usepackage{slashed} 
\makeatletter

\catcode`\@=11
\newdimen\ex@
\font\dozeb=cmmib10 scaled \magstep1
\font\dozesyb=cmbsy10 scaled \magstep1
\font\dezb=cmmib10
\def\beq{\begin{equation}}
\def\eeq{\end{equation}}
\def\beqa{\begin{eqnarray}}
\def\eeqa{\end{eqnarray}}
\newcommand\BA{\begin{array}}
\newcommand\EA{\end{array}}

\renewcommand{\theequation}{\arabic{chapter}.\arabic{section}.
\arabic{equation}}

\newcommand{\m}{\mu}
\newcommand{\n}{\nu}

\newcommand{\C}{{\sf C
\hspace*{-0.5ex}\rule{0.01ex}
{1.5ex}\hspace*{0.9ex}}}

\newcommand{\I}{\mathbb{I}}

\begin{document}
\def\thefootnote{\fnsymbol{footnote}}

\title{\bf Algebro-geometric approach to a fermion self- \\ 
consistent field theory on coset space 
$\frac{SU_{m+n}}{S(U_{m} \times U_{n})}$
\vskip-0.4cm
}

\author
{Seiya NISHIYAMA\footnotemark[1] $\!$,~
Jo\~ao da PROVID\^{E}NCIA\footnotemark[2]\\
\\[-10pt]
CFisUC,
Department of Physics, University of Coimbra.\\
\\[-10pt]
3004-516 Coimbra, Portugal\\
\\[-10pt]
}

\maketitle

\vspace{0.5cm}

\footnotetext[1]
{Corresponding author.
~E-mail address: seikoceu@khe.biglobe.ne.jp}
\footnotetext[2]
{E-mail address: }  

\vspace{-0.4cm}

\begin{abstract}
$\!\!\!\!\!\!\!\!\!\!\!\!$
The integrability-condition method is regarded as a mathematical tool
to describe the symmetry of collective sub-manifold.$\!$
We here adopt the particle--hole representation.
In the conventional 
time-dependent (TD) self-consistent field (SCF) theory,
we take the one-form 
linearly composed of the TD SCF Hamiltonian and
the infinitesimal generator induced by 
the collective-variable differential of canonical transformation
on a group.$\!$
Standing on 
the differential geometrical viewpoint,
we introduce a 
Lagrange-like manner familiar to fluid dynamics to
describe collective coordinate systems.
We construct a geometric equation,
noticing the structure of coset space
$\frac{SU_{m+n}}{S(U_{m} \times U_{n})}$.$\!$
To develop a perturbative method with the use of the 
collective variables, we aim at constructing  
a new fermion SCF theory, i.e., 
renewal of TD Hartree-Fock (TDHF) theory 
by using the canonicity condition
under the existence 
of invariant subspace in the whole HF space.$\!$
This is due to a natural consequence
of the {\em maximally decoupled} theory because
there exists an invariant subspace, 
if the 
{\em invariance principle of Schr\"{o}dinger equation} 
is realized.$\!$
The integrability condition of the TDHF equation 
determining a collective sub-manifold
is studied,
standing again on the differential geometric viewpoint.$\!$
A geometric equation works well 
{\em over a wide range} of physics
beyond the random phase approximation.
\end{abstract}

\vspace{0.2cm}

~~~Keywords:Particle-hole representation;TDHF theory; 
$U(\!N\!)$ Lie algebra;

~~~Lagrange-like manner;

~~~Integrability condition;Geometric equation;
Coset space $\frac{SU_{m+n}}{S(U_{m} \times U_{n})}$;

~~~Random phase approximation

\vspace{0.4cm} 
 ~~~Mathematics Subject Classification 2010: 81R05, 81R12, 81R30, 81S10, 81V35

\def\thesection{\arabic{section}}
\setcounter{equation}{0}
\renewcommand{\theequation}{\arabic{section}.\arabic{equation}}

\newpage

%%%%%%%%%%%%%%%%%%%%%%%%%%%%%%%
%                                                                                                 %
%  1 Short History of Theory for  Nuclear Collective Motion   %     
%                                                                                                 %
%%%%%%%%%%%%%%%%%%%%%%%%%%%%%%%

\section{$\!\!$Short history of theory for nuclear collective motion}

According to Yamamura-Kuriyama (YK)
\cite{YK.87}, 
we view simply a short history for the model 
and the theory:$\!$
In the Bohr-Mottelson (BM) model 
\cite{BM.53}, 
a liquid drop model is taken 
for the collective motion and 
the independent particle motion
is described by the shell model.$\!$
However, microscopically, the constituents of the liquid drop
are nucleons themselves, which move in a nucleus independently.$\!$ 
In the year 1960, Arvieu-Veneroni, Baranger and Marumori
proposed, independently, a collective-motion theory 
for spherical {\em even}-mass nuclei
\cite{AM.60} 
called
the quasi-particle random phase approximation (Q-RPA). 
There exist the two types of the fundamental
correlations, that is, 
a short-range correlation and a long-range one 
\cite{Mo.59}.
The former is expressed in terms of the pairing interaction
and generates a superconducting mode.$\!$
The excited states are classified by a seniority coupling
scheme and well described in terms of 
the quasi-particles given by 
the Hartree-Bogoliubov (HB) theory 
\cite{Bogo.59}. 
The latter is expressed in terms of 
the particle and hole operators
and give rise to the collective motions
related to the density fluctuations around the equilibrium state.
The particle-hole random phase approximation (ph-RPA)
is a method for collective motions such as rotational
and vibrational motions around the equilibrium state.

The Q- and ph-RPA theories are the systematic methods for treating
phenomena of coexistence of both the correlations.$\!$
However, these theories are essentially
a harmonic oscillator approximation.$\!$ 
They can be extended to take into account
of the nonlinear terms contained in the equations of motion.$\!$
To solve such a problem, the boson expansion theory
based on the HB theory 
(abbreviated as BEHB) 
has been proposed by Belyaev and Zelevinsky 
\cite{BZ.62},
and Marumori, Yamamura and Tokunaga 
\cite{MYT.64}. 
The essence of the BEHB theory is to express
the fermion-pairs obeying the $SO(2N)$ algebra
with $N$ single-particle states
as the functions of the boson operators, 
to estimate the deviation of the fermion-pairs 
from the pure boson character.
Keeping the original idea, 
up to now, the various BEHBs have been proposed.$\!$
These methods are classified into mainly the two classes:
the first class: 
the boson representation
is constructed to reproduce the Lie algebra 
which the fermion-pairs obey; 
the second class:
in so-called the Marumori-type,
let the state vectors in the fermion Fock space correspond to
the state vectors in the boson Fock space by 
the one-to-one mapping
and the boson representation is constructed to achieve
the coincidence of  
the transition matrix-values of any physical quantity
for between the boson-state vectors 
and 
the vectors in the original fermion states.
In the former, 
there are Holstein-Primakoff (HP)-type,
Schwinger-type and Dyson-type.
The HP expresses
the fermion-pairs in terms of 
an infinite series of the boson operators,
the other two do that in terms of the finite series ones. $\!$ 
The algebraic structure governing the fermion-pairs
and the BEHB theories have been investigated.$\!$
The boson operators
proposed by  Providencia and Weneser 
and Marshalek 
\cite{PWM.68} 
are based on 
the  boson representation of particle-hole pairs forming
$SU(N)$ algebra.$\!$
Fukutome, Yamamura and 
one of the present authors S. N.
\cite{FYN.77, YN.76, Fu.81}, 
found the fermions 
to obey the 
$SO(2N\!\!+\!\!1)$, 
$\!SO(2N\!\!+\!\!2)\!$ and $\!U(N\!\!+\!\!1)\!$ 
Lie algebras, respectively.
The BEHB theory formulated by the Schwinger type
boson representation has been greatly developed by 
Fukutome and S. N. 
However, 
the BEHB theories themselves do not contain 
any scheme under which
the collective degree of freedom is selected from the
whole degrees of freedom. 

Meanwhile, there exists another approach 
to the microscopic theory of the collective motion,
that is, the time dependent Hartree-Fock (TDHF) or
time dependent Hartree-Bogoliubov (TDHB) theory.
At an early stage, the idea of the TDHF theory was
proposed by Nogami 
\cite{Nogami.55} 
and soon later by Marumori 
\cite{AM.60}
for the small amplitude vibrational motions. 
With the help of this method, we determine
the time dependence of any physical quantity,
function of the density matrix. 
The frequency of the small fluctuation 
around the static HF field 
and the equation for the frequency are the same forms 
as
those given by the RPA theory.
The RPA theory is quantal and the frequency
given by this method means the excitation energy of 
the first excited state.
Then, 
the RPA theory is a possible quantization
of the TDHF theory in the small amplitude limit.
In fact, as proved by Malshalek-Horzwarth 
\cite{MH.72},
the BEHB theory reduces to the TDHB theory
at any order, under the replacement of the boson operators
by the classical canonical variables.
Using the technique analogous to the canonical
transformation in the classical mechanics,
it is expected to obtain a scheme for choosing
the collective degree of freedom in the TDHF theory.
Historically, 
there is a stream,
whose origin
is the cranking model given by Inglis 
\cite{Inglis.56}.
The basic standpoint of this model is a kind of 
the adiabatic perturbation theories.
It starts from the assumption 
that the speed of the collective
motion is much slower than that of 
any other non-collective motions.
The TDHF theory with the adiabatic treatment (ATDHF)
was presented by Shono-Tanaka 
and 
Thouless-Valatin,
respectively,
at the early stage of the study of such an adiabatic treatment 
\cite{ST.59,TV.62}. 

At the middle of 1970, 
the TDHF theory was revived
not only for the study of anharmonic vibration 
but also for studies of heavy ion-reaction
and nuclear fission.$\!$
These problems have a common feature from the viewpoint
of the large amplitude collective motions.$\!$
In this new situation, 
the ATDHF theory
was developed mainly
by Baranger-Veneroni
\cite{BV.78},
Brink
\cite{BGV.78}
and 
Goeke-Reinhard
and 
Mukherjee-Pal
\cite{GR.78}.
Especially, the most important point 
of the ATDHF theory by Villars
\cite{Villars.77}
is  
the introduction of the concept of {\em collective path}
into the phase space.
A collective motion corresponds to a trajectory
in the phase space which moves along the {\em collective path}.
Along the same spirit, 
Holtzwarth-Yukawa
\cite{HY.74}, 
Rowe-Bassermann
\cite{RB.76}, 
and
Marumori, Maskawa, Sakata and Kuriyama gave the TDHF theory
so called the {\em maximal decoupling} method 
\cite{Ma.80}.
This theory is formulated in a canonical form framework.
The various techniques of the classical mechanics are useful
and the canonical quantization is expected. 
By solving the equation, we obtain the corrections
whose order is higher than the RPA order.
The collective sub-manifold is, in some sense,
a possible extension from the collective path.  
This theory has a potentiality to give
not only the collective modes but also the intrinsic modes.

$\!\!\!\!\!$In the BM model
the degrees of$\!$ freedom of
collective motion and 
independent-particle one are overestimated.$\!$
This problem has been inquired
from the microscopical theoretic viewpoint.$\!$
See our ``exact canonical momenta approach"
\cite{NishiProvi.14}.$\!$ 
At an early stage of the study, Marumori,Yukawa and Tanaka
\cite{MYT.55} 
and Tomonaga
\cite{Tomonaga.55,NishiProvi.15}
proposed independently the remarkable theories 
which, however, are, in certain sense,
kinematical 
because the collective motion is given {\em a priori}.
In this concern, there exist three points
to be solved dynamically:
i) to determine the microscopic structure of the collective motion,
which is the ensemble of the individual particles motion,
in relation to the dynamics under consideration,
ii) to determine independent-particle motions which should be
{\em orthogonal} to the collective motion
and 
iii) to give a coupling between these two types of motions.$\!$
The TDHF theory in the canonical form enables us to select
the collective motion in relation to the dynamics.
However, it gives us no scheme to take into
account the effect of the quasi-fermion,
because the TD Slater determinant (S-det)
contains only variables to represent the collective motion.$\!$
Then, for example, an odd-particle system cannot be described.$\!$
Along the same spirit as the TDHF spilit,
YK extended the TDHF theory 
to that on a fermion coherent state constructed 
on the TD S-det.$\!$    
The coherent state contains not only the usual
canonical variables but also the Grassmann variables.
Candlin
\cite{Candlin.56},
Berezin
\cite{Berezin.66}
and
Casalbuoni obtained
a classical image of the fermions by regarding
Grassmann variables as canonical 
\cite{Grass variable}.$\!$ 
The constraints governing the variables to remove 
the overestimated degrees of freedom were decided
under a physical consideration.
With the use of Dirac's canonical theory
\cite{Dirac.58} 
for a constrained system,
the TDHF theory was successfully solved by YK,
for a unified description of the collective and independent-particle
motions in the classical mechanics 
\cite{YK.81}.

\newpage

%%%%%%%%%%%%%%%%%%%%%%%
%                                                                      %    
%   2  Symmetry of the Evolution Equations  %
%                                                                      %
%%%%%%%%%%%%%%%%%%%%%%%

\section{Symmetry of the evolution equation}

\vspace{-0.1cm}

The historical evolution is summarized in order to present
the optimal coordinate system to describe 
the group manifold itself, based on the Lie algebra 
in which pairs of the finite-dimensional fermion obey, 
or to do dynamics on the manifold.$\!$
The boson operator in the BEHB %theory 
is the operator arising from
the coordinate system of the tangent spaces on the manifold 
in the fermion Fock space.$\!$
The BEHB %theories 
itself does not contain any scheme
under which
the collective degree of freedom can be selected from the
whole degrees of freedom.$\!$ 
While, approaches to collective motions
by the TDHF theory suggest that
the coordinate system on which the collective motions can be
described is deeply related not only to the global symmetry
of the finite-dimensional (FD) group manifold itself
but is also behind the {\em local symmetry} 
besides the Hamiltonian
\cite{Olver}.$\!$
Therefore the various collective motions are understood 
by taking the local symmetry into account.$\!$
The local symmetry may have close connection with
an infinite-dimensional (ID) Lie algebra.$\!$
The TDHF leads to nonlinear dynamics
owing to the SCF character.$\!$
However there have not been enough attempts to {\em manifestly}
understand
the collective motions in relation to the local symmetry.$\!$  
Then from the viewpoint of the symmetry of the evolution equations
in nonlinear problems,
we should study an algebro-geometric structure
toward unified a understanding of both the motions.$\!$
The first theme of our study 
is to investigate the
{\em curvature equation} as a geometric equation
to extract the collective sub-manifolds out of
the TDHF manifold.
We show that 
the zero-curvature equation 
in a particle-hole frame ($\!$PHF$\!$)
leads to the nonlinear RPA theory
which is the natural extension of the usual RPA. 
We denote the RPA/QRPA simply as RPA.
We had started from a question 
whether soliton equations exist or not in the TDHF manifold,
in spite of the difference that 
the former is described in terms of the
infinite degrees of freedom
and the latter uses finite ones.$\!$
We had met with AKNS formulation in the Inverse 
Scattering Transform ($\!$IST$\!$) method 
\cite{AKH.74} and 
the differential geometric approach 
on group manifold
developed by Sattinger
\cite{Satt.82}.$\!$
The essential point in the geometric viewpoint 
has attracted us:AKNS stands on 
 group manifold $\!SL(2)\!$
and then
integrable system 
can be explained by the zero-curvature
$\!$(integrability condition)$\!$
of connection on the corresponding
Lie group of the system.

In various approaches 
to $\!$the collective motions,$\!$ 
those from $\!$the$\!$ viewpoint of curvature were scarce.$\!$
In studying the {\em maximal decoupling} method 
by Marumori %et al. 
$\!$and$\!$ YK, 
we had an image 
for the large amplitude collective motions:$\!$
If a collective sub-manifold is the collection of
collective paths, the infinitesimal condition to switch from
a path to another 
is nothing but an {\em integrability condition} 
for the sub-manifold with respect to time $t$ 
describing the trajectory under the SCF Hamiltonian
and 
the parameters specifying any point on the sub-manifold.$\!$
However, 
the trajectory
is unable to exactly remain on the manifold.$\!$
The curvature is
able to work as a criterion of the effectiveness 
of the collective sub-manifold.$\!$
The RPA is
considered as an {\em integrability condition} 
under a linear approximation
so that the idea existing behind the theory 
can be contained in the idea of the curvature.$\!$
From the wide viewpoint of the symmetry, 
the RPA must be 
extended to any point on the manifold because 
an equilibrium state
which we select as a starting point must be equivalent 
to other points on the manifold 
among one another.$\!$
The
RPA was introduced 
as a linear approximation to treat excited states
around the equilibrium ground state
which is essentially a harmonic oscillator approximation.   
The amplitude of oscillation becomes larger
and then an anharmonicity appears so that 
we treat the anharmonicity 
by taking into account the nonlinear terms
in the equation of motion.$\!$ 
We present {\em a set of equations defining
the curvature} of collective sub-manifold
which become 
{\em a geometric equation to treat the anharmonicity}
called the {\em formal RPA}.
It is useful to understand 
the algebro-geometric meanings of
the large amplitude collective motions.
The solution procedure by the perturbative method
\cite{GPM}
suggests 
to study an ID Lie algebra working behind it. 

$\!\!\!$To develop a perturbative method, 
we study the relation $\!$between$\!$  
the method extracting {\em collective motions} 
in the SCF method
and the $\tau$-functional method ($\!$$\tau$-FM$\!$)$\!$
\cite{JM.83}(a)
constructing {\em integrable equations in solitons},$\!$
Dickey$\!$
\cite{Dickey}.$\!$ 
This is the second theme.$\!$
The relation
between the $\tau$-function $\!$and$\!$ the coherent state 
has been pointed out first  
$\!$by$\!$ D'Ariano $\!$and$\!$ Rasetti 
\cite{AR.85} 
for $\!$an$\!$ ID harmonic charged Fermi gas$\!$
\cite{Wiegmann.07}.$\!$
If we stand on the observation,$\!$
we$\!$ assert that
the SCF method presents $\!$the$\!$ theoretical $\!$scheme$\!$ for 
integrable sub-dynamics 
$\!$on$\!$ a certain ID $\!$fermion$\!$ Fock space.$\!$
Up $\!$to$\!$ now, $\!$however,$\!$ 
it $\!$has$\!$ been $\!$insufficiently$\!$ investigated 
$\!$the$\!$ relation between the SCF method in 
the {\em finite-dimensional (FD) fermion system} 
and the $\tau$-FM 
in the {\em ID fermion system}
because the description of 
dynamical fermion systems 
by them have looked very differently.$\!$
We investigate the relation between
the collective sub-manifold and various subgroup orbits
in the $\!$SCF$\!$ manifold.$\!$
Using the $\!$TDHF$\!$ theory on 
$\!U(N)\!$ group,
we study the relation
between both the methods, according to 
an essential point of the picture in the 
conventional SCF method.$\!$
To tackle this problem we have to solve
the following problems:first, 
how to imbed the FD fermion system 
into the ID one
and to rebuilt the TDHF theory on it;$\!$
second, to make sure that any algebraic mechanism working behind
the particle and collective motions, 
and any relation between the collective variable
and the spectral parameter in soliton theory remains present;$\!$
last, to$\!$ select from the SCF hamiltonian 
various subgroup orbits and to
generate a collective sub-manifold of them and$\!$
how to relate the above 
to the formal RPA in the first theme.

We obtain a unified aspect
for both the methods.$\!$
The HF theory is made 
by a variational method to
optimize the energy expectation value 
by S-det and 
to obtain a$\!$ variational equation 
for orbitals in S-det$\!$ 
\cite{RS.80}.$\!$
The particle-hole pair operators 
of fermion with $\!N\!$ 
single-particle states are closed 
under the Lie multiplication and
forms the basis of Lie algebra
$u_{\!N}\!$ 
\cite{Fu.Int.J.Quantum Chem.81}.$\!$ 
It generates 
the Thouless transformation
\cite{Th.60} 
which induces a representation (rep) of 
the corresponding $U(N)$ group.$\!$ 
The $U(N)$ canonical transformation 
changes the S-det with $m$ particles 
into another S-det.$\!$
Any S-det is obtained 
by such a transformation 
of a given reference S-det 
(Thouless theorem).$\!$
The Thouless transform provides 
an exact generator coordinate (GC)
rep of the fermion state vectors.$\!$ 
The GC is 
a $U(N)$ group and 
generating wave function (WF) is 
an independent-particle WF.$\!$ 
This is the generalized coherent state
rep (GCS rep)
\cite{Pere.72}.

In soliton theory on a group,    
transformation group to cover the solution for the soliton
equation is the ID Lie group 
whose infinitesimal generator of 
the corresponding Lie algebra is expressed as
infinite-order differential operator of 
the associative affine Kac-Moody (KM) algebra. 
The operator 
is represented in terms of the ID fermion.$\!$
A space of complex polynomial algebra 
is realized in terms of 
the {\em ID fermion} Fock space.$\!$ 
The soliton equation becomes 
nothing else than 
the differential equation defining  
the group orbit of the highest weight
vector in the ID 
Fock space $F_{\!\infty}$.$\!$
The generating WF, 
GCS rep is just the S-det 
(Thouless theorem) and provides a key
to elucidate the interrelation of 
the HF WF to the $\tau$-function
in soliton theory 
\cite{JM.83}(a).

In abstract fermion Fock spaces,   
we find the common features in the SCF method and the $\tau\!$-FM:$\!$
{\em Each solution space} is described as
the ID Grassmannian (${\bf Gr}$) 
that is the
group orbit of the corresponding vacuum state.$\!$
The former implicitly explains 
the Pl\"{u}cker relation
in terms of 
not bilinear differential equations 
defining finite-dimensional $\!{\bf Gr}\!$ 
but 
physical concept of quasi-particle and vacuum and
mathematical language of coset space-variable.$\!$
The various boson expansion methods
are built $\!$on$\!$ the Pl\"{u}cker relation $\!$to$\!$ hold 
the $\!{\bf Gr}\!$.
The latter asserts that 
the soliton equations are nothing else 
the bilinear differential equations.$\!$  
This fact gives {\em the boson rep of the Pl\"{u}cker relation}.$\!$
We study them and show 
both the methods stand on the common feature
to be the Pl\"{u}cker relation or 
the bilinear differential equation
defining the ${\bf Gr}$.

Meanwhile, 
we observe 
the two different points 
between both the methods:
(1) The former is built 
on the {\em FD} Lie algebra but the latter 
on the {\em ID} one.
(2) The former has the SCF {\em Hamiltonian} 
consisting of the fermion one-body operator, 
derived from 
the functional derivative of expectation value of 
fermion Hamiltonian by ground-state WF
but
the latter introduces artificially the one-body type
{\em fermion Hamiltonian}   
as the {\em boson mapping operator}
from the states 
on fermion Fock space to the corresponding ones 
on $\tau$-functional space ($\tau$-FS). 

Overcoming the difference due to 
the dimension of fermion,
we aim at having the {\em close connection} between 
the {\em concept of mean-field (MF) potential} 
and 
the {\em gauge of fermion}
inherent in the SCF method and to making a role of
the loop group 
\cite{PS.86} 
clear.
Through the observation, 
we make the ID fermion operators 
with finite-dimension
by the Laurent expansion in terms of a circle $S^1$.$\!$
Then with the use of 
affine KM algebra owing to
the idea of Dirac's electron-positron theory,
we rebuilt a TDHF theory in $F_\infty$.$\!$
The TDHF result in a gauge theory of fermions
and the collective motion 
(motion of MF potential) appears as 
the motion of fermion gauges with a common factor.$\!$
The physical concept of particle-hole and vacuum in 
the SCF method on $S^1$ connects to 
the {\em Pl\"{u}cker relations} according to the idea of 
Dirac, saying,
the algebraic mechanism extracting
various sub-group orbits consisting of {\em loop}
path out of the TDHF manifold
is just the {\em Hirota's bilinear form}
\cite{Hi.76}
which is $su_N(\!\in\! sl_N)$ reduction to $gl_N$ 
in the $\tau$-FM.
For the $sl_2$ case,
see 
Lepowsky-Wilson
\cite{LW.78}.$\!$
As a result, 
it is shown that the algebraic structure of
ID fermion system is also realizable
in the finite-dimension.
In such a constructive way, 
the roles of the soliton equation 
(Pl\"{u}cker relation) and the TDHF equation are made clear
and we also understand 
a non-dispersive property of 
the ${\bf Gr}$ and 
the SCF dynamics through the gauge of interacting
ID fermions.$\!$
Thus we have a 
{\em simple unified aspect for both the methods}.
 
%Studying the manifest roles of the ID
%shift operators (bosons) in $\tau$-FM,
We derive an algebraic mechanism arising
the concept of 
the particle and collective motions
and induce a close connection 
between  the collective variable 
and the spectral parameter, 
but we clear the relation
more explicitly.
For the last problem, 
further the research must be made.

As the last theme, basing on 
the above viewpoint of the TDHF theory on circle $S^1$, 
we show the algebraic mechanism bringing
both the motions.
The mechanism can be elucidated from the following points:
first, the $su_N$-condition for HF Hamiltonian; 
second, the vacuum state (highest weight vector) according to
the idea of Dirac;
and last, {\em the phase of fermion gauge is separated 
into the particle mode and  the collective one}.
We propose a new theory for unified description of 
both the motions,
beyond the static HF equation and the RPA equation 
in the usual manner.
This theory simply and clearly elucidates not only 
the collective motion
as that of the MF potential but also 
the symmetry breaking and the occurrence
of the collective motion due to the recovery of the symmetry.

Tajiri has suggested significant problems
to be inquired about why soliton solution for
classical wave equation shows fermion-like behavior in
quantum dynamics and about what symmetry $\!$is$\!$ hidden in soliton
equation$\!$ 
\cite{TW.97}.$\!$
This is a very interesting problem.
As we start from a quasi-classical dynamics in
the TDHF and do not notice the fermion systems behind them,
we inquire why their solution space is Grassmannian.$\!$   
However, 
we have not been concerned with this problem yet 
but gave a few observations to future problems.$\!$   
The anti-commutation relations
of fermions are regarded as quantal orthogonal
conditions among the canonical coordinate
variables described by the Grassmann number$\!$
\cite{YK.81}.$\!$
The tangent space has no norm as like
as that of bosons,
since 
the Grassmann number has only anti-commutative
property but no measure as a classical number.$\!$
On the contrary, $\!SO(2N\!\!+\!\!1)\!$ theory$\!$
\cite{Nishi.88,Nishi.98,FYN.77} 
describes
all degrees of freedom with respect to
pair and unpaired fermions with the use of a classical number.$\!$
Then we inquire how both systems of
boson and fermion to relate
with each other from algebro-geometric viewpoint.$\!$ 
We, further, inquire  whether the above questions 
are concerned with
the fermion-like and boson-like behaviors of solitons  
suggested by Tajiri-Watanabe 
\cite{TW.97}.$\!$
In which the idea of nonlinear superposition
principle so called imbricate series 
\cite{Boyd90}
is similar to
that of GC method in the SCF method.$\!$  
The SCF method and soliton theory
have some similar features.
However, the relation between them has been 
insufficiently investigated yet.$\!$
The explicit approaches from the viewpoint of local symmetry
are not sufficient in the SCF method.

\newpage

%%%%%%%%%%%%%%%%%%%%%%%
%                                                                      % 
%      3    Group theoretical preliminaries        %
%                                                                      %
%%%%%%%%%%%%%%%%%%%%%%%

\setcounter{equation}{0}
\section{Group theoretical preliminaries}

Let $c_\alpha$ and $c_\alpha^\dagger$
($\alpha \!=\! 1,\cdots, \!N$)
be the annihilation-creation operators of the fermion.
Owing to the anti-commutation relations
among the fermion operators,
the fermion operators 
${E}^\alpha_{~\beta}
\!=\!
c_\alpha^\dagger c_\beta
$
forming the $U(N)$ Lie algebra
$
[{E}^\alpha_{~\beta},{E}^\gamma_{~\delta}]
\!=\!
\delta_{\gamma \beta}{E}^\alpha_{~\delta}
\!-\!
\delta_{\alpha \delta}{E}^\gamma_{~\beta}
$
generates a canonical transformation $U(G)$ 
(Thouless transformation
\cite{Th.60}) 
specified by a matrix $g$
belonging to a $U(N)$ unitary group.
We decompose
the creation operator
${[c^\dag}_{\!\alpha}]$
as
$
{[c^\dag}_{\!\alpha}]
\!=\!
{[\Hat{c}^\dag_a, \Check{c}^\dag_i}]
$
and
the annihilation operator
${[c_{\!\alpha}}]$
as
$
{[c_{\!\alpha}]
\!\!=\!\!
{[\Hat{c}_a, \Check{c}_i}}]
$.$\!$
Then,
the canonical transformation is expressed as\\[-14pt]
\begin{eqnarray}
\!\!\!\!
\left.
\begin{array}{l}
[\Hat{d}^\dag,\Check{d}^\dagger]
\!=\!
U(g)
[\Hat{c}^\dag,\Check{c}^\dagger]
U^{\!-1}(g)
\!=\!
[\Hat{c}^\dag,\Check{c}^\dagger]g, ~
g
\!=\!
\left[ \!\!
\BA{cc} 
\Hat{a} & \acute{b} \\ 
\grave{b} & \Check{a} 
\EA \!\!
\right] \! ,~
gg^\dagger \!=\! g^\dagger \! g \!=\! 1_{\!N}, \\
\\[-4pt]
U^{\!-1}(g) \!\!=\!\! U^\dag(g),
U^\dag(g) \!\!=\!\! U(g^\dagger),
U(g)U(g^\prime) \!\!=\!\! U(gg^\prime),
(1_{\!N} \!:\! N\mbox{-dimensional~unit~matrix}) 
\end{array} \! 
\right\}
\label{SO(2N) Canonical Trans}
\end{eqnarray}\\[-4pt]
where 
$
[\Hat{c}^\dag,\Check{c}^\dagger]
\!\!=\!\!
[(\Hat{c}^\dag_a), (\Check{c}^\dagger_i)]$
and 
$
[\Hat{d}^\dag,\Check{d}^\dagger]
\!\!=\!\! 
[(\Hat{d}^\dag_a), (\Check{d}^\dagger_i)]$
are row vectors 
and 
$\Hat{a} \!\!=\!\! (\Hat{a}_{ab})~m \!\times\! m$ 
matrix,
$a, b \!=\! 1,2 \cdots m \mbox{(occupied~states)}$,
$\acute{b}
\!\!=\!\!
(\acute{b}_{ ai})~m \!\times\! (\! N \!-\! m)$ 
matrix,
$i \!\!=\!\! m \!+\! 1, \!\cdots\! N$,
$\grave{b} 
\!\!=\!\! 
(\grave{b}_{ia}) (\! N \!-\! m) \!\times\! m$ 
matrix
and
$\Check{a} \!=\! (\Check{a}_{ ij})$ 
$(\! N \!-\! m) \!\times\! (\! N \!-\! m)$
matrix
$i, j \!\!=\!\! m  \!+\! 1, \!\cdots\! N
~\mbox{(unoccupied~states)}$
and
$c_\alpha|0 \rangle \!=\! 0$
(vacuum),
$\Check{c}_i|0 \rangle \!=\! 0$
and
$\Hat{c}^\dagger_a|0 \rangle \!=\! 0$.
The symbols $\dagger, \star$ and \mbox{\scriptsize T} 
mean hermitian conjugate, complex conjugation 
and transposition,
respectively. 
The set of fermion operators 
$(\bar{E}^i_{~j})$ 
are constructed from the operators
$\Hat{d}^\dag$ and $\Check{d}^\dagger$ in 
(\ref{SO(2N) Canonical Trans}) 
by the same way as
the one to define the set 
$(E^\alpha_{~\beta})$.
Using the second equation of
(\ref{SO(2N) Canonical Trans3}),
the set $(\bar{E})$ 
in the quasi particle-hole frame (PHF) is transformed 
into the set $(E)$ in the original particle frame as,\\[-4pt] 
\begin{equation}
\left[ \!
\begin{array}{c}
\Hat{d} \\
\Check{d}
\end{array} \!
\right] \!
[\Hat{d}^\dagger,\Check{d}^\dagger]
\!=\!
\left[ \!
\begin{array}{cc}
\bar{E}^{\Hat{\bullet}\dagger}_{~\Hat{\bullet}} &
\bar{E}^{\Hat{\bullet}\dagger}_{~\Check{\bullet}}  \\[2pt]
\bar{E}^{\Check{\bullet} \dagger}_{~\Hat{\bullet}}  
& 
\bar{E}^{\Check{\bullet} \dagger}_{~\Check{\bullet}}
\end{array} \!\!
\right]
\!=\!
g^\dag\!
\left[ \!
\begin{array}{c}
\Hat{c} \\
\Check{c}
\end{array} \!
\right] \!
[\Hat{c}^\dagger,\Check{c}^\dagger]
g
\!=\!
g^\dagger \!
\left[ \!
\begin{array}{cc}
E^{\Hat{\bullet}\dagger}_{~\Hat{\bullet}} &
E^{\Hat{\bullet}\dagger}_{~\Check{\bullet}}  \\[2pt]
E^{\Check{\bullet} \dagger}_{~\Hat{\bullet}}  
& E^{\Check{\bullet} \dagger}_{~\Check{\bullet}}
\end{array} \!\!
\right] \!
g 
\stackrel{\mbox{PHF}}{\Longrightarrow} 
\left[  \!\!\!
\begin{array}{cc}
(hh)  &\!\! (hp) \\
 (ph)       &\!\!  (pp)
\end{array}  \!\!\!
\right] .
\label{E-E}
\end{equation}\\[-2pt]
$
\bar{E}^{\Hat{\bullet}}_{~\Hat{\bullet}}
\!\!=\!\!
(\bar{E}^a_{~b}),
\bar{E}^{\Hat{\bullet}}_{~\Check{\bullet}}
\!\!=\!\!
(\bar{E}^a_{~i}),
\bar{E}^{\Check{\bullet}}_{~\Hat{\bullet}}
\!\!=\!\!
(\bar{E}^i_{~a}),
\bar{E}^{\Check{\bullet}}_{~\Check{\bullet}}
\!\!=\!\!
(\bar{E}^i_{~j})
$
: 
$
E^{\Hat{\bullet}}_{~\Hat{\bullet}} 
\!\!=\!\!
(E^a_{~b}),
E^{\Hat{\bullet}}_{~\Check{\bullet}} 
\!\!=\!\!
(E^a_{~i}),
E^{\Check{\bullet}}_{~\Hat{\bullet}}
\!\!=\!\!
(E^i_{~a}),
E^{\Check{\bullet}}_{~\Check{\bullet}}
\!\!=\!\!
(E^i_{~j})
$
are the $m \!\times\! m,~
m \times (N - m),~(N - m) \!\times\! m,~
(N - m) \!\times\! (N - m)$
matrices,
respectively.

The HF energy functional
is defined in 
(\ref{HFenergyfunc}) 
and
the Fock operator $H_{\mbox{\scriptsize HF}}$
is given as\\[-16pt]
\begin{eqnarray}
\left.
\begin{array}{c}
E_{\mbox{\scriptsize HF}}
\!\stackrel{d}{=}\!
h_{\beta\alpha} \! {\cal Q}_{\alpha \beta}
\!+\!
{\displaystyle \frac{1}{2}}
[\gamma\alpha|\delta\beta]
{\cal Q}_{\alpha \gamma} 
{\cal Q}_{\beta \delta} ,~
{\cal Q}
\!\!\equiv\!\!
\left[ \!\!
\begin{array}{l}
\Hat{a} \\
\grave{b}
\end{array} \!\! 
\right] \!\!
\left[ \Hat{a}^\dag, \grave{b}^\dag \right] \!
\!\!=\!\!
\left[  \!\!\!
\begin{array}{cc}
\Hat{a} \Hat{a}^\dag  &\!\! 
\Hat{a} \grave{b}^\dag \\
\grave{b} \Hat{a}^\dag  &\!\! 
\grave{b} \grave{b}^\dag
\end{array}  \!\!\!
\right]
\!\!\equiv\!\!
\left[  \!\!
\begin{array}{cc}
\Hat{\cal Q}  &\!\! \acute{\cal Q} \\
 \grave{\cal Q}  &\!\!  \Check{\cal Q}
\end{array}  \!\!
\right] \! ,\\
\\  [-6pt]
H_{\mbox{\scriptsize HF}} 
\!\equiv\! 
[\Hat{c}^\dag, \Check{c}^\dag]
{\cal F} \!
\left[ \!
\begin{array}{l}
\Hat{c} \\
\Check{c}
\end{array} \! 
\right] 
\!=\!
\left(
h_{\alpha\beta}
\!+\!
[\alpha\beta|\gamma\delta]
\langle
E^\gamma_{~\delta} 
\rangle
_g
\right)
c^\dag_\alpha c_\beta ,~
{\cal F} 
\!=\! 
\left[  \!\!
\begin{array}{cc}
\Hat{F}  & \acute{F} \\
 \grave{F}       &  \Check{F}
\end{array}  \!\!
\right] ,~
{\cal F}^\dagger
\!=\! 
{\cal F} , 
\end{array} \!
\right\}
\label{HB-matrix}
\end{eqnarray}\\[-10pt]
where
${\cal Q}$
is the HF density matrix
given in detail in
(\ref{DM_Q})
and
${\cal F}$
is the HF matrix.
The set
$(F) \!=\! (F_{\alpha\beta})$ 
is a particle-hole
vacuum expectation value of the Lie operator,
expressed,
using the 
dummy index
convention to take summation over the repeated index,
as,\\[-14pt]
\begin{eqnarray}
\!\!
\begin{array}{cc}
F_{\alpha\beta}
\!\!=\!\!
\left[  \!\!\!\!
\begin{array}{cc}
\Hat{F}_{ab}
\!\!=\!\!
h_{ab}
\!\!+\!\!
[ab|cd] 
\langle E^c_{~d} \rangle_{\!g} , &\!\!\!
\acute{F}_{ai}
\!\!=\!\!
h_{ai}
\!\!+\!\!
[ai|bj] 
\langle E^b_{~j} \rangle_{\!g} \\
\\[-6pt] 
\grave{F}_{ia}
\!\!=\!\!
h_{ia}
\!\!+\!\!
[ia|jb] 
\langle E^j_{~b} \rangle_{\!g} ,&\!\!\!
\Check{F}_{ij}
\!\!=\!\!
h_{ij}
\!\!+\!\!
[ij|kl] 
\langle E^k_{~l} \rangle_{\!g} 
\end{array} \!\!\!\!
\right] \!,\!
\left[  \!\!\!\!
\begin{array}{cc}
\langle E^a_{~b} \rangle_{\!g}
\!\!=\!\!
( \grave{b}^\star \acute{b}^\dag )_{a b}, &\!\!\! 
\langle E^a_{~i} \rangle_{\!g}
\!\!=\!\!
( \grave{b}^\star \acute{b}^\dag )_{a i} \\
\\[-6pt]
 \langle E^i_{~a} \rangle_{\!g}
\!\!=\!\!
( \grave{b}^\star \acute{b}^\dag )_{i a}, &\!\!\! 
\langle E^i_{~j} \rangle_{\!g}
\!\!=\!\!
( \grave{b}^\star \acute{b}^\dag )_{i j}
\end{array}  \!\!\!\!
\right]  \!.
\end{array}
\label{HB-element}
\end{eqnarray}\\[-8pt]
The quantities 
$h_{\alpha\beta}$ and $[\alpha\beta|\gamma\delta]$
are the matrix element of the one-body Hamiltonian
and the antisymmterized one of the interaction potential.
Here,
it should be emphasized that
the TDHB equation has also been derived using
a path integral on the coset space of $SO(2N)$ group,
$\frac{SO(2N)}{U(N)}$,
by one of the present authors (S. N.)
\cite{Nishi.81,Nishi.82.83}.

\newpage

        %%%%%%%%%%%%%%%%%%%%%%%
        %                                                                      %
        %  Section 4 Integrability conditions and       %
        %                                                                      %
        %            collective sub-manifold                    %           
        %                                                                     %
        %%%%%%%%%%%%%%%%%%%%%%%

\setcounter{equation}{0} 
\section{Integrability condition and collective sub-manifold}

We consider $\!$an$\!$ evolution equation 
$\partial_t u \!\!=\!\! K(u)\!$ for $\!u(x,\!t)$.$\!$
$K(u)$ 
is $\!$an$\!$ operator acting on a function of $\!u\!$ dependent on $\!x\!$
and is expressed as
a polynomial of differential to $\!u\!$ with respect to $x$.$\!$
Regarding $\!\partial_t u \!\!=\!\! K(u)\!$ as an equation to give
$\!$an$\!$ infinitesimal transformation of function $u$,
we search for a symmetry included in 
the evolution equation.$\!$
We introduce another evolution equation, 
$\partial_s u(x,\!t,\!s) \!\!=\!\! \widehat{K} (u(x,\!t,\!s))$ 
for which we should search.$\!$
$\widehat{K}$
also consists of the polynomial operator 
of differential.$\!$
This does not necessarily mean 
$[K, \widehat{K}]\!\!=\!\! 0$
to seek $\!$for$\!$ the required symmetry
$\!$of$\!$ the present subject, $\!$e.g,,
rotational symmetry.$\!$ 
The infinitesimal condition for the existence of the symmetry
appears as the well-known integrability condition
$
\partial_s K\!(u(x,\!t,\!s)) 
\!\!=\!\! 
\partial_t \widehat{K}\!(u(x,\!t,\!s))
$$\!$
\cite{JM.83}(b).$\!$
The {\em maximal decoupling} method by Marumori
based on
the invariance principle of
Schr\"{o}dinger equation and the canonicity condition
\cite{Ma.80}
are considered to adopt the integrability condition into
%multiply 
parametrized symmetries.$\!$
This method is seen as a description of the symmetries 
of the collective submanifold
with respect to time %variable 
and collective variables,
%(other parameters else t),
in which the canonicity condition
makes the collective variables roles of
the orthogonal coordinate of a system.

In differential geometrical approach to nonlinear problem,
integrability condition is regarded as
zero curvature of connection on the corresponding
Lie group of system.$\!$
The nonlinear evolution equations, $\!$famous KdV
$\!$and$\!$ sine/sinh-Gordan equations etc.,
come from the well-known Lax equation 
\cite{Lax.68}
arisen as zero curvature 
\cite{Satt.82}.$\!$
These soliton equations describe motions of
tangent space of local gauge fields 
dependent on time $t$ and space $x$: 
equations of  Lie-valued-function 
arising from the integrability condition of the gauge field 
with respect to $t$ and $x$.$\!$
We also get a two-dimensional soliton solution
\cite{BLMP.88},$\!$
 i.e.,$\!$
{\it dromion} of the Davey-Stewartson equation
\cite{FS.88,DS.74,HH.90,HMM.91}.$\!$
On the other hand, in the SCF method,$\!$
the corresponding Lie groups 
are the unitary transformation groups of their orthonormal bases  
dependent on $t$ but not on $x$.
Although at this point the construction of means for both 
the dynamical systems are different from each other,
the RPA theory also describes a motion  of {\em tangent space} on
the group manifold.

$\!$We aim at concept of curvature unfamiliar in the nuclear physics.
The reason why is the following:
let us consider a description of motions of systems 
on a group manifold.
An arbitrary state of the system induced by a transitive
group action corresponds to any one point in the full
group parameter space and its time evolution
is given by an integral curve.
In the whole rep space adopted, assume the existence 
of $2\m$ parameters specifying 
a proper subspace in which the original motion
of the system is well defined;
saying, the existence of certain symmetry.
Suppose we start from a given point on a space
consisting of time and $2\m$ parameters, and end 
at the same point again along a closed curve.$\!$
Then we have in general value
of the group parameter different from the one at an initial point
on the subspace.
We search for some quantity characterizing such a difference.$\!$
To achieve our aim, standing on 
the differential geometrical viewpoint,
we introduce a sort of
Lagrange manner familiar to fluid dynamics to
describe collective coordinate systems
and take a one-form $\Omega$
which is linearly composed of  
the MF Hamiltonian and 
infinitesimal generators induced by 
collective variable differentials of 
certain canonical transformation.
The integrability conditions of our system read
the zero curvature.
We study curvature equations of 
the TDHF equation to determine collective sub-manifolds 
from group theoretical viewpoint.
The non-zero curvature with respect to time and 
collective variable are shown to be gradient 
of the expectation value of the residual Hamiltonian along 
the direction of the collective coordinate.$\!$ 
The set of expectation value of the zero-curvature equation 
for vacuum state function is shown to be nothing but 
the set of equations of motion, 
imposing restrictions of 
{\em weak} canonical commutation relations.
We study the relation between Marumori's
$\ll$maximal decoupled$\gg$ theory
and YK and ours.$\!$  
We discuss the role of the non zero-curvature equation 
not appearing in the former theory.
We investigate the nonlinear time evolution equation
arising from the zero-curvature equation.
We find that 
the {\em expression for equation in 
a quasi PHF
is nothing but the formal} 
RPA {\em equation} and 
show it to be regarded as the nonlinear RPA theory.
We give a solution procedure by the power expansion 
with respect to the collective variables.
Our methods are constructed manifesting itself
the structure of group under consideration 
in order to make easy to understand physical
characters at any point on the group manifold.

%%%%%%%%%%%%%%%%%%%
%                                                         %
%    4.1   A Lagrange-like manner      %
%                                                         %
%%%%%%%%%%%%%%%%%%%

\subsection{Lagrange-like manner}

The $U\!(\!N\!)\!$ WF 
$\!|\phi(\Breve{g})\rangle$
is constructed by a transitive action of $\!U\!(\!N\!)$
canonical transformation 
$U^{\!-1}\!(\Breve{g})\!$ on $\!|0 \rangle$;
$\!|\phi(\Breve{g})\rangle$
\!\!=\!\!
$U^{\!-1}\!(\Breve{g})|0 \rangle$,
$\!\Breve{g} \!\!\in\!\! U\!(\!N\!)$. 
In the conventional TDHF,
TD WF $|\phi(\Breve{g})\rangle$
is given through the group $U(N)$ parameters 
$\Hat{a},\grave{b},\acute{b}$ and $\Check{a}$.
They characterize the TD SC
mean HF field $F$ whose dynamical change induces
the collective motions of many fermion systems.
As in the TDHF by Marumori et al.$\!$ 
\cite{Ma.80} 
$\!$and$\!$
YK$\!$ 
\cite{YK.81}, 
we introduce the following TD 
$U(N)$ canonical transformation (TR) to derive 
collective motions
within the TDHF framework:\\[-8pt]
\begin{equation}
U(\Breve{g})
 \!=\!
U[\Breve{g} (\Breve{\Lambda}(t), \Breve{\Lambda}^\star (t))],~
\Breve{g}(\Breve{\Lambda}(t), \Breve{\Lambda}^\star (t))
\!=\!
\left[ \!\!
\BA{cc} 
\Hat{a} (\Breve{\Lambda}(t), \Breve{\Lambda}^\star (t))& 
\acute{b} (\Breve{\Lambda}(t), \Breve{\Lambda}^\star (t))\\
\\[-6pt] 
\grave{b} (\Breve{\Lambda}(t), \Breve{\Lambda}^\star (t))& 
\Check{a} (\Breve{\Lambda}(t), \Breve{\Lambda}^\star (t))
\EA \!\!
\right] \! ,
\label{Collective trans}
\end{equation}\\[-6pt]
where a set of TD complex functions  
$\!
(\!\Breve{\Lambda}(t), \!\Breve{\Lambda}^\star(t)\!)
\!\!=\!\!
(\Breve{\Lambda}_\n (t), \!\Breve{\Lambda}_\n^\star(t);
\n\!=\!1,\!\cdots\!,\!\m)
$\!
associated with the collective motions specifies 
the group parameters
$\Breve{g}$.$\!$
The number $\m$ is assumed to be much smaller
than the degree of the $U(N)$ Lie algebra, which
means there exist only a few {\em collective degrees of freedom}.$\!$
The $U(\Breve{g})$ is a natural extension of the method 
in the simple TDHF case to the TDHF on the
$2\m$-dimensional ($2\m$D) collective sub-manifold$\!$
\cite{NK.84.87}.
For our aim,
taking $\!$a$\!$ {\em Lagrange-like manner},
it is convenient 
to introduce complex parameters
$
\{\Lambda, \Lambda^\star\}
$
which
are regarded as local coordinates to specify any point
of $2\m$D collective sub-manifold. 
Instead of 
$(\Breve{\Lambda}(t), \Breve{\Lambda}^\star(t))$,
we use the functions of 
$\{\Lambda, \Lambda^\star\}$ and $t$,\\[-20pt]
\begin{eqnarray}
\begin{array}{rl}
\Breve{\Lambda}_\n (t)
\!=\!
\Breve{\bf \Lambda}_\n 
(\Lambda,\Lambda^\star,t) ,~ 
\Breve{\Lambda}^\star_\n (t)
\!=\!
 \Breve{\bf \Lambda}^\star_\n 
(\Lambda,\Lambda^\star,t)  .
\end{array} \!
\label{Initial variable}
\end{eqnarray}\\[-20pt]
This means that the set
$\!\{\! \Lambda, \!\Lambda^\star\!, \!t \!\} \!\!\in\!\! \epsilon^3\!$
is mapped into the set
$\!\{ \!\Lambda, \!\Lambda^\star \!\} \!\!\in\!\! \epsilon^2\!$,i.e.,$\!$
$\!\{ \!\Breve{\Lambda}, \!\Breve{\Lambda}^\star\!, \!t \!\} 
\!\!\rightarrow\!\!
\{ \!\Breve{\Lambda}, \!\Breve{\Lambda}^\star \!\}\!$
through $\!$the$\!$ functions
$\!\{ \!\Breve{\bf \Lambda}, \!\Breve{\bf \Lambda}^\star \!\}$.$\!$
This means that 
one is considering at $t$ a coordinate set $S_t$ 
that depends on $t$ 
and is labeled by the set of pairs 
$\!\{ \!(\!\Lambda, \!\Lambda^\star\!) \!\}$.$\!$ 
At $t^\prime$ the set $S_{t^\prime}$ is still labeled 
by the same set of pairs 
$\{ \!(\!\Lambda, \!\Lambda^\star\!) \!\}$.$\!$ 
However, the state which is labeled by the pair 
$(\!\Lambda, \!\Lambda^\star\!)$  
at $t$ is different from the state which is labeled by the same pair 
$(\!\Lambda, \!\Lambda^\star\!)$ 
at $t^\prime$.$\!$
This manner seems very analogous to the one
founded by Lagrange in the fluid dynamics.$\!$
An invariant subspace labeled by the parameters
$\{ \!\Breve{\Lambda}, \!\Breve{\Lambda}^\star \!\}$
is being assumed.$\!$
The collective subspace becomes defined 
only when such a TR as the one
which has been formulated is considered, but this is not enough.$\!$ 
Now comes the definition of collective subspace.$\!$ 
The invariant subspace is the collective subspace
if the evolution of the physical system determined by the TDHF theory is such
that its coordinates in $S_t$, namely $(\Lambda, \!\Lambda^\star)$, 
do not change with $t$.$\!$
Using 
(\ref{Initial variable}),
$U(N)$ canonical TR is rewritten as
$
U\!(\Breve{g})
\!\!=\!\!
U[g (\Lambda,\!\Lambda^\star\!,\!t)] 
\!\!\in\!\! U(N)
$.$\!$
Notice that the functional form 
$\Breve{g} (\Breve{\Lambda}(t),\!\Breve{\Lambda}^\star (t))$
in
(\ref{Collective trans})
changes into another functional form
$g (\Lambda,\!\Lambda^\star\!,\!t )$
due to the Lagrange-like manner.$\!$ 
This %manner 
enables us to take a one-form
$\Omega$,
%linearly 
composed of the infinitesimal
generators induced by time 
and collective variable differentials
$(\partial_t,\!\partial_\Lambda,\!\partial_{\Lambda^\star})$
of 
canonical TR 
$U[g (\Lambda,\!\Lambda^\star\!,\!t)]$.
Introducing the
one-form $\Omega$,
we search for collective path
and collective Hamiltonian
separated from other remaining degrees
of freedom.

%%%%%%%%%%%%%%%%%%
%                                                      %
%    4.2   Integrability conditions    %
%                                                     %
%%%%%%%%%%%%%%%%%%

\subsection{Integrability conditions}

Following YK 
\cite{YK.81}, 
we define Lie-algebra-valued
infinitesimal generators
for collective Hamiltonian $H_c$
and collective coordinates 
$(O_\n^\dagger,O_\n)~(\n=1,\cdots,\m) $ 
in collective sub-manifolds
as follows:\\[-6pt]
\begin{equation}
H_c 
\stackrel{d}{=}
(i\hbar\partial_t U(g))U^{-1}(g) , 
\label{Hc def}
\end{equation}
\vspace{-0.8cm} 
\begin{eqnarray}
\begin{array}{cc}
\begin{array}{c}
~~~~O_\n^\dagger 
\stackrel{d}{=}
(i\partial_\Lambda{}_\n U(g))U^{-1}(g) ,
~~~~O_\n\stackrel{d}{=}
(i\partial_{\Lambda_\n^\star}U(g))U^{-1}(g) ,
\end{array} \!\!
\end{array}
\label{Oc def}
\end{eqnarray}\\[-16pt]
We abbreviate
$g(\Lambda,\!\Lambda^\star\!,\!t)$
simply as $g$.
In the TDHF theory, 
the Lie-algebra-valued infinitesimal
generators are expressed by the trace form as\\[-14pt]
\begin{eqnarray}
\!\!\!\!
H_c 
\!=\!
-
\mbox{Tr}
\left\{
(i\hbar\partial_t g.g^\dagger) \!
\left[ \!
\begin{array}{cc}
E^{\Hat{\bullet}\dagger}_{~\Hat{\bullet}} &
E^{\Hat{\bullet}\dagger}_{~\Check{\bullet}}  \\[2pt]
E^{\Check{\bullet} \dagger}_{~\Hat{\bullet}}  
& E^{\Check{\bullet} \dagger}_{~\Check{\bullet}}
\end{array} \!\!
\right] \!
\right\} 
\!=\!
[\Hat{c}^\dagger,\Check{c}^\dagger]
(i\hbar\partial_t g.g^\dagger) \!
\left[ \!
\begin{array}{c}
\Hat{c} \\
\Check{c}
\end{array} \!
\right] \! ,
\label{Hc Tr-express}
\end{eqnarray}
\vspace{-0.60cm}
\begin{eqnarray}
\left.
\begin{array}{c}
O_\n^\dagger
\!=\!
-
\mbox{Tr}
\left\{
(i\partial_{\Lambda_n}g.g^\dagger) \!
\left[ \!
\begin{array}{cc} 
E^{\Hat{\bullet}\dagger}_{~\Hat{\bullet}} &
E^{\Hat{\bullet}\dagger}_{~\Check{\bullet}}  \\[2pt]
E^{\Check{\bullet} \dagger}_{~\Hat{\bullet}}  
& E^{\Check{\bullet} \dagger}_{~\Check{\bullet}}
\end{array} \!\! 
\right] 
\right\} 
\!=\!
[\Hat{c}^\dagger,\Check{c}^\dagger] 
(i\partial_{\Lambda_\n}g.g^\dagger) \!
\left[ \!
\begin{array}{l}
\Hat{c} \\
\Check{c}
\end{array} \!
\right], \\
\\[-6pt]
O_\n 
\!=\!
-
\mbox{Tr}
\left\{
(i\partial_{\Lambda_n^\star}g.g^\dagger) \!
\left[ \!
\begin{array}{cc} 
E^{\Hat{\bullet}\dagger}_{~\Hat{\bullet}} &
E^{\Hat{\bullet}\dagger}_{~\Check{\bullet}}  \\[2pt]
E^{\Check{\bullet} \dagger}_{~\Hat{\bullet}}  
& E^{\Check{\bullet} \dagger}_{~\Check{\bullet}}
\end{array} \!\!
\right] 
\right\} 
\!=\!
[\Hat{c}^\dagger,\Check{c}^\dagger]
(i\partial_{\Lambda_\n^\star}g.g^\dagger) \!
\left[ \!
\begin{array}{l}
\Hat{c} \\
\Check{c}
\end{array} \!
\right] .
\end{array}
\right \}
\label{Oc Tr-express}
\end{eqnarray}\\[-8pt]
$\!\!\!\!\!\!\!\!\!\!\!\!$
The direct derivation of 
(\ref{Oc Tr-express}) 
using
(\ref{Hc def})
and
(\ref{Oc def})
is given in detail 
in Appendix A.
We, however, for the moment,
remove the factor
$e^\Xi$
in
(\ref{Hc Tr-express})
and
(\ref{Oc Tr-express}). 
Multiplying $|\phi(g)\rangle$
on the both sides of 
(\ref{Hc def}) and (\ref{Oc def}),
we get a set of equations on the $U(N)$ Lie algebra:\\[-20pt]
\begin{eqnarray}
\begin{array}{cc}
\left.
\begin{array}{c}
\!\!\!\!D_t|\phi(g)\rangle
\!\stackrel{d}{=}\!
(i\hbar\partial_t \!\!-\!\! H_c)
|\phi(g)\rangle \!=\! 0 ,\\
\\[-10pt]
D_{\Lambda_\n}|\phi(g)\rangle
\!\stackrel{d}{=}\!
(\partial_{\Lambda_\n} \!\!+\!\! iO_\n^\dagger)
|\phi(g)\rangle \!=\! 0 ,~
D_{\Lambda_\n^\star}|\phi(g)\rangle
\!\stackrel{d}{=}\!
(\partial_{\Lambda_\n^\star} \!\!+\!\! iO_\n)
|\phi(g)\rangle \!=\! 0 .
%( \n \!\!=\!\! 1,\!\cdots\!, \!\m) .
\end{array}
\right\}
\end{array}
\label{D-t D-Lambda}
\end{eqnarray}\\[-12pt]
We regard these equations 
(\ref{D-t D-Lambda}) 
as
partial differential equations for $|\phi(g)\rangle$.
In order to discuss the conditions for that the
differential equation 
(\ref{D-t D-Lambda}) 
can be solved,
the mathematical method well known as integrability 
conditions is useful.
For this aim, we take the following one-form $\Omega$
linearly composed of the infinitesimal generators 
(\ref{Hc def})
and 
(\ref{Oc def}):\\[-10pt]
\begin{equation}
\Omega
=
H_c \cdot dt+\hbar O_\n^\dagger\cdot d\Lambda_\n
+
\hbar O_\n\cdot d\Lambda_\n^\star .
\label{one-form eq}
\end{equation}\\[-16pt]
With the aid of the one-form $\Omega$, the integrability
conditions of the system read\\[-8pt]
\begin{equation}
C\stackrel{d}{=}d\Omega-\Omega\wedge \Omega=0 ,
\label{Curvature=0}
\end{equation}\\[-16pt]
where $d$ and $\wedge$ denote the exterior differentiation
and the exterior product, respectively.
From the differential geometrical viewpoint,
the quantity $\cal C$ defined in the above means
the curvature of a connection.
Then the integrability conditions is interpreted as 
the vanishing of the curvature of the connection
$(D_t, D_\Lambda{}_\n, D_\Lambda{}_\n^\star)$.
The detailed structure of the curvature is calculated to be\\[-16pt]
\begin{eqnarray}
\BA{c}
C 
=
C_{t,\Lambda_\n}d\Lambda_\n\wedge dt
+
C_{t,\Lambda_\n^\star}d\Lambda_\n^\star\wedge dt
+
C_{\Lambda_{\n^\prime},\Lambda_\n^\star}
d\Lambda_\n^\star\wedge d\Lambda_{\n^\prime}  \nonumber\\
\nonumber\\[-10pt]
+
\frac{1}{2} C_{\Lambda_{\n^\prime},\Lambda_\n}
d\Lambda_\n\wedge d\Lambda_{\n^\prime}
+
\frac{1}{2} C_{\Lambda_{\n^\prime}^\star,\Lambda_\n^\star}
d\Lambda_\n^\star\wedge d\Lambda_{\n^\prime}^\star ,
\EA
\label{C detail}
\end{eqnarray}\\[-14pt]
where\\[-30pt]
\begin{eqnarray}
\left.
\begin{array}{r}
C_{t,\Lambda_\n} 
\stackrel{d}{=}
[D_t,D_{\Lambda_\n}]
=
i\hbar\partial_tO_\n^\dagger
-
i\partial_{\Lambda_\n}H_c
+
[O_\n^\dagger,H_c] ,    \\
\\[-12pt]
C_{t,\Lambda_\n^\star} 
\stackrel{d}{=}
[D_t,D_{\Lambda_\n^\star}]
=
i\hbar\partial_tO_n
-
i\partial_{\Lambda_\n^\star} H_c
+
[O_\n,H_c]~,\\
\\[-12pt]
C_{\Lambda_{\n^\prime},\Lambda_\n^\star} 
\stackrel{d}{=}
[D_\Lambda{}_{\n^\prime},D_{\Lambda_\n^\star}]
=
i\partial_{\Lambda_{\n^\prime}}O_\n
-
i\partial_{\Lambda_\n^\star}O_{\n^\prime}^\dagger 
+
[O_\n,O_{\n^\prime}^\dagger] ,      \\
\\[-12pt]
C_{\Lambda_{\n^\prime},\Lambda_\n} 
\stackrel{d}{=} 
[D_{\Lambda_{\n^\prime}},D_{\Lambda_\n}]
=
i\partial_{\Lambda_{\n^\prime}}O_\n^\dagger
-
i\partial_{\Lambda_\n} O_{\n^\prime}^\dagger 
+
[O_\n^\dagger,O_{\n^\prime}^\dagger] , \\
\\[-12pt]
C_{\Lambda_{\n^\prime}^\star,\Lambda_\n^\star} 
\stackrel{d}{=}
[D_{\Lambda_{\n^\prime}^\star},D_{\Lambda_\n^\star}]
=
i\partial_{\Lambda_{\n^\prime}^\star} O_\n
-
i\partial_{\Lambda_\n^\star} O_{\n^\prime} 
+
[O_\n,O_{\n^\prime}] .
\end{array}
\right\}
\label{C.. explicit}
\end{eqnarray}\\[-10pt]
The vanishing of curvature $\cal C$ means 
$C_{\bullet , \bullet}=0$.
For basic study of differential geometry,
see the famous textbooks
\cite{Schutz.80}
and
\cite{HouHou.97}.

Finally using the expressions
of 
(\ref{Hc Tr-express}) 
and 
(\ref{Oc Tr-express}),
we get the following set of Lie-algebra-valued equations 
as the integrability conditions 
of the partial differential equations 
(\ref{D-t D-Lambda}).\\[-18pt]
\begin{eqnarray}
\!\!\!\!
\left.
\begin{array}{c}
C_{t,\Lambda_\n}
\!=\!
[\Hat{c}^\dagger,\Check{c}^\dagger] 
{\cal C}_{t,\Lambda_\n} \!\!
\left[ \!
\begin{array}{l}
\Hat{c} \\[-2pt]
\Check{c}
\end{array} \!
\right] \! ,~
C_{t,\Lambda_\n^{\star}}
\!=\!
[\Hat{c}^\dagger,\Check{c}^\dagger] 
{\cal C}_{t,\Lambda_\n^{\star}} \!\!
\left[ \!
\begin{array}{l}
\Hat{c} \\[-2pt]
\Check{c}
\end{array} \!
\right] \! , \\
\\[-8pt]
C_{\Lambda_{\n^\prime},\Lambda_\n^\star}
\!\!=\!\!
[\Hat{c}^\dagger,\Check{c}^\dagger] 
{\cal C}_{\Lambda_{\n^\prime},\Lambda_\n^\star} \!\!
\left[ \!
\begin{array}{l}
\Hat{c} \\[-2pt]
\Check{c}
\end{array} \!
\right] \! ,~~
C_{\!\Lambda_{\n^\prime},\Lambda_\n}
\!\!=\!\!
[\Hat{c}^\dagger,\Check{c}^\dagger] 
{\cal C}_{\!\Lambda_{\n^\prime},\Lambda_\n} \!\!
\left[ \!
\begin{array}{l}
\Hat{c} \\[-2pt]
\Check{c}
\end{array} \!
\right] , 
C_{\!\Lambda_{\n^\prime}^{\star},\Lambda_\n^{\star}}
\!\!=\!\!
[\Hat{c}^\dagger,\Check{c}^\dagger] 
{\cal C}_{\!\Lambda_{\n^\prime}^{\star},\Lambda_\n^{\star}} \!\!
\left[ \!
\begin{array}{l}
\Hat{c} \\[-2pt]
\Check{c}
\end{array} \!
\right], 
\end{array} \!\!
\right \}
\label{C.. ope}
\end{eqnarray}\\[-14pt]
where\\[-26pt]
\begin{eqnarray}
\!\!\!\!\!\!
\left.
\begin{array}{c}
{\cal C}_{t,\Lambda_\n}
\!=\!
i\hbar\partial_t\theta^{\dag }_\n
\!-\!
i\partial_{\Lambda_\n}
{\cal F}_c
\!+\!
[\theta^{\dag }_\n,{\cal F}_c],~
{\cal C}_{t,{}_{\Lambda_\n^{\star}}}
\!=\!
i\hbar\partial_t\theta_\n
\!-\!
i\partial_{\Lambda_\n^{\star}}
{\cal F}_c
\!+\!
[\theta_\n,{\cal F}_c],  \\
\\[-8pt]
{\cal C}_{\!\Lambda_\n, \Lambda_\n^{\star}}
\!=\!
i\partial_{\Lambda_{\n^\prime}}\theta_\n
\!-\!
i\partial_{\Lambda_\n^\star}\theta_{\n^\prime}^\dagger 
\!+\!
[\theta_\n,\theta_{\n^\prime}^\dagger] ,\\
\\[-8pt]
{\cal C}_{\Lambda_{\n^\prime},\Lambda_\n}
\!=\!
i\partial_{\Lambda_\n} 
\theta^{\dag }_\n
\!-\!
i\partial_{\Lambda_\n}
\theta^{\dag }_{\n^\prime} 
\!+\!
[\theta^{\dag }_\n,\theta^{\dag }_{\n^\prime}] ,~
{\cal C}_{\Lambda_{\n^\prime}^{\star},\Lambda_\n^{\star}}
\!=\!
i\partial_{\Lambda_\n^{\star}} 
\theta_\n
\!-\!
i\partial_{\Lambda_\n^{\star}}
\theta_{\n^\prime} 
\!+\!
[\theta_\n,\theta_{\n^\prime}] .
\end{array} \!\!\!
\right \}
\label{C..c-val}
\end{eqnarray}\\[-10pt]
The quantities ${\cal F}_c$, 
$\theta_\n^\dagger$ and $\theta_\n$
are defined through partial differential equations,\\[-22pt]
\begin{eqnarray}
\begin{array}{ll}
i\hbar\partial_t 
g
\!=\!
{\cal F}_c g ,& 
\end{array}
\label{TD Fc}
\end{eqnarray}
\vspace{-1.1cm}
\begin{eqnarray}
\begin{array}{ll}
\begin{array}{c}
i\partial_{\Lambda_\n} g
\!=\!
\theta_\n^\dagger g ,~
i\partial_{\Lambda_\n^\star}g
\!=\!
\theta_\n g .
%(\n \!=\!1,\cdots,\m) .
\end{array} 
\end{array}
\label{SD theta}
\end{eqnarray}\\[-22pt]
The quantities 
${\cal C}_{\bullet , \bullet}$
are naturally regarded as the curvature
of the connection on the group manifold.
The reason becomes clear
if we take the following procedure quite parallel
with the above one:
Starting from
(\ref{TD Fc}) 
and 
(\ref{SD theta}),
we are led to a set of partial differential equations 
on the 
\mbox{$U(N)$ Lie group $g$}
%for $(\n \!=\!1,\cdots, \m)$
as \\[-26pt]
\begin{eqnarray}
\begin{array}{cc}
\left.
\begin{array}{cc}
&{\cal D}_t g
\stackrel{d}{=}
(i\hbar\partial_t - {\cal F}_c) g
\!=\!
0 ,\\
\\[-12pt]
&{\cal D}_{\Lambda_\n} g
\stackrel{d}{=}
(\partial_{\Lambda_\n}+i\theta_\n^\dagger) g
\!=\!
0 ,~
{\cal D}_{\Lambda_\n^\star} g
\stackrel{d}{=}
(\partial_{\Lambda_\n^\star}+i\theta_\n)
g =0 .
\end{array}
\right \}
\end{array}
\label{D-t D-Lambda g}
\end{eqnarray}\\[-10pt]
The curvature $C_{\bullet ,\bullet}
(
\stackrel{d}{=}
[{\cal D}_\bullet, {\cal D}_\bullet]
)$ of the connection 
$\!
(
{\cal D}_t,\!{\cal D}_{\Lambda_\n},\!{\cal D}_{\Lambda_\n^\star} 
)\!$
is shown to be equivalent to 
the quantity $C_{\bullet , \bullet}$ in 
(\ref{C..c-val}).
The above set of the Lie-algebra-valued equations (\ref{C.. ope})
evidently leads us to putting all the curvatures
$C_{\bullet , \bullet}$ 
equal to be zero.

On the other hand,
TDHF Hamiltonian 
(\ref{HB-matrix}),i.e.,$\!$
Hamiltonian on $\!U(N)\!$
WF space, is represented 
in the same form as 
(\ref{Hc def}) \\[-10pt]
\begin{equation}
H_{\mbox{\scriptsize HF}}
\!=\!
(i\partial_t U(g^\prime))U^{-1}(g^\prime) , 
\label{HB-Hc}
\end{equation}\\[-14pt]
where $g^\prime$ is any point on the $U(N)$
group manifold.
The RHS of 
(\ref{HB-Hc}) 
is transformed into the same form as 
(\ref{Hc Tr-express}). 
This fact leads us to
the well-known TDHF equation,\\[-14pt]
\begin{equation}
i\hbar\partial_t g^\prime
\!=\!
{\cal F}g^\prime .
\label{G'-t}
\end{equation}\\[-18pt]
The TDHF Hamiltonian is decomposed
into two components at a reference
point $g^\prime \!\!=\!\! g$ on the group manifold
as\\[-26pt]
\begin{eqnarray}
\begin{array}{c}
H_{\mbox{\scriptsize HF}}|_{U(g\prime)
=
U(g)}
\!=\!
H_c \!+\! H_{res} ,~~
{\cal F}|_{{g^\prime}=g}
\!=\!
{\cal F}_c \!+\! {\cal F}_{res} ,
\end{array}
\label{Hc+Hres}
\end{eqnarray}\\[-20pt]
where the second parts 
$H_{res}$
and 
${\cal F}_{res}$
mean residual components 
extracted out of a {\em well-defined}
collective sub-manifold for which
we must search now.
For our aim, let us introduce another curvatures
${\cal C}^\prime_{t,\Lambda_{}\n}$
and
${\cal C}^\prime_{t,\Lambda_{}\n^{\star}}$
with the same forms as those 
in 
(\ref{C.. ope}) 
and 
(\ref{C..c-val}),
in which 
instead of ${\cal F}_c$,
it is replaced
by 
${\cal F}|_{{g^\prime}=g}
~(=\!{\cal F}_c\!+\!{\cal F}_{res})$.
The vacuum expectation values
of another Lie-algebra-valued curvatures 
are easily calculated %for $\n \!=\!1,\!\cdots\!,\m$
as \\[-18pt]
\begin{eqnarray}
\begin{array}{c}
~~
\langle C_{t,\Lambda_{}\n} \rangle \!=\! 
\langle C_{t,\Lambda_{}\n^{\star}} \rangle \!=\! 0,~~~
{\langle {\cal C}^\prime_{t,\Lambda} \rangle}_{\!\!g} 
\!=\!
-i\partial_{\Lambda} {\langle H_{res} \rangle}_{g} 
~\mbox{and}~
{\langle C^\prime_{t,\Lambda_\n^\star} \rangle}_g 
\!=\!
-i\partial_{\Lambda_\n^\star} {\langle H_{res} \rangle}_g ,\\
\end{array} 
\label{C-res}
\end{eqnarray}\\[-30pt]
\begin{equation}
\BA{c}
{\langle H_{res} \rangle}_g
\!=\!
-
\mbox{Tr} \!
\left\{
g \!
\left[ \!\!\!
\begin{array}{cc}
1_m &\!\!\!\!\! 0 \\
\\[-14pt]
0 &\!\!\!\!\! -1_{\!N \!-\! m}
\end{array} \!\!\!
\right] \!
g^\dagger
({\cal F}_g -i\hbar\partial_t g.g^\dagger)
\right\} ,
\EA
\label{Hres-expectation}
\end{equation}\\[-6pt]
The first Eq. of
(\ref{C-res})
is derived through the relations
$
i \hbar
\dot{\theta}_\n
\!=\!
i\partial_{\Lambda_\n}
{\cal F}_c
$
and
$
i \hbar
\dot{\theta}_\n^{\dagger }
\!=\!
i\partial_{\Lambda_{}\n^{\star}}
{\cal F}_c
$
which
are obtained by
(\ref{TD Fc}) 
and
(\ref{SD theta}).$\!$
Eqs.(\ref{C-res}) and (\ref{Hres-expectation})
mean that values of 
${\cal C}^\prime_{t,\Lambda_{}\n}$
and
${\cal C}^\prime_{t,\Lambda_{}\n^{\star}}$
are the {\em gradient} of energy of the residual Hamiltonian
in $2\m$D manifold.
Suppose the existence of a {\em well-defined\/} 
collective sub-manifold.
Then it is not so wrong to deduce the following result:
The energy of the residual Hamiltonian has
almost constant value on the collective sub-manifold, \\[-18pt]
\begin{eqnarray}
\delta_g
\langle H_{res} \rangle_g 
\!\cong\! 
0 ,~~
\partial_{\Lambda_\n}
\langle H_{res} \rangle_g 
\!\cong\! 
0
 ~\mbox{and}~
\partial_{\Lambda_\n^{\star}}
\langle H_{res} \rangle_g 
\!\cong\! 
0 ,
\label{Hres nearly 0 for Lambda}
\end{eqnarray}\\[-18pt]
where $\delta_g$ denotes the $g$-variation, regarding $g$
as functions of $(\Lambda, \Lambda^\star)$ and $t$.
It may be achieved if we should determine $g$
(collective path) and ${\cal F}_c$
(collective Hamiltonian)
through the auxiliary quantity $(\theta, \theta^\dagger)$
so as to satisfy 
$H_c+const.\!=\!H_{\mbox{\scriptsize HF}}$ 
as far as possible.
Then putting ${\cal F}_c \!=\! \cal F$ in 
(\ref{C..c-val}), 
we seek for $g$ and ${\cal F}_c$ satisfying\\[-18pt] 
\begin{eqnarray}
\begin{array}{c}
{\cal C}_{t,\Lambda_\n}
\!\cong\!
0 
~\mbox{and}~
{\cal C}_{t,\Lambda_\n^{\star}}
\!\cong\!
0 ,~~
{\cal C}_{\Lambda_{\n^\prime},\Lambda_\n^\star}
\!=
0 , ~              
{\cal C}_{\!\Lambda_{\n^\prime}, \Lambda_\n}
\!=
0 
~\mbox{and}~
{\cal C}_{\Lambda_{\n^\prime}^{\star},
\Lambda_\n^{\star}}
\!=
0 .     
\end{array}
\label{C nearly 0}
\end{eqnarray}\\[-20pt]
The set of the equations ${\cal C}_{\bullet,\bullet}
\!\!=\!
0\!$
makes an essential role to determine the collective sub-manifold
in the TDHF method.
The set of the Eqs. 
(\ref{C nearly 0}) 
and 
(\ref{D-t D-Lambda g}) 
becomes an our geometric equation 
for describing the collective motions,
under the restrictions given later, i.e.,
(\ref{restriction}).

%%%%%%%%%%%%%%%%%%%%%%%%%%%%%%%
%                                                                                                 %
%    4.3   The correspondence of the Lagrange-like manner      %
%                                  to the usual one                                      %
%                                                                                                 %
%%%%%%%%%%%%%%%%%%%%%%%%%%%%%%%

\subsection{Correspondence of Lagrange-like manner 
to the usual one}

If we hope to describe collective motions 
through TD complex variables 
$(\Breve{\Lambda}(t), \Breve{\Lambda}^\star(t))$, 
we must know 
explicit forms of $\Breve{\Lambda}$ 
and $\Breve{\Lambda}^\star$ in 
(\ref{Initial variable})
in terms of
$(\Lambda,  \Lambda^\star)$ and $t$. For this aim, 
it is necessary to discuss the correspondence of 
the Lagrange-like manner to the usual one.

We define the Lie-algebra-valued infinitesimal
generators of collective sub-manifolds as\\[-16pt]
\begin{eqnarray}
\left.
\begin{array}{c}
\Breve{O}_\n^\dagger 
\stackrel{d}{=}
(i\partial_{\Breve{\Lambda}_n} \!
U(\Breve{g}))U^{-1}(\Breve{g}) ,\\
\\[-6pt]
\Breve{O}_\n
\stackrel{d}{=}
(i\partial_{\Breve{\Lambda}_\n^\star}
U(\Breve{g}))U^{-1}(\Breve{g}) , 
\end{array}
\right\}
\label{checkOc def}
\end{eqnarray}\\[-10pt]
whose form is the same as the one in 
(\ref{Oc def}).
In order to guarantee the variables 
$\Breve{\Lambda}_\n(=\Breve{\Lambda}_\n (t))$ and  
$\Breve{\Lambda}_\n^\star(=\Breve{\Lambda}_\n^\star(t))$
to be canonical, following Marumori
\cite{Ma.80}
and YK
\cite{YK.81},
we set up the following expectation values with use of
the $U(N)$ WF $|\phi(\Breve{g})\rangle$:\\[-16pt]
\begin{eqnarray}
\left.
\begin{array}{l}
\langle\phi(\Breve g)|i\partial_{{\Breve \Lambda}_\n} 
|\phi(\Breve g)\rangle
\!=\!
\langle\phi(\Breve g)|{\Breve O}_\n^\dagger 
|\phi(\Breve g)\rangle
\!=\!
~~i
{\Breve \Lambda}_\n^\star ~,\\
\\[-6pt]
\langle\phi(\Breve g)|i\partial_{{\Breve \Lambda}_\n^\star} 
|\phi(\Breve g)\rangle
\!=\!
\langle\phi(\Breve g)|{\Breve O}_\n 
|\phi(\Breve g)\rangle
\!=\!
-i
{\Breve \Lambda}_\n .     
\end{array}
\right \}
\label{canonikal condition}
\end{eqnarray}\\[-10pt]
The above relation leads us to the {\em week }
canonical commutation relation\\[-16pt]
\begin{eqnarray}
\left.
\begin{array}{c}
\begin{array}{c}
\langle\phi(\Breve{g})|[\Breve{O}_\n, 
\Breve{O}_{\n^\prime}^\dagger]
|\phi(\Breve{g})\rangle
\!=\!
\delta_{\n \n^\prime} ,\\
\\[-6pt]
\langle\phi(\Breve{g})|[\Breve{O}_\n^\dagger, 
\Breve{O}_{\n^\prime}^\dagger]
|\phi(\Breve{g})\rangle
\!=\!
0 ,~
\langle\phi(\Breve{g})|[\Breve{O}_\n, 
\Breve{O}_{\n^\prime}]
|\phi(\Breve{g})\rangle
\!=\!
0 ,
\end{array}
\end{array} \!
\right\}
\label{weak orthgonality}
\end{eqnarray}\\[-10pt]
proof of which is, shown by Marumori
and YK
\cite{Ma.80,YK.81},
due to the integrability conditions.

Using 
(\ref{D-t D-Lambda}),
the collective Hamiltonian $H_c$ and 
the infinitesimal generators $O_\n^\dagger$ and  $O_\n$
in the Lagrange-like manner are expressed by
$\Breve{O}_\n^\dagger$ and  $\Breve{O}_\n$
in the usual manner as follows:\\[-16pt]
\begin{eqnarray}
\left.
\begin{array}{c}
H_c
\!=\!
\hbar\partial_t 
\Breve{\bf \Lambda}_\n \!
\Breve{O}_\n^\dagger
\!+\!
\hbar\partial_t 
{\bf \Breve{\Lambda}_\n^\star}
\Breve{O}_\n ,\\
\\[-6pt]
O_\n^\dagger
\!=\!
\partial_{\Lambda_\n} \!
\Breve{\bf \Lambda}_{\n^\prime}
\Breve{O}_{\n^\prime}^\dagger
\!+\!
\partial_{\Lambda_\n} \!
{\Breve{\bf \Lambda}_{\n^\prime}^\star}
\Breve{O}_{\n^\prime} ,~
O_\n
\!=\!
\partial_{\Lambda_\n^\star}
\Breve{\bf \Lambda}_{\n^\prime}
\Breve{O}_{\n^\prime}^\dagger
\!+\!
\partial_{\Lambda_\n^\star}
{\Breve{\bf \Lambda}_{\n^\prime}^\star}
\Breve{O}_{\n^\prime} .
\end{array}
\right \}
\label{Lag H_ ets to the usual one}
\end{eqnarray}\\[-10pt]
Substituting  
(\ref{Lag H_ ets to the usual one})
into 
(\ref{C.. explicit}),
it is carried out to evaluate the expectation value
of Lie-algebra-valued curvature 
$C_{\bullet,\bullet}$ by the $\!U(N)\!$ WF
$|\phi[\Breve{g}(\Breve{\Lambda}(t),\Breve{\Lambda}^\star(t))]
\rangle(\!=\! |\phi[g(\Lambda,\Lambda^\star, t)]\rangle\/)$.  
A {\em weak\/} integrability condition requiring 
the expectation value 
$\langle
\phi(\Breve{g})|C_{\bullet,\bullet}|\phi(\Breve{g})
\rangle
\!\!=\!\!
0$ 
yields partial differential equations with aid of
the quasi particle-hole vacuum property, 
$
d|\phi(g)\rangle
\!=\!
0
$:\\[-16pt]
\begin{eqnarray}
\left.
\begin{array}{l}
\partial_{\Lambda_\n}\Breve{\bf \Lambda}_{\n^\prime}
\partial_t\Breve{\bf \Lambda}_{\n^\prime}^\star
\!-\!
\partial_{\Lambda_\n}\Breve{\bf \Lambda}_{\n^\prime}^\star
\partial_t\Breve{\bf \Lambda}_{\n^\prime}
\!=\!
-
\mbox{Tr}
\{\slashed{\cal Q}(g)[\theta_\n^\dagger,{\cal F}_c /\hbar]\} ,\\
\\[-6pt]
\partial_{\Lambda_\n^\star}\Breve{\bf \Lambda}_{\n^\prime}
\partial_t\Breve{\bf \Lambda}_{\n^\prime}^\star
\!-\!
\partial_{\Lambda_\n^\star}\Breve{\bf \Lambda}_{\n^\prime}^\star
\partial_t\Breve{\bf \Lambda}_{\n^\prime}
\!=\!
-
\mbox{Tr}
\{\slashed{\cal Q}(g)[\theta_\n^\dagger,{\cal F}_c /\hbar]\} ,
\end{array}
\right\}
\label{Lagrange bracket  and weak condition 1}
\end{eqnarray}
\vspace{-0.8cm}
\begin{eqnarray}
\left.
\begin{array}{l}
\partial_{\Lambda_\n^\star}
\Breve{\bf \Lambda}_{\n^{\prime\prime}}
\partial_{\Lambda_{\n^\prime}}
\Breve{\bf \Lambda}_{\n^{\prime\prime}}^\star
\!-\!
\partial_{\Lambda_\n^\star}
\Breve{\bf \Lambda}_{\n^{\prime\prime}}^\star
\partial_{\Lambda_{\n^\prime}}
\Breve{\bf \Lambda}_{\n^{\prime\prime}}
\!=\!
\mbox{Tr}
\{\slashed{\cal Q}(g)[\theta_\n,\theta_{\n^\prime}^\dagger]\} ,\\
\\[-10pt]
\partial_{\Lambda_\n}
\Breve{\bf \Lambda}_{\n^{\prime\prime}}
\partial_{\Lambda_{\n^\prime}}
\Breve{\bf \Lambda}_{\n^{\prime\prime}}^\star
\!-\!
\partial_{\Lambda_\n}
\Breve{\bf \Lambda}_{\n^{\prime\prime}}^\star
\partial_{\Lambda_{\n^\prime}}
\Breve{\bf \Lambda}_{\n^{\prime\prime}}
\!=\!
\mbox{Tr}
\{\slashed{\cal Q}(g)[\theta_\n^\dagger,\theta_{\n^\prime}^\dagger]\} ,\\
\\[-8pt]
\partial_{\Lambda_\n^\star}
\Breve{\bf \Lambda}_{\n^{\prime\prime}}
\partial_{\Lambda_{\n^\prime}^\star}
\Breve{\bf \Lambda}_{\n^{\prime\prime}}^\star
\!-\!
\partial_{\Lambda_\n^\star}
\Breve{\bf \Lambda}_{\n^{\prime\prime}}^\star
\partial_{\Lambda_{\n^\prime}^\star}
\Breve{\bf \Lambda}_{\n^{\prime\prime}}
\!=\!
\mbox{Tr}
\{\slashed{\cal Q}(g)[\theta_\n,\theta_{\n^\prime}]\} ,
\end{array}
\right \}
\label{Lagrange bracket  and weak condition 2}
\end{eqnarray}\\[-10pt]
where another $U(N)$ HF density matrix $\slashed{\cal Q}(g)$
is defined as\\[-18pt]
\begin{eqnarray}
~~~~
\begin{array}{lll}
\slashed{\cal Q}(g)
\stackrel{d}{=}
g \!
\left[ \!\!\!
\begin{array}{cc}
1_{\!m} &\!\!\!\! 0 \\
\\[-10pt]
0 &\!\!\!\! -1_{\!N \!-\! m}
\end{array} \!\!\!
\right] \!
g^\dagger 
\!\equiv\!
\left[ \!\!
\begin{array}{cc}
\Hat{\slashed{\cal Q}}&\! \acute{\slashed{\cal Q}} \\
\grave{\slashed{\cal Q}} &\! \Check{\slashed{\cal Q}}
\end{array} \!\!
\right] \!,
&\slashed{\cal Q}^\dagger \!=\! \slashed{\cal Q} ,
&\slashed{\cal Q}^{2} \!=\! 1_{\!N} ,
\end{array}
\label{density R(g)}
\end{eqnarray}\\[-14pt]
using the expression for $g$,
(\ref{matrix-g}),
the density matrix $\slashed{\cal Q}$
is explicitly expressed as\\[-16pt]
\begin{eqnarray}
\!\!\!\!
\begin{array}{ll}
\slashed{\cal Q}
&\!\!
\!\!=\!\!
\left[ \!\!\!
\BA{cc} C(\xi ) e^{ \upsilon } &\!\!\!\! 
-S^\dag \! (\xi ) e^{ \upsilon^\star } \\[2pt] 
S(\xi )e^{ \upsilon }  &\!\!\!\! 
\tilde{C}(\xi ) e^{ \upsilon^\star }
\EA \!\!\!\!
\right] \!\!
\left[ \!\!\!
\begin{array}{cc}
1_m &\!\!\!\! 0 \\[2pt]
0 &\!\!\!\! -1_{\!N \!-\! m}
\end{array} \!\!\!
\right] \!\!
\left[ \!\!\!
\BA{cc} e^{- \upsilon }C(\xi )  &\!\!\! 
e^{- \upsilon } S^\dag \! (\xi ) \\[2pt]
-e^{- \upsilon^\star } \!\! S (\xi )  &\!\!\! 
e^{-\upsilon^\star } \! \tilde{C}(\xi ) 
\EA \!\!\!
\right]
%\!\!=\!
%2 \!\!
%\left[ \!\!\!
%\BA{cc} C(\xi ) ^2 &\!\!\! 
%C(\xi ) S^\dag \! (\xi )  \\[2pt] 
%S(\xi ) C(\xi )  &\!\!\! 
%S(\xi ) S^\dag \! (\xi )
%\EA \!\!\!
%\right] \!
%\!-\!
%1_{\!N} \\
%\\[-12pt]
%&\!\!
\!\!=\!
2
{\cal Q}
 \! - \!
1_N %,~
%(\mbox{due~to}~
%(\ref{DM_Q})) ,
%~\mbox{then},
%\left[ \!\!
%\begin{array}{cc}
%\Hat{\slashed{\cal Q}}&\! \acute{\slashed{\cal Q}} \\
%\grave{\slashed{\cal Q}} &\! \Check{\slashed{\cal Q}}
%\end{array} \!\!
%\right]
\!\equiv\!
\left[ \!\!\!
\begin{array}{cc}
2 \Hat{\cal Q}\!-\!1_{\!m}&\!\!\!\! 2 \acute{\cal Q} \\
2 \grave{\cal Q} &\!\!\!\! 2 \Check{\cal Q}\!-\!1_{\!N \!-\! m}
\end{array} \!\!\!
\right] \! .
\end{array} 
\label{densitymatrix-Q}
\end{eqnarray}\\[-10pt]
The parameters involved in $g$ are functions
 of the complex variables 
$(\Lambda,\Lambda^\star)$
and $t$.
In the derivation of 
(\ref{Lagrange bracket  and weak condition 1}) 
and 
(\ref{Lagrange bracket  and weak condition 2}),
we have used the transformation property 
(\ref{E-E}),
the trace formulas appeared in 
(\ref{Hc Tr-express})
and 
(\ref{Oc Tr-express}) 
and 
the differential formulae are expressed as\\[-18pt]
\begin{eqnarray}
\!\!\!\!\!\!\!\!
%\left.
\begin{array}{c}
\langle\phi(\!\Breve{g}\!)|
i\partial_{\Breve{\Lambda}_{\n^\prime}} \!
\Breve{O}_{\!\n}
|\phi(\!\Breve{g}\!)\rangle
\!\!=\!\!
-
\delta_{\!\n \n^\prime} ,%~
\langle\phi(\!\Breve{g}\!)|
i\partial_{\Breve{\Lambda}_{\n^\prime}
^\star}\Breve{O}_{\!n}^\dagger \!
|\phi(\!\Breve{g}\!)\rangle
\!\!=\!\!
\delta_{\!\n \n^\prime} ,%~%\\
%\\[-14pt]
\langle\phi(\!\Breve{g}\!)|
i\partial_{\Breve{\Lambda}_{\n^\prime}} \!
\Breve{O}_{\!\n}
^\dagger|\phi(\!\Breve{g}\!)\rangle
\!\!=\!\!
0 ,%~
\langle\phi(\!\Breve{g}\!)|
i\partial_{\Breve{\Lambda}_{\n^\prime}^\star} \!
\Breve{O}_{\!\n}
|\phi(\!\Breve{g}\!)\rangle
\!\!=\!\!
0  .
\end{array} \!\!
%\right \}
%(\mbox{due~to
%(\ref{canonikal condition})
%$\!$-$\!$
%(\ref{weak orthgonality})
%$\!\!$}). 
\label{expectation of Partial O etc}
\end{eqnarray}\\[-20pt]
This is the consequence of canonicity condition
 and 
{\em weak\ } canonical commutation relation.

Through the above procedure, finally,
we get the correspondence of Lagrange-like manner
to the usual one.$\!$
We have no unknown quantities in the RHS of
(\ref{Lagrange bracket  and weak condition 1}) 
and 
(\ref{Lagrange bracket  and weak condition 2}),
if we could completely solve the geometric equations.$\!$
Then we become able to know in principle
the explicit forms of the functions 
$\Breve{\bf \Lambda}$
and $\Breve{\bf \Lambda}^\star$ 
in terms of
$(\Lambda,\Lambda^\star)$ and $t$ 
by solving the set of partial differential equations
(\ref{Lagrange bracket  and weak condition 1}) 
and 
(\ref{Lagrange bracket  and weak condition 2}).$\!$
However we should take enough notice of the roles
different from each other made by
(\ref{Lagrange bracket  and weak condition 1}) 
and
(\ref{Lagrange bracket  and weak condition 2}), 
respectively,
to construct the solutions.
It turns out that the LHS in 
(\ref{Lagrange bracket  and weak condition 2})
has a close connection with Lagrange bracket
often appeared in analytical dynamics.
Since we have set up from the outset 
the canonicity conditions 
to guarantee the complex variables 
$(\Breve{\Lambda},\Breve{\Lambda}^\star)$
in the usual manner being canonical, 
the functions
$\Breve{\bf \Lambda}$ and $\Breve{\bf \Lambda}^\star$
in 
(\ref{Initial variable}) 
can be interpreted as functions giving
a canonical transformation from 
$(\Breve{\Lambda},\Breve{\Lambda}^\star)$ to 
another complex variables 
$(\Lambda,\Lambda^\star)$
in the Lagrange-like manner.
From this interpretation, 
we see that 
requirement of the canonical invariance imposes the following
restrictions on the RHS of 
(\ref{Lagrange bracket  and weak condition 2}):\\[-20pt]
\begin{eqnarray}
\begin{array}{c}
\begin{array}{c}
\mbox{Tr} 
\{ 
\slashed{\cal Q}(g)
[\theta_\n,\theta_{\n^\prime}^\dagger] 
\}
\!=\!
\delta_{\n \n^{\prime}} ,~
 \mbox{Tr}
\{ 
\slashed{\cal Q}(g)
[\theta_\n^\dagger,\theta_{\n^\prime}^\dagger] 
\}
\!=\!
0 ,~
\mbox{Tr} \!
\left\{ 
\slashed{\cal Q}(g)[\theta_\n,\theta_{\n^\prime}] 
\right\}
\!=\!
0 .
%(\n,\n^\prime \!=\! 1,\!\cdots\!, \m).
\end{array}
\end{array}
\label{restriction}
\end{eqnarray}\\[-18pt]
It is quite self-evident that, combining 
(\ref{Lagrange bracket  and weak condition 2}) 
with 
(\ref{restriction}),
we get the Lagrange bracket for the canonical transformation 
from 
$(\Breve{\Lambda},\Breve{\Lambda}^\star)$ 
to 
$(\Lambda,\Lambda^\star)$.
From the correspondence arguments,
it is reasonable to add the restrictions (\ref{restriction})
to our geometric equations
in order to describe the collective motion in terms
of the canonical coordinate variables.

We have studied the integrability conditions of
the TDHF equation to determine the collective
sub-manifolds from the group theoretical viewpoint.
Our idea lies in 
the adoption of the Lagrange-like manner to
describe the collective coordinates.
It should be noted that
the variables are nothing but the parameters
to describe the symmetry of the TDHF equation.
Introducing the one-form, 
we gave the integrability conditions,
the vanishing of the curvatures of the connection,
expressed as the Lie-algebra-valued equations.$\!$
The TDHF Hamiltonian 
$H_{\mbox{\scriptsize HF}}$ 
is decomposed
into a collective Hamiltonian $H_c$ and a
residual one $H_{res}$.$\!$ 
To search for 
{\em well-defined} collective sub-manifold, 
we demand that the expectation value
of the curvature is minimized to satisfy 
$\!H_{res}\!\!\cong\!\!$ const./  
$\!\!\!H_c$ \!\!+\!\! const. 
\!\!=\!\! 
$H_{\mbox{\scriptsize HF}}\!$ 
as far as possible.
We also
require the Lagrange bracket for the usual 
and Lagrange-like variables.$\!$
Our geometric equation together with 
the requirment describes 
the collective motion of a system.
It is expected to work well {\em over a wide range}
of physics
beyond the $U(N)$ RPA as the small amplitude limit, 
with certain boundary condition.

%%%%%%%%%%%%%%%%%%%%
%                                                            %
%        4.4    Geometric equation           %
%                in quasi-particle frame        %
%                                                            %
%%%%%%%%%%%%%%%%%%%%

\subsection{Geometric equation in quasi PHF}

To extract the collective sub-manifolds,
we demand the zero curvature of the connection 
on the TDHF manifold.
We transform the geometric equation 
into the equation in the quasi PHF.
The TDHF Hamiltonian
in the quasi PHF 
is expressed as\\[-18pt]
\begin{eqnarray}
\begin{array}{c}
U^{-1}(g)H_{\mbox{\scriptsize HF}}U(g) 
\!=\!
[\Hat{d}^\dagger\!,\Check{d}^\dagger]
{\cal F} \!
\left[ \!\!
\begin{array}{c}
\Hat{d} \\[-2pt]
\Check{d}
\end{array} \!\!
\right] 
\!=\!
[\Hat{c}^\dagger\!,\Check{c}^\dagger]
g^\dag \! {\cal F} \!g \!\!
\left[ \!\!
\begin{array}{c}
\Hat{c} \\[-3pt]
\Check{c}
\end{array} \!\!
\right] 
\!, ~
g^\dagger \! {\cal F} \! g
\!=\!
{\cal F}_{o}  .
\end{array}
\label{HB-matrix in qf}
\end{eqnarray}\\[-14pt]
The infinitesimal generators of collective sub-manifolds
and integrability conditions for
$\n \!=\! 1,\!\cdots\!,\m$,
expressed as 
Lie-algebra-valued equation, 
are rewritten in the quasi PHF as,\\[-20pt]
\begin{eqnarray}
\begin{array}{c}
H_c 
\!=\! 
[\Hat{c}^\dagger\!,\Check{c}^\dagger]
{\cal F}_{o-c} \!
\left[ \!\!
\begin{array}{c}
\Hat{c} \\[-2pt]
\Check{c}
\end{array} \!\!
\right] \! ,~
{\cal F}_{o-c}
\!=\! 
\left[  \!\!
\begin{array}{cc}
\Hat{F}_{o-c}  & \acute{F}_{o-c} \\
 \grave{F}_{o-c}       &  \Check{F}_{o-c}
\end{array}  \!\!
\right] \! , 
\end{array}
\label{Hc def on qp-frame}
\end{eqnarray}
\vspace{-0.7cm}
\begin{eqnarray}
\begin{array}{c}
~~~
O_\n^\dagger 
\!=\! 
[\Hat{c}^\dagger\!,\Check{c}^\dagger]
\theta_{o-\n}^\dagger \!
\left[ \!\!
\begin{array}{c}
\Hat{c} \\[-3pt]
\Check{c}
\end{array} \!\!
\right] \! ,~
(\theta^\dagger_{o-\n} \!\equiv\! g^\dagger \theta_\n^\dagger g) ,~
O_\n 
\!=\! 
[\Hat{c}^\dagger\!,\Check{c}^\dagger]
\theta_{o-\n} \!
\left[ \!\!
\begin{array}{c}
\Hat{c} \\[-3pt]
\Check{c}
\end{array} \!\!
\right] \! ,
\end{array} \!
~
(\theta_{o-\n} \!\equiv\! g ^\dagger \theta_\n g) ,
\label{Oc def on qp-frame}
\end{eqnarray}
\vspace{-0.9cm}
\begin{eqnarray}
\!\!\!\!
\left.
\begin{array}{c}
U^{-1}(g)C_{t,\Lambda_\n}U(g)
\!=\!
[\Hat{c}^\dagger\!,\Check{c}^\dagger]
{\cal C}_{o-t,\Lambda_\n} \!\!
\left[ \!\!
\begin{array}{c}
\Hat{c} 
\\[-3pt]
\Check{c}
\end{array} \!\!
\right]
\!=\!
U^{-1}(g)C_{t,\Lambda_\n^{\star}}U(g)
\!=\!
[\Hat{c}^\dagger\!,\Check{c}^\dagger]
{\cal C}_{o-t,\Lambda_\n^{\star}} \!\!
\left[ \!\!
\begin{array}{c}
\Hat{c} 
\\[-3pt]
\Check{c}
\end{array} \!\!
\right]
\!=\!
0 , \\
\\[-12pt]
U^{\!-1}(g)C_{\Lambda_{\n^\prime},\Lambda_\n^\star}U(g)
\!=\!
[\Hat{c}^\dagger\!,\Check{c}^\dagger]
{\cal C}_{o-\Lambda_{\n^\prime},\Lambda_\n^\star} \!\!
\left[ \!\!
\begin{array}{c}
\Hat{c} \\[-3pt]
\Check{c}
\end{array} \!\!
\right]
\!=\!
0 ,\\
\\[-12pt]
U^{\!-1}(g)
C_{\!\Lambda_{\n^\prime},\Lambda_\n} \!
U(g)
\!\!=\!\!
[\Hat{c}^\dagger\!,\Check{c}^\dagger]
{\cal C}_{o-\Lambda_{\n^\prime},\Lambda_\n} \!\!
\left[ \!\!
\begin{array}{c}
\Hat{c} \\[-3pt]
\Check{c}
\end{array} \!\!
\right]
\!\!=\!\!
%\\[-12pt]
U^{\!-1}(g)
C_{\Lambda_{\n^\prime}^{\star},\Lambda_\n^{\star}} \!
U(g)
\!\!=\!\!
[\Hat{c}^\dagger\!,\Check{c}^\dagger]
{\cal C}_{o-\Lambda_{\n^\prime}^{\star},
\Lambda_\n^{\star}} \!\!
\left[ \!\!
\begin{array}{c}
\Hat{c} \\[-3pt]
\Check{c}
\end{array} \!\!
\right]
\!\!=\!
0 , 
\end{array}
\right \}
\label{C.. ope on qp-frame}
\end{eqnarray}\\[-16pt]
\\[-32pt]
\begin{eqnarray}
\begin{array}{cc}
\left.
\begin{array}{c}
U^{\!-1}(g){\cal C}_{t,\Lambda_\n}U(g)
\!=\!
i\hbar\partial_t\theta^{\dag }_{o-\n}
\!-\!
i\partial_{\Lambda_\n}{\cal F}_{o-c}
\!-\!
[\theta^{\dag }_{o-\n},{\cal F}_{o-c}] ,\\
\\[-6pt]
U^{\!-1}(g){\cal C}_{t,{}_{\Lambda_\n^{\star}}}U(g)
\!=\!
i\hbar\partial_t\theta_{o-\n}
\!-\!
i\partial_{\Lambda_\n^{\star}}{\cal F}_{o-c}
\!-\!
[\theta_{o-\n},{\cal F}_{o-c}] ,\\
\\[-6pt]
U^{\!-1}(g)
{\cal C}_{\Lambda_{\n^\prime},\Lambda_\n^\star}
U(g)
\!=\!
i\partial_{\Lambda_{\n^\prime}}\theta_{o-\n}
\!-\!
i\partial_{\Lambda_\n^\star} \! \theta_{o-\n^\prime}^\dagger 
\!-\!
[\theta_{o-\n},\theta_{o-\n^\prime}^\dagger] ,\\
\\[-8pt]
U^{\!-1}(g)
{\cal C}_{\!\Lambda_{\n^\prime},\Lambda_\n}
U(g)
\!=\!
i\partial_{\Lambda_\n} 
\theta^{\dag }_{o-\n}
\!-\!
i\partial_{\Lambda_\n}
\theta^{\dag }_{o-\n^\prime} 
\!-\!
[\theta^{\dag }_{o-\n},\theta^{\dag }_{o-\n^\prime}] ,\\
\\[-6pt]
U^{\!-1}(g)
{\cal C}_{\Lambda_{\n^\prime}^{\star},\Lambda_\n^{\star}}
U(g)
\!=\!
i\partial_{\Lambda_\n^{\star}} 
\theta_{o-\n}
\!-\!
i\partial_{\Lambda_\n^{\star}}
\theta_{o-\n^\prime} 
\!-\!
[\theta_{o-\n},\theta_{o-\n^\prime}] ,
\end{array}
\right\}
\end{array}
\label{explicit C.. on qp-frame}
\end{eqnarray}\\[-10pt]
where 
all the curvatures ${\cal C}_{o-\bullet ,\bullet}$ should be vanished.
The
${\cal F}_{o-c},~\theta_{o-\n}^\dagger
$ and 
$\theta_{o-\n}$
satisfy the  partial differential equations
on the $U(N)$ Lie group manifold $g$,\\[-20pt]
\begin{eqnarray}
-i\hbar\partial_t g^\dagger
=
{\cal F}_{o-c} g^\dagger  ,~ 
-i\partial_{\Lambda_\n} g^\dagger
=
\theta_{o-\n}^\dagger g^\dagger ,~
-i\partial_{\Lambda_\n^\star} g^\dagger
=
\theta_{o-\n} g^\dagger . 
\label{SD theta on qp-frame}
\end{eqnarray}\\[-22pt]
The TDHF Hamiltonian in a quasi PHF is decomposed into collective
and residual ones as\\[-22pt]
\begin{eqnarray}
\begin{array}{rl}
U^{-1}(g)H_{\mbox{\scriptsize HF}}U(g)
\!=\!
H_c\!+\!H_{res},
&{\cal F}_o
\!=\!
{\cal F}_{o-c}\!+\!{\cal F}_{o-res},
\end{array}
\label{Hc+Hres in qf}
\end{eqnarray}\\[-22pt]
at a reference point $\!g\!$ on $\!U\!(N)\!$ group.$\!$
The curvatures 
${\cal C}^\prime_{t,\Lambda_\n}\!$
and
$\!{\cal C}^\prime_{t,{}_{\Lambda_\n^{\star}}}\!$ 
introduced previously 
have the same forms as those in 
(\ref{C.. ope on qp-frame}) 
and 
(\ref{explicit C.. on qp-frame})
except that ${\cal F}_{o-c}$ is replaced 
by ${\cal F}_o$.
Then the corresponding curvatures      
${\cal C}^\prime_{o-t,\Lambda_\n}$
and
${\cal C}^\prime_{o-t,\Lambda_\n^{\star}}$ 
are also decomposed into as,\\[-20pt]
\begin{eqnarray}
\begin{array}{r}
{\cal C}^\prime_{o-t,\Lambda_\n}
\!=
{\cal C}^c_{o-t,\Lambda_\n}
\!+
 {\cal C}^{res}_{o-t,\Lambda_\n},~
{\cal C}^\prime_{o-t,{}_{\Lambda_\n^{\star}}}
\!=
{\cal C}^c_{o-t,\Lambda_\n^{\star}}
\!+
 {\cal C}^{res}_{o-t,\Lambda_\n^{\star}} .
\end{array}
\label{C..=C..c+C..res}
\end{eqnarray}\\[-22pt]
The collective curvatures 
${\cal C}^c_{o-t,\Lambda_\n}$
and
${\cal C}^c_{o-t,{\Lambda_\n^{\star}}}$ 
arising from ${\cal F}_{o-c}$ are given in the same forms as
the ones in 
(\ref{explicit C.. on qp-frame}).$\!$
The residual ones 
${\cal C}^{res}_{o-t,\Lambda_\n}\!$
and
${\cal C}^{res}_{o-t,{\Lambda_\n^{\star}}}$
arising from ${\cal F}_{o-res}$ are defined as\\[-20pt]
\begin{eqnarray}
\begin{array}{r}
{\cal C}^{res}_{o-t,\Lambda_\n}
\!\!=
\!-\!
i\partial_{\Lambda_\n}{\cal F}_{o-res}
\!\!-\!\!
[\theta^{\dag }_{o-\n},{\cal F}_{o-res}] ,~
{\cal C}^{res}_{o-t,{\Lambda_\n^{\star}}}
\!\!=
\!-\!
i\partial_{\Lambda_\n^{\star}}{\cal F}_{o-res}
\!\!-\!\!
[\theta_{o-\n},{\cal F}_{o-res}].
\end{array}
\label{explicit C..res}
\end{eqnarray}\\[-20pt]
Using 
(\ref{SD theta on qp-frame}) 
and 
(\ref{Hc+Hres in qf}), Lie-algebra-valued forms of
curvatures are calculated to be\\[-20pt]
\begin{eqnarray}
\begin{array}{c}
C^{res}_{t,\Lambda_\n}
=
-i\partial_{\Lambda_\n} \!\!
<\!\!H_{res}\!\!>_g ,
~\mbox{and}~
C^{res}_{t,{}_{\Lambda_\n^{\star}}}
=
-i\partial_{\Lambda_\n^{\star}} \!\!
<\!\!H_{res}\!\!>_g .
\end{array}
\label{C-res=grad H}
\end{eqnarray}\\[-22pt]
Supposing there exist a well-defined collective sub-manifold
satisfying 
(\ref{SD theta on qp-frame}),
we should demand that the following curvature are made equal
to zero:\\[-8pt]
\begin{equation}
\BA{c}
{\cal C}^c_{o-t,\Lambda_\n}
=
{\cal C}^c_{o-t,\Lambda_\n^{\star}}
=
0 , ~
{\cal C}_{o-\Lambda_{\n^\prime},\Lambda_\n^\star}
=
0 ,~
{\cal C}_{o-\Lambda_{\n^\prime},\Lambda_\n}
=
0 
~\mbox{and}~
{\cal C}_{o-\Lambda_{\n^\prime}^{\star},
\Lambda_\n^{\star}}
=
0 ,
\EA
\label{C..=0 on collective}
\end{equation}\\[-16pt]
the first equation 
of which lead us to the Lie-algebra-valued relations,\\[-20pt]
\begin{eqnarray}
\begin{array}{r}
{\cal C}^\prime_{t,\Lambda_\n}
=
C^{res}_{t,\Lambda_\n}
=
-i\partial_{\Lambda_\n} \!\!
<\!\!H_{res}\!\!>_g 
~\mbox{and}~
{\cal C}^\prime_{t,\Lambda_\n^{\star}}
=
C^{res}_{t,{}_{\Lambda_\n^{\star}}}
=
-i\partial_{\Lambda_\n^{\star}} \!\!
<\!\!H_{res}\!\!>_g .
\end{array}
\label{C..'=C..res=grad Hres}
\end{eqnarray}\\[-22pt]
Then the curvatures 
${\cal C}^\prime_{t,\Lambda_\n}\!\!$
and
$\!{\cal C}^\prime_{t,\Lambda_\n^{\star}}$
are regarded as the {\em gradient} of
quantum-mechanical potentials due to the existence 
of the residual Hamiltonian $H_{res}$ on 
the collective sub-manifolds.
The potentials become almost flat 
on the collective sub-manifolds, 
\mbox{$H_{\mbox{\scriptsize HF}} \!\!=\!\! H_c \!+\! const.$}, 
if the proper subspace determined
is almost invariant subspace of full TDHF Hamiltonian.
This collective subspace is the almost 
degenerate eigenstate of the residual Hamiltonian.$\!$
The residual curvature at a point 
on the invariant subspace 
is extremely small.$\!$ 
The way of extracting 
collective sub-manifolds out of the TDHF manifold
is possible by the minimization of 
the residual curvature.
Therefore, 
a deeper insight into 
(\ref{C..'=C..res=grad Hres}) 
becomes necessary.

Finally, restrictions to assure the Lagrange bracket
for the usual collective variables and Lagrange-like ones are 
transformed into the forms represented 
in quasi PHF as\\[-14pt]
\begin{eqnarray}
\!\!\!\!
\begin{array}{rl}
\begin{array}{c}
\mbox{Tr} \!\!
\left\{ \!
\left[ \!\!\!\!
\begin{array}{cc}
1_m &\!\!\!\!\!\! 0 \\[-2pt]
 0 &\!\!\!\!\!\! -1_{\!N\!-\!m}
\end{array} \!\!\!\!
\right] \!\!
[\theta_{\!o-\n},\theta_{\!o-\n^\prime}^\dagger] \!
\right\}
\!\!=\!\!
\delta_{\n \n^{\prime}} ,
\mbox{Tr} \!\!
\left\{ \!
\left[ \!\!\!\!
\begin{array}{cc}
1_m &\!\!\!\!\!\! 0 \\[-2pt]
 0 &\!\!\!\!\!\! -1_{\!N\!-\!m}
\end{array} \!\!\!\!
\right] \!\!
[\theta_{\!o-\n},\theta_{\!o-\n^\prime}]
\right\}
\!\!=\!\!
\mbox{Tr} \!\!
\left\{ \!
\left[ \!\!\!\!
\begin{array}{cc}
1_m &\!\!\!\!\!\! 0 \\[-2pt]
 0 &\!\!\!\!\!\! -1_{\!N\!-\!m}
\end{array} \!\!\!\!
\right] \!\!
[\theta^{\dag }_{\!o-\n},\theta^{\dag }_{\!o-\n^\prime}] \!
\right\}
\!\!=\!\!
0 .
\end{array}
\end{array}
\label{restriction in qf}
\end{eqnarray}%\\[-34pt]

%%%%%%%%%%%%%%%%%%%%%%%%%%%%%
%                                                                                           %
%    4.5   The transformation of the Lagrange-like picture  %
%                                to the usual one                                 %
%                                                                                          %
%%%%%%%%%%%%%%%%%%%%%%%%%%%%%

\subsection{Transform of Lagrange-like picture 
to picture in quasi PHF}

We discuss how the Lagrange-like picture
is transformed into the picture in the quasi PHF.
We regard any point on the collective sub-manifold
as an initial point (initial value).$\!$
Suppose a time evolution of system
with various initial value.$\!$ 
Then we have a relation
between the Lagrange-like way
and the usual one by through a similar way to 
the previous one as\\[-20pt]
\begin{eqnarray}
\begin{array}{c}
{\cal F}_{o-c}
\!=\!
\hbar\partial_t\Breve{\bf \Lambda}_\n
\Breve{\theta}_{o-\n}^\dagger
\!+\!
\hbar\partial_t{\Breve{\bf \Lambda}_\n^\star}
\Breve{\theta}_{o-\n} ,
\end{array}
\label{transfered Fo-c} 
\end{eqnarray} \\[-44pt]
\begin{eqnarray}
\begin{array}{c}
\theta_{o-\n}^\dagger
\!=\!
\partial_{\Lambda_\n}\Breve{\bf \Lambda}_{\n^\prime}
\Breve{\theta}_{o-\n^\prime}^\dagger
\!+\!
\partial_{\Lambda_n}{\bf \Breve{\Lambda}_{\n^\prime}^\star}
\Breve{\theta}_{o-\n^\prime} ,~
\theta_{o-\n}
\!=\!
\partial_{\Lambda_\n^\star}\Breve{\bf \Lambda}_{\n^\prime}
\Breve{\theta}_{o-\n^\prime}^\dagger
\!+\!
\partial_{\Lambda_\n^\star}{\Breve{\bf \Lambda}_{\n^\prime}^\star}
\Breve{\theta}_{o-\n^\prime} ,
\end{array}
\label{theta o-n}
\end{eqnarray}\\[-18pt]
in which the transformation functions are set up by 
the initial conditions,\\[-20pt]
\begin{eqnarray}
\left.
\begin{array}{rl}
\Breve{\Lambda}_\n(t)|_{t=0}
\!=\!
\Breve{\bf \Lambda}_\n(\Lambda,\Lambda^\star,t)|_{t=0}
\!=\!
\Lambda_\n , \\
\\[-8pt]
\Breve{ \Lambda}_\n^\star(t)|_{t=0}
\!=\!
\Breve{\bf \Lambda}_\n^\star
(\Lambda,\Lambda^\star,t)|_{t=0}
\!=\!
\Lambda_\n^\star ,
\end{array} 
\begin{array}{ll}
\partial_{\bf \Lambda_\n}
\Breve{\bf \Lambda}_{\n^\prime}|_{t=0}
\!=\!
\delta_{\n \n^\prime},
&
\partial_{\Lambda_\n^\star}
\Breve{\bf \Lambda}_{\n^\prime}^\star|_{t=0}
\!=\!
\delta_{\n \n^\prime},\\
\\[-8pt]
\partial_{\Lambda_\n}
\Breve{\bf \Lambda}_{\n^\prime}^\star|_{t=0}
\!=\!
0 ,
&\partial_{\Lambda_\n^\star}
\Breve{\bf \Lambda}_{\n^\prime}|_{t=0}
\!=\!
0,
\end{array} \!\!
\right\}
\label{Initial condition2}
\end{eqnarray}\\[-12pt]
to guarantee both the pictures to coincide at 
$t \!=\! 0$.
The Hamiltonian ${\cal F}_{o-c}$
is expressed as\\[-8pt]
\begin{equation}
{\cal F}_{o-c}
\!\!=\!\!
\hbar v_\n(\Lambda,\Lambda^\star,t)\theta_{o-\n}^\dagger
\!+\!
\hbar v_\n^\star(\Lambda,\Lambda^\star,t)\theta_{o-\n} .
\label{velocity expless}
\end{equation}\\[-14pt]
Expansion coefficients $\!(v_\n,\!v_\n^\star)\!$
are interpreted as velocity field 
in the Lagrange-like manner.
Substituting 
(\ref{theta o-n}) 
into 
(\ref{velocity expless}) 
and 
comparing with 
(\ref{transfered Fo-c}), 
we get the relations\\[-20pt]
\begin{eqnarray}
\begin{array}{l}
\dot{\Breve \Lambda}_\n
\!\!=\!\!
\partial_t \! \Breve {\bf \Lambda}_\n
\!\!=\!\!
v_{\n^\prime}\!
\partial_{\Lambda_{\n^\prime}}
\Breve {\bf \Lambda}_\n
\!\!+\!\!
v_{\n^\prime}^\star \!
\partial_{\Lambda_{\n^\prime}^\star}
\Breve {\bf \Lambda}_\n ,~~
\dot{\Breve \Lambda}_\n^\star
\!\!=\!\!
\partial_t \! \Breve {\bf \Lambda}_\n^\star
\!\!=\!\!
v_{\n^\prime} \!
\partial_{\Lambda_{\n^\prime}}
\Breve {\bf \Lambda}_\n^\star
\!\!+\!\!
v_{\n^\prime}^\star \!
\partial_{\Lambda_{\n^\prime}^\star}
\Breve {\bf \Lambda}_\n^\star ,
\end{array}
\label{relations}
\end{eqnarray}\\[-18pt]
from which the initial conditions of 
the velocity fields are given as\\[-22pt]
\begin{eqnarray}
\begin{array}{l}
\dot{\Breve \Lambda}_\n (t)|_{t=0}
\!\!=\!\!
\partial_t \Breve {\bf \Lambda}_\n|_{t=0}
\!\!=\!\!
v_\n(\Lambda,\!\Lambda^\star\!,\!t)|_{t=0} ,~
\dot{\Breve \Lambda}_\n^\star (t)|_{t=0}
\!\!=\!\!
\partial_t \Breve {\bf \Lambda}_\n^\star |_{t=0}
\!\!=\!\!
v_\n^\star (\Lambda,\!\Lambda^\star\!,\!t)|_{t=0} .
\end{array}
\label{Initial velocity}
\end{eqnarray}\\[-18pt]
Then we obtain the correspondence of the time derivatives
of the collective coordinates in the usual manner to the
velocity fields in the Lagrange-like one.

Finally we impose the canonicity conditions 
in the usual manner,\\[-18pt]
\begin{eqnarray}
\begin{array}{ll}
\langle\phi(\check g)|{\Breve O}_\n^\dagger 
|\phi(\Breve g)\rangle
=
i
{\Breve \Lambda}_\n^\star ,
&\langle\phi(\Breve g)|{\Breve O}_\n 
|\phi(\Breve g)\rangle
=
-i
{\Breve \Lambda}_\n ,      
\end{array}
\label{canonical condition in qf}
\end{eqnarray}\\[-16pt]
which leads to the {\em weak\/} canonical 
commutation relation  with the aid of 
(\ref{C..=0 on collective})
and 
(\ref{restriction}).\\[-26pt]

\newpage

%%%%%%%%%%%%%%%%%%%%%
%                                                               %    
%     5    On Validity of the Maximally     %
%                    Decoupled Theory              %
%                                                               %
%%%%%%%%%%%%%%%%%%%%%

\setcounter{equation}{0}
\renewcommand{\theequation}{\arabic{section}.
\arabic{equation}}
\section{On the validity of $\ll$maximally decoupled$\gg$ theory}

The our basic concept lies in
the {\em invariance principle of Schr\"{o}dinger
equation}, and the TDHF equation is solved under
the canonicity condition and vanishing of non-collective
dangerous terms.$\!$ 
We have no justification
on the validity of {\em maximally decoupled} method, 
we must give a criterion how it extract the collective
sub-manifolds effectively out of the TDHF manifold.$\!$
To establish the criterion, 
we express the collective Hamiltonian ${\cal F}_{o-c}$ 
and the residual one ${\cal F}_{o-res}$
in the same form as those of the TDHF Hamiltonian
${\cal F}_o$
(\ref{HB-matrix in qf}).$\!$
We also represent the quantities 
$\theta_{\!o-\n}^\dagger~\!(\!\n \!\!=\!\! 1,\!\cdots\!,\!\m\!)$, 
$\!{\cal C}_{\!o-t,\Lambda_n}^{res}\!$ and 
${\cal C}_{\!o-t,\Lambda_n^\star}^{res}\!$
in the form consisting of $m \!\times\! m,~\!\!
m \!\times\! (\!N \!\!-\!\! m\!),~\!\!(\!N \!\!-\!\! m\!) \!\times\! m,~\!\!
(\!N \!\!-\!\! m\!) \!\times\! (\!N \!\!-\!\! m\!)$
block matrices,
respectively 
as follows:\\[-16pt]
\begin{eqnarray}
\begin{array}{c}
\theta_{o-\n}^\dagger
\!=\!
\left[ \!\!\!
\begin{array}{rr}
\xi_o & \varphi_o\\[1pt]
\psi_o & 
\zeta_o 
\end{array} \!\!\!
\right]_\n,~
\begin{array}{llll}
{\cal C}_{o-t,\Lambda_\n}^{res}
\!=\!
\left[ \!\!
\begin{array}{cc}
{\cal C}_\xi^{res}&{\cal C}_\varphi^{res}\\[2pt]
{\cal C}_\psi^{res}&
{\cal C}_\zeta^{res} 
\end{array} \!\!
\right]_\n,
&
{\cal C}_{o-t,\Lambda_n^\star}^{res}
\!=\!
\left[ \!\!
\begin{array}{cc}
{\cal C}_{\xi^\star}^{res}&{\cal C}_{\varphi^\star}^{res}\\[2pt]
{\cal C}_{\psi^\star}^{res}&
{\cal C}_{\zeta^\star}^{res} 
\end{array} \!\!
\right]_\n.
\end{array}
\end{array} \!\!\!\!\!\!\!\!
\label{matrix of thetaC..res}
\end{eqnarray}\\[-10pt]
Substitution of the explicit form of ${\cal F}_{o-res}$ and
(\ref{matrix of thetaC..res})
into 
(\ref{explicit C..res}) 
yields\\[-16pt]
\begin{eqnarray}
\!\!\!\!\!
\left.
\begin{array}{l}
{\cal C}_{\xi,\n}^{res}
\!=\!
-i\partial_{\Lambda_\n} \Hat{F}_{o-res}
\!-\!
[\xi_{o,\n},\Hat{F}_{o-res}]
\!-\!
\varphi_{o,\n}\grave{F}_{o-res}
\!+\!
\acute{F}_{o-res} \psi_{o,\n} ,\\
\\[-6pt]
{\cal C}_{\psi,\n}^{res}
\!=\!
-i\partial_{\Lambda_\n}\grave{F}_{o-res}
-
\zeta_{o,\n}\grave{F}_{o-res}
\!+\!
\grave{F}_{o-res} \xi_{o,\n}
\!-\!
\psi_{o,\n}\Hat{F}_{o-res}
\!+\!
\Check{F}_{o-res} \psi_{o,\n} ,\\
\\[-6pt]
{\cal C}_{\varphi,\n}^{res}
\!=\!
-i\partial_{\Lambda_\n}\acute{F}_{o-res}
\!-\!
\xi_{o,\n}\acute{F}_{o-res}
\!+\!
\acute{F}_{o-res} \zeta_{o,\n}
\!-\!
\varphi_{o,\n}\Check{F}_{o-res}
\!+\!
\Hat{F}_{o-res} \varphi_{o,\n} ,\\
\\[-6pt]
{\cal C}_{\zeta,\n}^{res}
\!=\!
-i\partial_{\Lambda_\n} \Check{F}_{o-res}
\!-\!
[\zeta_{o,\n},\Check{F}_{o-res}]
\!-\!
\psi_{o,\n} \acute{F}_{o-res}
\!+\!
\grave{F}_{o-res}\varphi_{o,\n} .
\end{array} \!\!\!
\right \}
\label{explicit C..res2}
\end{eqnarray}\\[-6pt]
Another ${\Breve{\cal \theta}}_{0-\n}^\dagger$ in
(\ref{transfered Fo-c})
and
(\ref{theta o-n}) 
is expressed in the same form as the one 
given in 
(\ref{matrix of thetaC..res}).    
Substituting this expression for
${\Breve{\cal \theta}}_{0-\n}^\dagger$
into 
(\ref{transfered Fo-c}) 
and using
(\ref{Hc+Hres in qf}) , 
we obtain the relations\\[-16pt]
\begin{eqnarray}
\left.
\begin{array}{l}
\Hat{F}_{o-res}
\!=\!
\Hat{F}_o
\!-\!
\hbar\partial_t \Breve{\bf \Lambda}_\n
\Breve{\xi}_{o,\n}
\!-\!
\hbar\partial_t \Breve{\bf \Lambda}_\n^\star
\Breve{\xi}_{o,\n}^\dag ,~~
\acute{F}_{o-res}
\!=\!
\acute{F}_o
\!-\!
\hbar\partial_t \Breve{\bf \Lambda}_\n
\Breve{\varphi}_{o,\n}
\!-\!
\hbar\partial_t \Breve{\bf \Lambda}_\n^\star
\Breve{\psi}_{o,\n}^\dag ,\\
\\[-6pt]
\grave{F}_{o-res}
\!=\!
\grave{F}_o
\!-\!
\hbar\partial_t \Breve{\bf \Lambda}_\n
\Breve{\psi}_{o,\n}
\!-\!
\hbar\partial_t \Breve{\bf \Lambda}_\n^\star
\Breve{\varphi}_{o,\n}^\dag ,~
\Check{F}_{o-res}
\!=\!
\Check{F}_o
\!-\!
\hbar\partial_t \Breve{\bf \Lambda}_\n
\Breve{\zeta}_{o,\n}
\!-\!
\hbar\partial_t \Breve{\bf \Lambda}_\n^\star
\Breve{\zeta}_{o,\n}^\dag ,
\end{array}
\right \}
\label{Fo-res Do-res}
\end{eqnarray}\\[-10pt]
together with their c.c. 
Further using
(\ref{canonical condition in qf}),
we can get
the relations
\\[-8pt]
\begin{equation}
\mbox{Tr}
\Breve{\xi}_{o,\n}
=
i\Breve{\Lambda}_\n^\star ,~~
\mbox{Tr}
\Breve{\xi}_{o,\n}^\dagger
=
-i\Breve{\Lambda}_\n .
\label{TR xi=iLamba}
\end{equation}\\[-30pt]

The way of extracting the collective sub-manifolds 
out of the full TDHF manifold is made possible 
by the minimization of the residual curvature.
This is achieved if we require at least 
the expectation values of the residual curvatures 
be minimized as far as possible,\\[-18pt]
\begin{eqnarray}
\begin{array}{l}
\langle\phi(\Breve g)
|C_{t,\Lambda_\n}^{res}|
\phi(\Breve g)\rangle
\!=\!
\mbox{Tr}
{\cal C}_{\xi,\n}^{res}
\cong 0 ,~
\langle\phi(\Breve g)
|C_{t,\Lambda_\n^\star}^{res}|
\phi(\Breve g)\rangle
\!=\!
\mbox{Tr}
{\cal C}_{\xi^\star,\n}^{res}
\cong 0 .
\end{array}
\label{<C..res>->0}
\end{eqnarray}\\[-18pt]
$C_{t,\Lambda_\n }^{res}\!$
and
$C_{t,\Lambda_\n^\star}^{res}\!$
are given in the same way as the one in
(\ref{C.. ope on qp-frame}). 
We adopt the condition of
the stationary HF method (YK)
\cite{YK.81}. 
The so-called dangerous term in 
${\cal F}_{\!o-res}$ must vanish,\\[-8pt]
\begin{equation}
\grave{F}_{o-res}=0 ,~~\acute{F}_{o-res}=0 .
\label{Do-res=0}
\end{equation}\\[-12pt]
With the aid of
(\ref{Fo-res Do-res}) 
and 
(\ref{TR xi=iLamba}), 
(\ref{Do-res=0}) 
and 
(\ref{<C..res>->0}) 
are rewritten 
as\\[-16pt]
\begin{eqnarray}
\grave{F}_o
\!=\!
\hbar \dot{\Breve{{\bf \Lambda}}}_\n
\Breve{\psi}_{o,\n}
\!+\!
\hbar \dot{\Breve{{\bf \Lambda}}}^\star_\n
\Breve{\varphi}_{o,\n}^\dag ,~
\acute{F}_{o}
\!=\!
\hbar \dot{\Breve{{\bf \Lambda}}}_\n
\Breve{\varphi}_{o,\n}
\!+\!
\hbar \dot{\Breve{{\bf \Lambda}}}^\star_\n
\Breve{\psi}_{o,\n}^\dagger ,
\label{Do= ...}
\end{eqnarray}
\vspace{-0.9cm}
\begin{eqnarray}
\left.
\begin{array}{l}
i\partial_{\Lambda_\n} \!
\mbox{Tr}
\Hat{F}_{o-res}
\!=\!
i\partial_{\Lambda_\n} \!
\mbox{Tr}
\Hat{F}_o
\!+\!
\partial_{\Lambda_\n}
(
\hbar \dot{\Breve{{\bf \Lambda}}}_{\n^\prime}
\Breve{ \Lambda}_{\n^\prime}^\star
\!-\!
\hbar \dot{\Breve{{\bf \Lambda}}}_{\n^\prime}^\star
\Breve{ \Lambda}_{\n^\prime}
)
\cong 0 ,\\
\\[-6pt]
i\partial_{\Lambda_\n^\star} \!
\mbox{Tr}
\Hat{F}_{o-res}
\!=\!
i\partial_{\Lambda_\n^\star} \!
\mbox{Tr}
\Hat{F}_o
\!+\!
\partial_{\Lambda_\n^\star}
(
\hbar \dot{\Breve{{\bf \Lambda}}}_{\n^\prime}
\Breve{ \Lambda}_{\n^\prime}^\star
\!-\!
\hbar \dot{\Breve{{\bf \Lambda}}}_{\n^\prime}^\star
\Breve{ \Lambda}_{\n^\prime}
)
\cong 0 .
\end{array}
\right \}
\label{id-Lambda Tr Fo-res}
\end{eqnarray}\\[-34pt]

%%%%%%%%%%%%%%%%%%%%%%%%%%
%                                                                                % 
%  5.1    Geometric equation in the fluctuating         %
%                        quasi-particle frame.                       %
%                                                                                %
%%%%%%%%%%%%%%%%%%%%%%%%%%

\subsection{Geometric equation in fluctuating quasi PHF}

The functional form
$\!g(\Lambda,\!\Lambda^\star\!,\!t)\!$ is divided into 
the stationary and fluctuating components, 
$g \!\!=\!\! g_0  \tilde{g}$,
product of stationary 
$g_0$ 
and 
fluctuating $\tilde{g}
(\Lambda,\! \Lambda^\star\!,\! t)$.
The $g_0$ satisfies the HF eigenvalue equation. 
A density matrix 
$\slashed{\cal Q}(\Lambda,\! \Lambda^\star\!,\! t)$
is decomposed as 
$\!
\slashed{\cal Q}
\!\!=\!\!
g_0 \widetilde{\slashed{\cal Q}} g_0^\dagger
$.
The
fluctuating
$\tilde{\theta}_\n\!$
and
$\!\tilde{\theta}_\n^\dagger$
are given as
$
\tilde{\theta}_{\!\n}
\!\!=\!\!
g_{\!0}^\dagger \theta_{\!\n} g_{\!0}
\!$
and
$\!
\tilde{\theta}_{\!\n}^\dagger
\!\!=\!\!
g_{\!0}^\dagger \theta_{\!\n}^\dagger g_{\!0}.
\!$
Under the decomposition  
$\!g\!\!=\!\!g_{\!0} \tilde{g}$,
zero curvatures$\!$
(\ref{C..=0 on collective})
are transformed to\\[-20pt]
\begin{eqnarray}
\begin{array}{l}
i\hbar\partial_t \tilde{\theta}^{\dag }_\n
\!-\!
i\partial_{\Lambda_\n}
\widetilde{\cal F}_c
\!+\!
[\tilde{\theta}_\n^{\dag },\widetilde{\cal F}_c]
\!=\!
0 ,~
i\hbar\partial_t \tilde{\theta}_\n
\!-\!
i\partial_{\Lambda_\n^{\star}}
\widetilde{\cal F}_c
\!+\!
[\tilde{\theta}_\n,\widetilde{\cal F}_c]
\!=\!
0 ,
\end{array}
\label{fluctuating C..t Lamda}
\end{eqnarray}
\vspace{-1.0cm}
\begin{eqnarray}
\left.
\begin{array}{c}
i\partial_{\Lambda_{\n^\prime}}\tilde{\theta}_\n
\!-\!
i\partial_{\Lambda_\n^\star}\tilde{\theta}_{\n^\prime}^\dagger 
\!+\!
[\tilde{\theta}_\n,\tilde{\theta}_{\n^\prime}^\dagger]
\!=\!
0 ,
%\right.
\\
\\[-8pt]
%\left.
i\partial_{\Lambda_{\n^\prime}}
\tilde{\theta}^{\dag }_\n
\!-\!
i\partial_{\Lambda_\n}
\tilde{\theta}^{\dag }_{\n^\prime} 
\!+\!
[\tilde{\theta}^{\dag }_\n,\tilde{\theta}^{\dag }_{\n^\prime}]
\!=\!
0 ,~
i\partial_{\Lambda_{\n^\prime}^{\star}}
\tilde{\theta}_\n
\!-\!
i\partial_{\Lambda_\n^{\star}}
\tilde{\theta}_{\n^\prime} 
\!+\!
[\tilde{\theta}_\n,\tilde{\theta}_{\n^\prime}]
\!=\!
0 ,
\end{array}
\right\} 
\label{fluctuating C..Lambda  Lamda}
\end{eqnarray}
\vspace{-0.6cm}
\begin{eqnarray}
~
%\left.
\begin{array}{c}
\mbox{Tr} \!
\left\{ \!
\widetilde{\slashed{\cal Q}}(\tilde{g})
[\tilde{\theta}_\n,\tilde{\theta}_{\n^\prime}^\dagger] \!
\right\}
\!=\!
\delta_{\n \n^{\prime}} ,~
\mbox{Tr} \!
\left\{ \!
\widetilde{\slashed{\cal Q}}(\tilde{g})
[\tilde{\theta}_\n^\dagger,\tilde{\theta}_{\n^\prime}^\dagger] \!
\right\}
\!=\!
0  ,~
\mbox{Tr} \!
\left\{ \!
\widetilde{\slashed{\cal Q}}(\tilde{g})
[\tilde{\theta}_\n,\tilde{\theta}_{\n^\prime}] \!
\right\}
\!=\!
0 ,
\end{array}
%\right\}
\label{fluctuating weak orthgonality}
\end{eqnarray}\\[-16pt]
where 
$\widetilde{\cal F}_c$, 
$\tilde{\theta}_\n^\dagger$ and $\tilde{\theta}_\n$
satisfy the partial differential equations 
given below,\\[-10pt]
\begin{equation}
i\hbar\partial_t\tilde{g}
\!=\!
\widetilde{\cal F}_c\tilde{g} ,
\label{fluctuating TDHB}
\end{equation}
\vspace{-0.9cm}
\begin{eqnarray}
\begin{array}{lr}
i\partial_{\Lambda_\n}\tilde{g}
\!=\!
\tilde{\theta}_\n^\dagger\tilde{g} ,
&i\partial_{\Lambda_\n^\star}\tilde{g}
\!=\!
\tilde{\theta}_\n\tilde{g} .
\end{array}
\label{fluctuating SDHB}
\end{eqnarray}\\[-20pt]
Putting 
$\!\widetilde{\cal F}_{\!c} \!\!=\!\! \widetilde{\cal F}\!$
of$\!$
(\ref{tilde RandF}) $\!$in$\!$ 
(\ref{fluctuating C..t Lamda}),
we look for collective-path $\!\tilde{g}\!$
and collective-Hamiltonian $\!\widetilde{\cal F}_{\!c}$ 
under the minimization of the residual curvature arising 
from residual Hamiltonian $\widetilde{\cal F}_{\!res}$.

Next, for our convenience of further discussion, 
we also introduce modified
fluctuating auxiliary quantities with the same forms as those in
the previous section, through,\\[-10pt]
\begin{equation}
\tilde{\theta}_{o-\n}^\dagger
\!=\!
\tilde{g}^\dagger\tilde{\theta}_\n^\dagger\tilde{g} ,
~~~\tilde{\theta}_{o-\n}
\!=\!
\tilde{g}^\dagger\tilde{\theta}_\n\tilde{g} .
\label{modified fluctuating theta}
\end{equation}\\[-16pt]
We rewrite 
a set of equations
(\ref{fluctuating C..t Lamda}),
(\ref{fluctuating C..Lambda  Lamda}) 
and
(\ref{fluctuating weak orthgonality})
in terms of the above quantities as\\[-22pt]
\begin{eqnarray}
\begin{array}{l}
i\hbar\partial_t\tilde{\theta}_{o-\n}^\dagger
\!-\!
i\tilde{g}^\dagger(\partial_{\Lambda_\n}
\tilde{\cal F}_c)\tilde{g}
\!=\!
0 ,~~
i\hbar
\partial_t\tilde{\theta}_{o-\n}
\!-\!
i\tilde{g}^\dagger(\partial_{\Lambda_\n^\star}
\tilde{\cal F}_c)\tilde{g}
\!=\!
0 ,
\end{array}
\label{C..t Lambda on quasi-frame0}
\end{eqnarray}
\vspace{-1.0cm}
\begin{eqnarray}
\!\!\!\!\!\!\!
\left.
\begin{array}{c}
i\partial_{\Lambda_{\n^\prime}}
\tilde{\theta}_{o-\n}
\!-\!
i\partial_{\Lambda_\n^{\star}}
\tilde{\theta}_{o-\n^\prime}^\dagger 
\!-\!
[\tilde{\theta}_{o-\n},\tilde{\theta}_{o-\n^\prime}^\dagger] 
\!=\!
0 , \\
\\[-10pt]
i\partial_{\Lambda_\n}
\tilde{\theta}^{\dag }_{o-\n}
\!-\!
i\partial_{\Lambda_\n}
\tilde{\theta}^{\dag }_{o-\n^\prime} 
\!-\!
[\tilde{\theta}^{\dag }_{o-\n},\tilde
{\theta}^{\dag }_{o-\n^\prime}] 
\!=\!
0 ,~~
i\partial_{\Lambda_\n^{\star}}
\tilde{\theta}_{o-\n}
\!-\!
i\partial_{\Lambda_\n^{\star}}
\tilde{\theta}_{o-\n^\prime} 
\!-\!
[\tilde{\theta}_{o-\n},\tilde
{\theta}_{o-\n^\prime}] 
\!=\!
0 .
\end{array}
\right\}
\label{C..Lambda Lamda on quasi-frame}
\end{eqnarray}\\[-10pt]
In the derivation of Eqs.$\!$ 
(\ref{C..t Lambda on quasi-frame0}) 
and (\ref{C..Lambda Lamda on quasi-frame}),$\!$
we have used 
(\ref{fluctuating TDHB}) 
and 
(\ref{fluctuating SDHB}), 
respectively.
While, 
$
\Breve{\theta}_{\!o-\n}^\dagger 
\left( 
\!=\!\!
\left[ \!\!\!
\begin{array}{rr}
\Breve{\xi}_o &\!\!\! 
\Breve{\varphi}_o\\[-1pt]
\Breve{\psi}_o &\!\!\! 
\Breve{\zeta}_o 
\end{array} \!\!\!
\right]_{\!\n} 
\right)
$
obeys
$\!
-i\partial_{\Breve{\Lambda}_\n} \Breve{g}^\dagger
\!=\!
\Breve{\theta}_{o-\n}^\dagger \Breve{g}^\dagger 
$
by which 
(\ref{Do= ...}) 
is changed
to the equation of path for collective motion.
Using the rep.
(\ref{SO(2N) Canonical Trans}) 
of $g$,
we get the partial differential equations\\[-20pt]
\begin{eqnarray}
\left.
\begin{array}{cc}
\Breve{\xi}_{0,\n}
=
-i(
\partial_{\Breve{\Lambda}_\n} \!
\Breve{\Hat{a}}^\dagger ~\!
\Breve{\Hat{a}}
+
\partial_{\check{\Lambda}_\n} \!
\Breve{\grave{b}}^\dagger ~\!
\Breve{\grave{b}}
) ,&\!\!
\Breve{\varphi}_{0,\n}
=
-i(
\partial_{\Breve{\Lambda}_\n} \!
\Breve{\Hat{a}}^\dagger ~\!
\Breve{\acute{b}}
+
\partial_{\check{\Lambda}_\n} \!
\Breve{\grave{b}}^\dagger ~\!
\Breve{\Check{a}}
) ,~\\
\\[-8pt]
\Breve{\psi}_{0,\n}
=
-i(
\partial_{\Breve{\Lambda}_\n} \!
\Breve{\acute{b}}^{\dag} ~\!
\Breve{\Hat{a}}
+
\partial_{\Breve{\Lambda}_\n} \!
\Breve{\Check{a}}^{\dag} ~\! 
\Breve{\grave{b}}
) ,&\!\!
\Breve{\zeta}_{0,\n}
=
-i(
\partial_{\Breve{\Lambda}_\n} \!
\Breve{\acute{b}}^{\dag} ~\!
\Breve{\acute{b}}
+
\partial_{\check{\Lambda}_\n} \!
\Breve{\Check{a}}^{\dag} ~\!
\Breve{\Check{a}}
) .
\end{array} \!\!
\right\}
\label{explicit psi and varph}
\end{eqnarray}\\[-10pt]
Substituting 
$
\Breve{\cal F}_o 
\!=\! 
\Breve{g}^\dagger \! \Breve{\cal F} \! \Breve{g}
$
and 
(\ref{explicit psi and varph}) 
into 
(\ref{Do= ...}), 
we obtain for
$\Breve{\grave{F}}_o$
and
$\Breve{\acute{F}}_o$
approximately as \\[-4pt] 
\begin{equation}
\!\!\!%\!\!
\left.
\BA{cc}
&\!\!\!\!
\Breve{\Check{a}}^{\dag} \!
(
\Breve{\grave{F}} \Breve{\Hat{a}}
\!+\!
\Breve{\Hat{F}}\Breve{\grave{b}}
)
\!+\!
\Breve{\acute{b}}^{\dag} \!
(
\Breve{\Hat{F}} \Breve{\Hat{a}}
\!+\!
\Breve{\acute{F}}\Breve{\grave{b}}
)
\!+\!
\hbar
i\dot{\Breve{\Lambda}}_\n
\partial_{\Breve{\Lambda}_\n} \!
\Breve{\acute{b}}^{\dag} ~\!
\Breve{\Hat{a}}
\!-\!
\hbar
i\dot{\Breve{\Lambda}}_\n^\star
\Breve{\Check{a}}^{\dag} 
\partial_{\Breve{\Lambda}_\n^\star} \! 
\Breve{\grave{b}}  
\!+\!
\hbar
i\dot{\Check{\Lambda}}_\n
\partial_{\Check{\Lambda}_n} \!
\Breve{\Check{a}}^\dag ~\!
\Breve{\grave{b}}
\!-\!
\hbar
i\dot{\Breve{\Lambda}}_\n^\star
\Breve{\acute{b}}^{\dag} ~\!
\partial_{\Breve{\Lambda}_\n^\star} 
\Breve{\Hat{a}}
\!=\!
0 ,\\
\\[-10pt]
\!=&\!\!\!\!\!\!\!\!\!\!\!\!
\Breve{\Check{a}}^{\dag} \!\!
\left\{ \!
( \!
\Breve{\grave{F}} \Breve{\Hat{a}}
\!+\!
\Breve{\Hat{F}}\Breve{\grave{b}}
)
\!-\!
\hbar
i\dot{\Breve{\Lambda}}_\n^\star
\partial_{\Breve{\Lambda}_\n^\star} \! 
\Breve{\grave{b}}  
\right\}
\!+\!
\Breve{\acute{b}}^{\dag} \!\!
\left\{ \!
( \!
\Breve{\Hat{F}} \Breve{\Hat{a}}
\!+\!
\Breve{\acute{F}}\Breve{\grave{b}}
)
\!-\!
\hbar
i\dot{\Breve{\Lambda}}_\n^\star
\partial_{\Breve{\Lambda}_\n^\star} 
\Breve{\Hat{a}}
\right\}
\!+\!
\hbar
i\dot{\Breve{\Lambda}}_\n
\partial_{\Breve{\Lambda}_\n} \!
\Breve{\acute{b}}^{\dag} ~\!
\Breve{\Hat{a}}
\!+\!
\hbar
i\dot{\Check{\Lambda}}_\n
\partial_{\Check{\Lambda}_n} \!
\Breve{\Check{a}}^\dag ~\!
\Breve{\grave{b}} ,\\
\\[-10pt]
\!=&\!\!\!\!\!\!\!\!\!\!\!\!
\Breve{\Check{a}}^{\dag} \!\!
\left\{ \!
\left( \!
\Breve{\grave{F}} \Breve{\Hat{a}}
\!+\!
\Breve{\Hat{F}}\Breve{\grave{b}} \!
\right)
\!-\!
\hbar \!
\left( \!
i\dot{\Check{\Lambda}}_\n
\partial_{\Check{\Lambda}_n} \!
\Breve{\grave{b}}
\!+\!
i\dot{\Breve{\Lambda}}_\n^\star
\partial_{\Breve{\Lambda}_\n^\star} \! 
\Breve{\grave{b}}
\right)  \!
\right\}
\!+\!
\Breve{\acute{b}}^{\dag} \!
\left\{ \!
\left( \!
\Breve{\Hat{F}} \Breve{\Hat{a}}
\!+\!
\Breve{\acute{F}}\Breve{\grave{b}} \!
\right)
\!-\!
\hbar \!
\left( \!
i\dot{\Breve{\Lambda}}_\n
\partial_{\Breve{\Lambda}_\n} \!
\Breve{\Hat{a}}
\!+\!
i\dot{\Breve{\Lambda}}_\n^\star
\partial_{\Breve{\Lambda}_\n^\star} 
\Breve{\Hat{a}}
\right) \!
\right\} ,
\end{array} \!\!
\right\}
\label{Do rep on g parameters1}
\end{equation}
\vspace{0.1cm}
\begin{equation}
\left.
\!\!\!\!\!\!
\BA{ll}
&\!\!\!\!
\Breve{\Hat{a}}^{\dag} \!
(
\Breve{\acute{F}} \Breve{\Check{a}}
\!+\!
\Breve{\Hat{F}} \Breve{\acute{b}}
)
\!+\!
\Breve{\grave{b}}^\dag 
(
\Breve{\Check{F}} \Breve{\Check{a}}
\!+\!
\Breve{\grave{F}}\Breve{\acute{b}}
)
\!+\!
\hbar
i\dot{\Breve{\Lambda}}_\n
\partial_{\Breve{\Lambda}_\n} \!
\Breve{\Hat{a}}^\dag ~\!
\Breve{\acute{b}} 
\!-\!
\hbar
i\dot{\Breve{\Lambda}}_\n^\star
\Breve{\Hat{a}}^{\dag} 
\partial_{\Breve{\Lambda}_\n^\star} \! 
\Breve{\grave{b}}  
\!+\!
\hbar
i\dot{\Breve{\Lambda}}_\n^\star
\partial_{\Breve{\Lambda}_\n} \!
\Breve{\grave{b}}^{\dag} ~\!
\Breve{\Check{a}}
\!-\!
\hbar
i\dot{\Check{\Lambda}}_\n
\Breve{\Hat{a}}^\dag ~\!
\partial_{\Check{\Lambda}^\star_\n} 
\Breve{\acute{b}}
\!=\!
0 ,\\
\\[-10pt]
\!=&\!\!\!\!
\Breve{\Hat{a}}^{\dag} \!\!
\left\{ \!
(
\Breve{\acute{F}} \Breve{\Check{a}}
\!+\!
\Breve{\Hat{F}} \Breve{\acute{b}}
)
\!-\!
\hbar  
i\dot{\Breve{\Lambda}}_\n^\star
\partial_{\Breve{\Lambda}_\n^\star}  
\Breve{\acute{b}}
\right\}  
\!+\!
\Breve{\grave{b}}^\dag \!
\left\{ \! 
(
\Breve{\Check{F}} \Breve{\Check{a}}
\!+\!
\Breve{\grave{F}}\Breve{\acute{b}}
)
\!-\!
\hbar
i\dot{\Check{\Lambda}}^\star_\n 
\partial_{\Check{\Lambda}^\star_\n}
\Breve{\Check{a}}
\right\}
\!+\!
\hbar
i\dot{\Breve{\Lambda}}_\n
\partial_{\Breve{\Lambda}_\n} \!\!
\Breve{\Hat{a}}^\dag ~\!
\Breve{\acute{b}} 
\!+\!
\hbar
i\dot{\Breve{\Lambda}}_\n
\partial_{\Breve{\Lambda}_\n} 
\Breve{\grave{b}}^{\dag} ~\!
\Breve{\Check{a}}\\
\\[-10pt]
\!=&\!\!\!\!
\Breve{\Hat{a}}^{\dag} \!\!
\left\{ \!
\left( \!
\Breve{\acute{F}} \Breve{\Check{a}}
\!+\!
\Breve{\Hat{F}} \Breve{\acute{b}} \!
\right)
\!-\!
\hbar \!
\left( \!
i\dot{\Breve{\Lambda}}_\n
\partial_{\Breve{\Lambda}_\n} \!\!
\Breve{\acute{b}} 
\!+\!
i\dot{\Breve{\Lambda}}_\n^\star
\partial_{\Breve{\Lambda}_\n^\star}  
\Breve{\acute{b}}
\right) \!
\right\}  
\!+\!
\Breve{\grave{b}}^\dag \!
\left\{ \! 
\left( \!
\Breve{\Check{F}} \Breve{\Check{a}}
\!+\!
\Breve{\grave{F}}\Breve{\acute{b}} \!
\right)
\!-\!
\hbar \!
\left( \!
i\dot{\Breve{\Lambda}}_\n
\partial_{\Breve{\Lambda}_\n} 
\Breve{\Check{a}}
\!+\!
i\dot{\Check{\Lambda}}^\star_\n 
\partial_{\Check{\Lambda}^\star_\n}
\Breve{\Check{a}} \!
\right) \!
\right\} ,  
\end{array} \!\!\!
\right\}
\label{Do rep on g parameters2}
\end{equation}\\[6pt]
where we have assumed the anti-commutatibity
$
\partial_{\Breve{\Lambda}_\n} \!
\Breve{\Hat{a}}^\dag ~\!
\!=\!
-
\Breve{\Hat{a}}^\dag 
\partial_{\Breve{\Lambda}_\n} ,
$
and
$
\partial_{\Breve{\Lambda}_\n} \!
\Breve{\grave{b}}^{\dag} ~\!
 \!=\!
-
\Breve{\grave{b}}^{\dag} 
\partial_{\Breve{\Lambda}_\n} \! .
$

From
(\ref{HB-matrix}),
we have the differential formulas for
${\cal Q}$
as follows:\\[-14pt]
\begin{eqnarray}
{\displaystyle 
\frac{\partial {\cal Q}}{\partial \Hat{a}}
}
\!\!\equiv\!\!
\left[ \!\!
\begin{array}{l}
\Hat{a} \\
\grave{b}
\end{array} \!\! 
\right] \!\!
\left[ \Hat{a}^\dag, \grave{b}^\dag \right] \!
\!\!=\!\!
\left[  \!\!\!
\begin{array}{cc}
\Hat{a}^\dag  &\!\! \grave{b}^\dag \\
0  &\!\! 0
\end{array}  \!\!\!
\right] \! ,~
{\displaystyle 
\frac{\partial {\cal Q}}{\partial \Check{a}}
}
\!\!\equiv\!\!
\left[ \!\!
\begin{array}{l}
\acute{b} \\
\Check{a}
\end{array} \!\! 
\right] \!\!
\left[ \acute{b}^\dag, \Check{a}^\dag \right] \!
\!\!=\!\!
\left[  \!\!\!
\begin{array}{cc}
0  &\!\! 
0 \\
\acute{b}^\dag, &\!\! 
\Check{a}^\dag
\end{array}  \!\!\!
\right] \! ,
\label{differential-formulasQ}
\end{eqnarray}\\[-10pt]
Let us denote 
the HF energy functional
(\ref{HB-matrix})
simply as 
$\langle H\rangle_{\!g}$.
Then 
we prove that the relations\\[-8pt]
\begin{equation}
\partial_{\Breve{\Hat{a}}}\langle H\rangle_{\Breve{g}}
\!=\!
\Breve{\Hat{F}} \Breve{\Hat{a}}^\dag 
+
\Breve{\acute{F}} \Breve{\grave{b}}^\dag,~~
\partial_{\Breve{\Check{a}}}\langle H\rangle_{\Breve{g}}
\!=\!
\Breve{\Check{F}} \Breve{\Check{a}}^\dag 
+
\Breve{\grave{F}} \Breve{\acute{b}}^\dag ,
\label{relation da<H>}
\end{equation}\\[-16pt]
and that show, through 
which the TDHF equation is rewritten as\\[-8pt]
\begin{equation}
i\hbar
\dot{\Breve{g}}
\!=\!
\left[ \!\!
\begin{array}{rr}
\partial_{\Breve{\Hat{a}}} &\!\! 
\partial_{\Breve{\acute{b}}}\\
\\[-14pt]
\partial_{\Breve{\grave{b}}} &\!\! 
\partial_{\Breve{\Check{a}}}
\end{array} \!\!
\right] \!
\langle H\rangle_{\Breve{g}} ,
\mbox{~where~}  
\begin{array}{c}
\partial_{\Breve{\acute{b}}}
\langle H\rangle_{\Breve{g}}
\!=\!
\Breve{\acute{F}} \Breve{\Check{a}}
\!\!+\!\!
\Breve{\Hat{F}} \Breve{\acute{b}} ,~~
\partial_{\Breve{\grave{b}}}  
\langle H\rangle_{\Breve{g}}
\!=\!
\Breve{\acute{F}} \Breve{\Check{a}}
\!\!+\!\!
\Breve{\Hat{F}} \Breve{\acute{b}} .
\end{array}
\label{rewritten TDHB}
\end{equation}\\[-10pt]
Using the relation 
similar to 
(\ref{relation da<H>}),
Eqs. 
(\ref{Do rep on g parameters1})
and
(\ref{Do rep on g parameters2}) 
are reduced respectively to\\[-16pt] 
\begin{eqnarray}
\!\!\!\!
\begin{array}{lr}
\Breve{\Check{a}}^{\dag} \!
\left\{ \!
\left( \!
\Breve{\grave{F}} \Breve{\Hat{a}}
\!+\!
\Breve{\Hat{F}}\Breve{\grave{b}} \!
\right)
\!-\!
\hbar \!
\left( \!
i\dot{\Check{\Lambda}}_\n
\partial_{\Check{\Lambda}_n} \!
\Breve{\grave{b}}
\!+\!
i\dot{\Breve{\Lambda}}_\n^\star
\partial_{\Breve{\Lambda}_\n^\star} \! 
\Breve{\grave{b}} \!
\right)  \!
\right\}
\!+\!
\Breve{\acute{b}}^{\dag} \!
\left\{ \!
\left( \!
\Breve{\Hat{F}} \Breve{\Hat{a}}
\!+\!
\Breve{\acute{F}}\Breve{\grave{b}} \!
\right)
\!-\!
\hbar \!
\left( \!
i\dot{\Breve{\Lambda}}_\n
\partial_{\Breve{\Lambda}_\n} \!
\Breve{\Hat{a}}
\!+\!
i\dot{\Breve{\Lambda}}_\n^\star
\partial_{\Breve{\Lambda}_\n^\star} 
\Breve{\Hat{a}} \!
\right) \!
\right\}
\!=\!
0 ,
\end{array}
\label{Do rep on g 2}
\end{eqnarray}\\[-38pt]
\begin{eqnarray}
\!\!\!\!
\begin{array}{lr}
\Breve{\Hat{a}}^{\dag} \!
\left\{ \!
\left( \!
\Breve{\acute{F}} \Breve{\Check{a}}
\!+\!
\Breve{\Hat{F}} \Breve{\acute{b}} \!
\right)
\!-\!
\hbar \!
\left( \!
i\dot{\Breve{\Lambda}}_\n
\partial_{\Breve{\Lambda}_\n} \!\!
\Breve{\acute{b}} 
\!+\!
i\dot{\Breve{\Lambda}}_\n^\star
\partial_{\Breve{\Lambda}_\n^\star}  
\Breve{\acute{b}}
\right) \!
\right\}  
\!+\!
\Breve{\grave{b}}^\dag \!
\left\{ \! 
\left( \!
\Breve{\Check{F}} \Breve{\Check{a}}
\!+\!
\Breve{\grave{F}}\Breve{\acute{b}} \!
\right)
\!-\!
\hbar \!
\left( \!
i\dot{\Breve{\Lambda}}_\n
\partial_{\Breve{\Lambda}_\n} \!
\Breve{\Check{a}}
\!+\!
i\dot{\Check{\Lambda}}^\star_\n 
\partial_{\Check{\Lambda}^\star_\n}
\Breve{\Check{a}} \!
\right) \!
\right\} 
\!=\!
0 .  
\end{array}
\label{Do rep on g 3}
\end{eqnarray}\\[-16pt]
As a way of satisfying 
(\ref{Do rep on g 2}), 
we adopt the following
partial differential equations:\\[-16pt]
\begin{eqnarray}
\begin{array}{c}
\partial_{\Breve{\Hat{a}}}\langle H\rangle_{\Breve{g}}
\!-\!
\hbar \!
\left( \!
i\dot{\Breve{\Lambda}}_\n
\partial_{\Breve{\Lambda}_\n} \!
\Breve{\Hat{a}}
\!+\!
i\dot{\Breve{\Lambda}}_\n^\star
\partial_{\Breve{\Lambda}_\n^\star} 
\Breve{\Hat{a}} \!
\right)
\!=\!
0 ,~
\partial_{\grave{b}}
\langle H\rangle_{\Check{g}}
\!-\!
\hbar \!
\left( \!
i\dot{\Check{\Lambda}}_\n
\partial_{\Check{\Lambda}_n} \!
\Breve{\grave{b}}
\!+\!
i\dot{\Breve{\Lambda}}_\n^\star
\partial_{\Breve{\Lambda}_\n^\star} \! 
\Breve{\grave{b}} \!
\right)
\!=\!
0 
\mbox{~and~c.c.}
\end{array}
\label{Do rep on g 3}
\end{eqnarray}\\[-40pt]
\begin{eqnarray}
\begin{array}{c}
\partial_{\Breve{\Check{a}}}
\langle H\rangle_{\Breve{g}}
\!-\!
\hbar \!
\left( \!
i\dot{\Breve{\Lambda}}_\n
\partial_{\Breve{\Lambda}_\n} \!\!
\Breve{\acute{b}} 
\!+\!
i\dot{\Breve{\Lambda}}_\n^\star
\partial_{\Breve{\Lambda}_\n^\star}  
\Breve{\acute{b}}
\right) \!
=
0,~
\partial_{\acute{b}}
\langle H\rangle_{\Check{g}}
\!-\!
\hbar \!
\left( \!
i\dot{\Breve{\Lambda}}_\n
\partial_{\Breve{\Lambda}_\n} \!
\Breve{\Check{a}}
\!+\!
i\dot{\Check{\Lambda}}^\star_\n 
\partial_{\Check{\Lambda}^\star_\n}
\Breve{\Check{a}} \!
\right) \!
\!=\!
0 
\mbox{~and~c.c.}  
\label{Do rep on g 4}
\end{array}
\end{eqnarray}\\[-18pt]
From
(\ref{C.. ope on qp-frame})
and
(\ref{C..=0 on collective}),
we have
$
\langle\phi(\Breve g)
|C_{t,\Lambda_\n}|
\phi(\Breve g)\rangle
\!=\!
0 
$
and
$
\langle\phi(\Breve g)
|C_{t,\Lambda_\n^\star}|
\phi(\Breve g)\rangle
=
0
$
in which
${\cal F}_c$
is replaced by
${\cal F}_o$.
Using the transformation property 
of the differentials\\[-20pt]
\begin{eqnarray}
\begin{array}{l}
\partial_{\Lambda_\n}
=
\partial_{\Lambda_\n}
\check{\Lambda}_{\n^\prime}
\partial_{\check{\Lambda}_{\n^\prime}}
+
\partial_{\Lambda_\n}
\check{\Lambda}_{\n^\prime}^\star
\partial_{\check{\Lambda}_{\n^\prime}^\star} ,~
\partial_{\Lambda_\n^{\star}}
=
\partial_{\Lambda_\n^{\star}}
\check{\Lambda}_{\n^\prime}
\partial_{\check{\Lambda}_{\n^\prime}}
+
\partial_{\Lambda_\n^{\star}}
\check{\Lambda}_{\n^\prime}^\star
\partial_{\check{\Lambda}_{\n^\prime}^\star} ,
\end{array}
\label{differential ope d-Lambda}
\end{eqnarray}\\[-22pt]
and differential formulas for  expectation values
of Hamiltonian $H$ and 
HF one $H_{\mbox{\scriptsize HF}}$\\[-18pt]
\begin{eqnarray}
\left.
\begin{array}{l}
\partial_{\check{\Lambda}_\n} \!
\langle H\rangle_{\check{g}} 
\!=\!
-
\mbox{Tr}
[\partial_{\check{\Lambda}_\n} \!
\check{\cal Q}(\check{g})
\check{\cal F}] ,\\
\\[-4pt]
\partial_{\check{\Lambda}_\n} \!
\langle H_{\mbox{\scriptsize HF}}\rangle_{\check{g}}
\!=\!
-
\partial_{\check{\Lambda}_\n} \!
\mbox{Tr}
\check{\cal F}_o
\!=\!
\partial_{\check{\Lambda}_\n} \!
\langle H\rangle_{\check{g}}
-
\mbox{Tr}
[\check{\cal Q}(\check{g})
\partial_{\check{\Lambda}_\n} 
\check{\cal F}] .
\end{array}
\right\}
\label{grad<H> on lambda}
\end{eqnarray}\\[-10pt] 
Due to
(\ref{relation da<H>})-(\ref{rewritten TDHB}),(\ref{TR xi=iLamba})
and %approximation
$
\Breve{\Lambda}_\n \!
\mbox{Tr} \!
{\cal F}_{\!o}
\!\!\approx\!\!
-\partial_{\Breve{\Lambda}_\n} \!\!
\langle \!H\! \rangle_{\!\Check{g}}
$,
{\em the invariance principle of Schr\"{o}dinger equation
and canonicity condition} leads
to the canonical forms of equations of motion\\[-8pt]
\begin{equation}
i\hbar\dot{\Breve{\Lambda}}_\n^\star
\!=\!
-\partial_{\Breve{\Lambda}_\n}\!
\langle H\rangle_{\Check{g}} ,~~
i\hbar\dot{\Breve{\Lambda}}_\n
\!=\!
\partial_{\Breve{\Lambda}_\n^\star}\!
\langle H\rangle_{\Check{g}} .
\label{collective equation}
\end{equation}\\[-14pt]
The structure 
$\!$of$\!$
(\ref{Do rep on g 3})
shows
that 
it 
becomes the equation
of path for collective motion under 
the substitution 
$\!$of$\!$ 
(\ref{collective equation}).$\!$
Then
it 
realizes a construction of the equation of path
in the TDHF 
\cite{SMHU.83,Ma.80,YK.81}.$\!$
Eqs. 
(\ref{Do rep on g 3}) 
and 
(\ref{collective equation}) 
determine the behavior of
{\em maximally decoupled} collective motions 
in the TDHF.$\!$ 
It is a renewal of the TDHF equation 
by using the canonicity condition
under the existence 
of invariant subspace in the TDHF.$\!$
This is due to a natural consequence
of the {\em maximally decoupled} theory because
there exists an invariant subspace, 
if the 
{\em invariance principle of Schr\"{o}dinger equation} 
is realized.$\!$
Then  
we can investigate the validity of 
{\em maximally decoupled} theory 
with the use of the condition.$\!$
The reason why the condition occurs in our theory,
which did not appear in the {\em maximally decoupled}
theory. %$\!$
Since the %{\em maximally decoupled} 
theory has 
no such terms, 
the condition is trivially fulfilled.$\!$ 
This is the essential difference between 
the {\em maximally decoupled} theory and the ours.$\!$
We stress that our theory has been formulated by 
{\em manifesting} the group structure 
which makes the present work applicable to the
$SO(2N\!+\!1)$ group 
\cite{FN.84}.

\newpage

%%%%%%%%%%%%%%%%%%%%%%%%%%
%                                                                                 %
%    6               Nonlinear RPA Theory                       %
%     arising from the Zero-Curvature Equations       %
%                                                                                 %
%%%%%%%%%%%%%%%%%%%%%%%%%%

\setcounter{equation}{0}
\renewcommand{\theequation}{
\arabic{section}.
\arabic{equation}}
\section{Nonlinear RPA arising from 
zero-curvature equation}

The integrability condition of TDHF equation 
determining a collective sub-manifold
has been studied
based on the differential geometric viewpoint.$\!$
A geometric equation works well 
{\em over a wide range} of physics
beyond the random phase approximation
(RPA).
A linearly approximated solution
of TDHF equation becomes the RPA.
Suppose we solved the geometric equation by expanding it
in power series of collective variables
$(\Lambda_\n,\Lambda^\star_\n)$ 
$(\n \!\!=\!\! 1,\!\cdots\!,\m; \m \!\ll\! N^2)$.$\!$
The geometric equation has
a RPA solution at the lowest power
of the collective variables.
The fluctuating density matrix
$\widetilde{\slashed{\cal Q}}$
and
HF matrix
$\widetilde{\cal F}$
in the quasi PHF
are expressed as follows: \\[-20pt]
\begin{eqnarray}
\!\!\!\!
\begin{array}{c}
\widetilde{\slashed{\cal Q}}(\tilde{g})
\!=\!
g_0^\dagger \slashed{\cal Q}(g)g_0,
\widetilde{\slashed{\cal Q}}
\!\equiv\!\!
\left[ \!\!
\begin{array}{cc}
\widetilde{\Hat{\slashed{\cal Q}}}& 
\widetilde{\acute{\slashed{\cal Q}}} \\[-1pt]
\widetilde{\grave{\slashed{\cal Q}}} & 
\widetilde{\Check{\slashed{\cal Q}}}
\end{array} \!\!
\right] \! ,~
\widetilde{\cal F}
\!=\!
g_0^\dagger \tilde{\cal F} g_0 ,~
\tilde{\cal F}
\!\equiv\!\!
\left[ \!\!
\begin{array}{cc}
\Hat{\varepsilon}_0\!+\!\Hat{\it f} &\!\!\!\!\!\! \acute{\it f}\\
\\[-6pt]
\grave{\it f} &\!\!\!\!\!\! \Check{\varepsilon}_0\!+\!\Check{\it f}
\end{array} \!\!
\right] \! ,~
g_0^\dagger {\cal F}_0 g_0
\!=\! 
\left[ \!\!
\begin{array}{cc}
\Hat{\varepsilon}_0 &\!\! 0 \\
\\[-2pt]
0 &\!\! \Check{\varepsilon}_0
\end{array} \!\!
\right] \! ,
\end{array}
\label{tilde RandF}
\end{eqnarray}\\[-12pt]
where 
${\cal F}_0$
is a stationary HF matrix
and
$\Hat{\varepsilon}_0$
and
$\Hat{\varepsilon}_0$
denote 
the hole-energy
and
the particle-energy.
The 
$\!\Hat{\it f},  \acute{\it f},\grave{\it f}\!$
and
$\!\Check{\it f}\!$
are given later.
A part of the geometric equation
(\ref{fluctuating weak orthgonality}) 
is rewritten as\\[-18pt]
\begin{eqnarray}
\left.
\begin{array}{c}
\mbox{Tr}
\left\{ \!
\left[ \!\!\!
\begin{array}{cc}
1_m &\!\!\!\! 0 \\
 0 &\!\!\!\! 1_{\!N\!-\!m}
\end{array} \!\!\!
\right] \!
[\tilde{\theta}_{o-\n},
\tilde{\theta}_{o-\n^\prime}^\dagger]
\right\}
\!=
\delta_{\n \n^{\prime}} ,\\
\\[-10pt]
\mbox{Tr}
\left\{ \!
\left[ \!\!\!
\begin{array}{cc}
1_m &\!\!\!\! 0 \\
 0 &\!\!\!\! -1_{\!N\!-\!m}
\end{array} \!\!\!
\right] \!
[\tilde{\theta}_{o-\n}^\dagger,
\tilde{\theta}_{o-\n^\prime}^\dagger]
\right\}
\!=
0 ,~
\mbox{Tr}
\left\{ \!
\left[ \!\!\!
\begin{array}{cc}
1_m &\!\!\!\! 0 \\
 0 &\!\!\!\! -1_{\!N\!-\!m}
\end{array} \!\!\!
\right] \!
[\tilde{\theta}_{o-\n},
\tilde{\theta}_{o-\n^\prime}]
\right\}
\!=
0 .
\end{array} \!\!\!
\right\}
\label{weak orthgonality on quasi-frame}
\end{eqnarray}\\[-12pt]
(\ref{weak orthgonality on quasi-frame})
is obtained with the aid of the fluctuating density matrix 
$\widetilde{\slashed{\cal Q}}$
defined in 
(\ref{tilde RandF}).
Using
(\ref{density R(g)})
and the first of
(\ref{tilde RandF}),
we have the relation between
$\slashed{\cal Q}(g)$
and
$\widetilde{\slashed{\cal Q}}(\tilde{g})$
used later as\\[-8pt]
\begin{equation}
\slashed{\cal Q}(g)
\!\equiv\!
\left[ \!\!\!
\begin{array}{cc}
 \Hat{ \slashed{\cal Q}}(g)
&\!\!\!
\acute{ \slashed{\cal Q}}(g)\\
 \\[-8pt]
\grave{ \slashed{\cal Q}}(g) 
&\!\!\!
\Check{ \slashed{\cal Q}}(g)
\end{array} \!\!\!
\right]
\!=\!
\slashed{\cal Q}_0
\!+\!
2 g_0 \!\!
\left[ \!\!\!
\begin{array}{cc}
\widetilde{\Hat{ \slashed{\cal Q}}}
(\tilde{g})
&\!\!\!
\widetilde{\acute{ \slashed{\cal Q}}}
(\tilde{g})\\
 \\[-10pt]
\widetilde{\grave{ \slashed{\cal Q}}}
(\tilde{g}) 
&\!\!\!
\widetilde{\Check{ \slashed{\cal Q}}}
(\tilde{g})
\end{array} \!\!\!
\right] \!\!
g_0^\dag ,~
\slashed{\cal Q}_0
\!\equiv\!
g_0 \!\!
\left[ \!\!
\begin{array}{cc}
1_m &\!\!\!\! 0 \\
 0 &\!\!\!\! -1_{N\!-\!m}
\end{array} \!\!
\right] \!\!
g_0^\dagger 
\!=\!
\left[ \!\!\!
\begin{array}{cc}
 \Hat{ \slashed{\cal Q}}_0
&\!\!\!
\acute{ \slashed{\cal Q}}_0\\
 \\[-8pt]
\grave{ \slashed{\cal Q}}_0 
&\!\!\!
\Check{ \slashed{\cal Q}}_0
\end{array} \!\!\!
\right] \! .
\label{fluctuating R}
\end{equation}\\[-8pt]
In order to investigate the matrix-valued
nonlinear time evolution equation 
(\ref{C..t Lambda on quasi-frame0}) arising 
from the zero curvature one, we give here
the $\partial_{\Lambda_\n}$ 
and $\partial_{\Lambda_\n^\star}$ differential 
forms of the TDHF density matrix 
and collective Hamiltonian.
First, using 
(\ref{fluctuating R})
and
$
\tilde{g}^\dagger\tilde{g}
\!\!=\!\!
\tilde{g}\tilde{g}^\dagger
\!\!=\!\!
1_{N}
$, 
we have\\[-18pt]
\begin{eqnarray}
\!\!\!\!\!\!
\begin{array}{ll}
&
\partial_{\Lambda_\n} \!
\widetilde{\slashed{\cal Q}}(\tilde{g})
\!=\!
\partial_{\Lambda_\n}{\tilde{g}}
\!\left[ \!\!\!
\begin{array}{cc}
1_{\!m} &\!\!\!\! 0 \\
 0 &\!\!\!\! -1_{\!N\!-\!m}
\end{array} \!\!\!
\right]\!
{\tilde{g}}^\dagger
\!+\!
{\tilde{g}} 
\!\left[ \!\!\!
\begin{array}{cc}
1_{\!m} &\!\!\!\! 0 \\
 0 &\!\!\!\! -1_{\!N\!-\!m}
\end{array} \!\!\!
\right]\!
\partial_{\Lambda_\n}\tilde{g}^\dagger \\
\\[-6pt]
&\!\!
=\!
\partial_{\Lambda_\n}\tilde{g}
(\!\tilde{g}^\dagger\tilde{g}\!) \!
\!\left[ \!\!\!\!
\begin{array}{cc}
1_{\!m} &\!\!\!\!\! 0 \\
 0 &\!\!\!\!\! -1_{\!N\!-\!m}
\end{array} \!\!\!\!
\right] \!\!
\tilde{g}^\dagger
\!\!+\!\!
\tilde{g} \!
\!\left[ \!\!\!\!
\begin{array}{cc}
1_{\!m} &\!\!\!\!\! 0 \\
 0 &\!\!\!\!\! -1_{\!N\!-\!m}
\end{array} \!\!\!\!
\right] \!\!
\partial_{\Lambda_\n}\tilde{g}^\dagger
(\!\tilde{g}\tilde{g}^\dagger\!) %, \\
%\\[-10pt]
%&\!\!
\!=\!
-i
(\!\tilde{g}\tilde{g}^\dagger\!)
\tilde{\theta}_\n^\dagger
\tilde{g}
\!\!\left[ \!\!\!\!
\begin{array}{cc}
1_{\!m} &\!\!\!\!\! 0 \\
 0 &\!\!\!\!\! -1_{\!N\!-\!m}
\end{array} \!\!\!\!
\right]\!\!
\tilde{g}^\dagger
\!\!+\!\!
i\tilde{g}
\!\!\left[ \!\!\!\!
\begin{array}{cc}
1_{\!m} &\!\!\!\!\! 0 \\
 0 &\!\!\!\!\! -1_{\!N\!-\!m}
\end{array} \!\!\!\!
\right]\!\!
\tilde{g}^\dagger 
\tilde{\theta}_\n^\dagger
(\!\tilde{g}\tilde{g}^\dagger\!) ,
\end{array}
\label{grad_Lambda tilde R}
\end{eqnarray}\\[-10pt]
where we have used
$
i\partial_{\Lambda_\n}\tilde{g}
\!=\!
\tilde{\theta}_\n^\dagger\tilde{g}
$.
Next, by using 
(\ref{modified fluctuating theta}), 
(\ref{grad_Lambda tilde R})
is transformed into\\[-6pt]
\begin{equation}
\partial_{\Lambda_\n}
\widetilde{\slashed{\cal Q}}(\tilde{g})
\!=\!
-i\tilde{g}
\left[
\tilde{\theta}_{o-\n}^\dagger,
\left[ \!\!\!
\begin{array}{cc}
1_m &\!\!\!\! 0 \\
 0 &\!\!\!\! -1_{\!N\!-\!m}
\end{array} \!\!\!
\right]
\right]
\tilde{g}^\dagger .
\label{grad_Lambda tilde R 3}
\end{equation}\\[-12pt]
From 
(\ref{tilde RandF})
and 
(\ref{fluctuating R}), 
we have\\[-12pt]
\begin{equation}
\partial_{\Lambda_\n}
\widetilde{\slashed{\cal Q}}(\tilde{g})
\!=\!
2
\partial_{\Lambda_\n} \!
\widetilde{\cal Q}(\tilde{g})
\!=\!
2 g_0 \!\!
\left[ \!\!
\begin{array}{cc}
\partial_{\Lambda_\n} \! \widetilde{\Hat{ \cal Q}}
(\tilde{g}) ,
&\!\!
\partial_{\Lambda_\n} \! \widetilde{\acute{ \cal Q}}
(\tilde{g}) \\
 \\[-10pt]
\partial_{\Lambda_\n} \! \widetilde{\grave{ \cal Q}}
(\tilde{g}) ,
&\!\!
\partial_{\Lambda_\n} \! \widetilde{\Check{ \cal Q}}
(\tilde{g})
\end{array} \!\!
\right] \!\!
g_0^\dag ,~~
g_0
\!\equiv\!\!
\left[ \!\!
\begin{array}{rr}
\Hat{a}_0 &\!\! \acute{b}_0 \\ 
\grave{b}_0 &\!\! \Check{a}_0 
\end{array} \!\!
\right] \!.
\label{grad_Lambda tilde R 4}
\end{equation}\\[-4pt]
Let us substitute the explicit reps for
$
\tilde{g}
\left( \!
\left[ 
\begin{array}{rr}
\tilde{\Hat{a}} & \tilde{\acute{b}} \\
\\[-20pt] 
\tilde{\grave{b}} & \tilde{\Check{a}} 
\end{array} \!
\right] 
\right)
$ 
and 
$
\tilde{\theta}_{o-\n}^\dagger
\left( \!
\left[ 
\begin{array}{rr}
\xi_{o-\n}&\varphi_{o-\n}\\
\psi_{o-\n}&\zeta_{o-\n}
\end{array} \!
\right] 
\right)
$
into the RHS of 
(\ref{grad_Lambda tilde R 3}) 
and combine it with 
(\ref{grad_Lambda tilde R 4}).
Then, we obtain the final $\partial_{\Lambda_n}$
differential form of the TDHF ($U(N)$) density matrix as
follows:\\[-16pt]
\begin{eqnarray}
\left[ \!\!
\begin{array}{cc}
\partial_{\Lambda_\n} \! \widetilde{\Hat{ \cal Q}}
(\tilde{g}) ,
&
\partial_{\Lambda_\n} \! \widetilde{\acute{ \cal Q}}
(\tilde{g})\\
 \\[-6pt]
\partial_{\Lambda_\n} \! \widetilde{\grave{ \cal Q}}
(\tilde{g}) ,
&
\partial_{\Lambda_\n} \! \widetilde{\Check{ \cal Q}}
(\tilde{g})
\end{array} \!
\right]
=
g_0^\dag \!\!
\left[ \!\!
\BA{ll}
i \!
\left(
\tilde{\acute{b}} \psi_{o-\n}\tilde{\Hat{a}}^\dagger
\!-\!
\tilde{\Hat{a}} \varphi_{o-\n}
\tilde{\acute{b}}^{\dag}
\right) \! ,
&\!\!
-i \!
\left(
\tilde{\Hat{a}}\psi_{o-\n}\tilde{\Check{a}}^\dagger
\!-\!
\tilde{\acute{b}}\varphi_{o-\n}
\tilde{\grave{b}}^{\dag}
\right) \! \\
\\[-12pt]
i \!
\left(
\tilde{\Check{a}}\psi_{o-\n}\tilde{\Hat{a}}^\dagger
\!-
\tilde{\grave{b}}\varphi_{o-\n}
\tilde{\acute{b}}^{\dag}
\right) \! ,
&\!\!
-i \!
\left(
\tilde{\grave{b}} \psi_{o-\n}\tilde{\Check{a}}^\dagger
\!-
\tilde{\Check{a}}\varphi_{o-\n}
\tilde{\grave{b}}^{\dag}
\right) 
\EA \!\!
\right] \!\!
g_0 .
\label{grad_Lambda R by use of theta}
\end{eqnarray}\\[-12pt]
$\partial_{\!\Lambda_\n^\star}$ differentiation of the
$U(N)$ density matrix is also made
analogously to the above case.
The explicit form of
(\ref{fluctuating R})
without constant matrices
is given as\\[-8pt]
\begin{equation}
\left[ \!\!\!\!
\begin{array}{ll}
\langle \! E^{\Hat{\bullet}}_{~\Hat{\bullet}} \! \rangle_{\!g}
&\!\!\!
\langle \! E^{\Hat{\bullet}}_{~\Check{\bullet}} \! \rangle_{\!g} \\
\\[-2pt]
\langle \! E^{\Check{\bullet}}_{~\Hat{\bullet}} \! \rangle_{\!g}
&\!\!\! 
\langle \! E^{\Check{\bullet}}_{~\Check{\bullet}} \! \rangle_{\!g}
\end{array} \!\!\!\!
\right] \!
\!\!=\!\!
g_0 \!
\left[ \!\!\!\!
\begin{array}{cc}
\widetilde{\Hat{ \slashed{\cal Q}}}
&\!
\widetilde{\acute{ \slashed{\cal Q}}}\\
\\[-8pt]
\widetilde{\grave{ \slashed{\cal Q}}}
&\!
\widetilde{\Check{ \slashed{\cal Q}}}
\end{array} \!\!\!\!
\right] \!
g_0^\dag
\!\!=\!\!
\left[ \!\!\!\!
\begin{array}{cc}
\Hat{a}_0 \!
( \!
\widetilde{\Hat{ \slashed{\cal Q}}}
\Hat{a}_0^\dag
\!+\!
\widetilde{\acute{ \slashed{\cal Q}}}
\acute{b}^\dag_0 \!
)
\!+\!
\acute{b}_0 \!
( \!
\widetilde{\grave{ \slashed{\cal Q}}}
\Hat{a}_0^\dag
\!+\!
\widetilde{\Check{ \slashed{\cal Q}}}
\acute{b}^\dag_0 \!
) ,~
&\!\!\!
\Hat{a}_0 \!
( \!
\widetilde{\Hat{ \slashed{\cal Q}}}
\grave{b}_0^\dag
\!+\!
\widetilde{\acute{ \slashed{\cal Q}}}
\Check{a}^\dag_0 \!
)
\!+\!
\acute{b}_0 \!
( \!
\widetilde{\grave{ \slashed{\cal Q}}}
\grave{b}_0^\dag
\!+\!
\widetilde{\Check{ \slashed{\cal Q}}}
\Check{a}^\dag_0 \!
) \\
\\[-8pt]
\grave{b}_0 \!
( \!
\widetilde{\Hat{ \slashed{\cal Q}}}
\Hat{a}_0^\dag
\!+\!
\widetilde{\acute{ \slashed{\cal Q}}}
\acute{b}^\dag_0 \!
)
\!+\!
\Check{a}_0 \!
( \!
\widetilde{\grave{ \slashed{\cal Q}}}
\Hat{a}_0^\dag
\!+\!
\widetilde{\Check{ \slashed{\cal Q}}}
\acute{b}^\dag_0 \!
) , 
&\!\!\!
\grave{b}_0 \!
( \!
\widetilde{\Hat{ \slashed{\cal Q}}}
\grave{b}_0^\dag
\!+\!
\widetilde{\acute{ \slashed{\cal Q}}}
\Check{a}^\dag_0 \!
)
\!+\!
\Check{a}_0 \!
( \!
\widetilde{\grave{ \slashed{\cal Q}}}
\grave{b}_0^\dag
\!+\!
\widetilde{\Check{ \slashed{\cal Q}}}
\Check{a}^\dag_0 \!
)
\end{array} \!\!\!\!
\right] \!\! .
\label{explicit_fluctuating R}
\end{equation}\\[-6pt]
From
(\ref{tilde RandF}),
we have
$
\tilde{\cal F}
\!=\!
\left[ \!\!\!
\begin{array}{cc}
\Hat{f} &\! \acute{f}\\
\\[-16pt]
\grave{ f} &\! \Check{f}
\end{array} \!\!\!
\right] 
(
\mbox{except diagonal}
~\Hat{\varepsilon}_0~
\mbox{and}~
\Check{\varepsilon}_0
~\mbox{terms}
)
$
and
$
\tilde{\cal F}
\!=\!
g_0 \widetilde{\cal F} g_0^\dagger 
$.

Substituting
(\ref{explicit_fluctuating R})
without constant matrices
into 
(\ref{HB-element}),
we have\\[-20pt]
\begin{eqnarray}
\!\!
\begin{array}{cc}
\left[  \!\!\!
\begin{array}{cc}
\Hat{f}_{ab}
\!\!=\!\!
[ab|cd] 
\langle E^c_{~d} \rangle_{\!g} , 
&\!
\acute{f}_{ai}
\!\!=\!\!
[ai|bj] 
\langle E^b_{~j} \rangle_{\!g} \\
\\[-8pt] 
\grave{f}_{ia}
\!\!=\!\!
[ia|jb] 
\langle E^j_{~b} \rangle_{\!g} ,&\!\!\!
\Check{f}_{ij}
\!\!=\!\!
[ij|kl] 
\langle E^k_{~l} \rangle_{\!g} 
\end{array} \!\!\!
\right]\\
\\[-10pt]
\!\!=\!\!\!\!
\begin{array}{cc}
\left[  \!\!\!\!
\begin{array}{cc}
[ab|cd] 
\{
\Hat{a}_0
( \!
\widetilde{\Hat{ \slashed{\cal Q}}}
\Hat{a}_0^\dag
\!+\!
\widetilde{\acute{ \slashed{\cal Q}}}
\acute{b}^\dag_0 \!
)
\!+\!
\acute{b}_0
( \!
\widetilde{\grave{ \slashed{\cal Q}}}
\Hat{a}_0^\dag
\!+\!
\widetilde{\Check{ \slashed{\cal Q}}}
\acute{b}^\dag_0 \!
) 
\}^{c}_{~d}, &\!
[ai|bj]
\{
\Hat{a}_0
( \!
\widetilde{\Hat{ \slashed{\cal Q}}}
\grave{b}_0^\dag
\!+\!
\widetilde{\acute{ \slashed{\cal Q}}}
\Check{a}^\dag_0 \!
)
\!+\!
\acute{b}_0
( \!
\widetilde{\grave{ \slashed{\cal Q}}}
\grave{b}_0^\dag
\!+\!
\widetilde{\Check{ \slashed{\cal Q}}}
\Check{a}^\dag_0 \!
)
\}^{b}_{~j}\\
\\[-10pt] 
[ia|jb]
\{
\grave{b}_0
( \!
\widetilde{\Hat{ \slashed{\cal Q}}}
\Hat{a}_0^\dag
\!+\!
\widetilde{\acute{ \slashed{\cal Q}}}
\acute{b}^\dag_0 \!
)
\!+\!
\Check{a}_0
( \!
\widetilde{\grave{ \slashed{\cal Q}}}
\Hat{a}_0^\dag
\!+\!
\widetilde{\Check{ \slashed{\cal Q}}}
\acute{b}^\dag_0 \!
)
\}^{j}_{~b},&\!
[ij|kl] 
\{
\grave{b}_0
( \!
\widetilde{\Hat{ \slashed{\cal Q}}}
\grave{b}_0^\dag
\!+\!
\widetilde{\acute{ \slashed{\cal Q}}}
\Check{a}^\dag_0 \!
)
\!+\!
\Check{a}_0
( \!
\widetilde{\grave{ \slashed{\cal Q}}}
\grave{b}_0^\dag
\!+\!
\widetilde{\Check{ \slashed{\cal Q}}}
\Check{a}^\dag_0 \!
)
\}^{k}_{~l}
\end{array} \!\!\!\!
\right] \! .
\end{array}
\end{array}
\label{HB-element2}
\end{eqnarray}\\[-10pt]
Then
the components of 
$g_0 \widetilde{\cal F} g_0^\dagger$ 
become linear functionals of
$\widetilde{\Hat{ \cal Q}}(\tilde{g}),
\widetilde{\acute{ \cal Q}}(\tilde{g}),
\widetilde{\grave{ \cal Q}}(\tilde{g})
$
and 
$\widetilde{\Check{ \cal Q}}(\tilde{g})$.
Using
(\ref{grad_Lambda R by use of theta}), 
we easily
calculate the $\partial_{\Lambda_n}$ differential of,
for example,
$\Hat{f}$
and
$\grave{f}$
as follows:\\[-22pt]
\begin{eqnarray}
\!\!\!\!\!\!\!\!\!
\left.
\begin{array}{l}
\partial_{\Lambda_\n} \! 
\Hat{f}_{ \!ab}
\!=\!
[ab|cd] 
\{
\Hat{a}_{\!0}
(
\partial_{\Lambda_\n} \!
\widetilde{\Hat{ \slashed{\cal Q}}}
\Hat{a}_0^\dag
\!+\!
\partial_{\Lambda_\n} \!
\widetilde{\acute{ \slashed{\cal Q}}}
\acute{b}^\dag_0 
)
\!+\!
\acute{b}_0
( 
\partial_{\Lambda_\n} \!
\widetilde{\grave{ \slashed{\cal Q}}}
\Hat{a}_0^\dag
\!+\!
\partial_{\Lambda_\n} \!
\widetilde{\Check{ \slashed{\cal Q}}}
\acute{b}^\dag_0 
) 
\}^{c}_{~d}
\!=\!
i({\bf F}\} \!
\psi_{o-\n}
\!+\!
i(\overline{\bf F}\} \!
\varphi_{o-\n},\\
\\[-8pt]
\partial_{\Lambda_\n} \! 
{\grave{f}}_{ia}
\!=\!
[ia|jb]
\{
\grave{b}_0
( 
\partial_{\Lambda_\n} \!
\widetilde{\Hat{ \slashed{\cal Q}}}
\Hat{a}_0^\dag
\!+\!
\partial_{\Lambda_\n} \!
\widetilde{\acute{ \slashed{\cal Q}}}
\acute{b}^\dag_0 
)
\!+\!
\Check{a}_0
( 
\partial_{\Lambda_\n} \!
\widetilde{\grave{ \slashed{\cal Q}}}
\Hat{a}_0^\dag
\!+\!
\partial_{\Lambda_\n} \!
\widetilde{\Check{ \slashed{\cal Q}}}
\acute{b}^\dag_0 
)
\}^{j}_{~b}
\!=\!
i({\bf D}\} \!
\psi_{o-\n}
\!+\!
i(\overline{\bf D}\} \!
\varphi_{o-\n}.
\end{array} \!\!\!\!
\right \}
\label{grad_Lambda d}
\end{eqnarray}\\[-8pt]
The  ({\bf D}) etc. 
have the same form as the one in 
\cite{FN.84}
and new ({\bf D}\} etc.\/ are also defined by\\[-20pt]
\begin{eqnarray}
\begin{array}{c}
({\bf D}\}
\!\equiv\!
\|({\it ij}|D|{\it kl}\}\|,
(\overline{\bf D}\}
\!\equiv\!
\|({\it ij}|\overline{D}|{\it kl}\}\| ,~~
({\bf F}\}
\!\equiv\!
\|({\it ij}|F|{\it kl}\}\| ,
(\overline{\bf F}\}
\!\equiv\!
\|({\it ij}|\overline{F}|{\it kl}\}\| .
\end{array}
\label{the difinitions of bold D,F}
\end{eqnarray}\\[-20pt]
Using the simplified notations
$a_0$ for  
$(\Hat{a}_0, \Check{a}_0)$
and $b_0$ for
$(\acute{b}_0, \grave{b}_0)$,
we define
$({\it ij}|D|{\it kl}\}$
etc. as\\[-14pt]
\begin{eqnarray}
\left.
\begin{array}{r}
({\it ij}|D|{\it kl}\}
=
(ij|D|k^\prime l^\prime)\tilde{a}_{k^\prime k}^\star
\tilde{a}_{l^\prime l}^\star
-
(ij|\overline{D}|k^\prime l^\prime)
\tilde{b}_{k^\prime k}^\star\tilde{b}_{l^\prime l}^\star
+
(ij|d|k^\prime l^\prime)
\tilde{b}_{k^\prime k}^\star\tilde{a}_{l^\prime l}^\star , \\
\\[-6pt]
-({\it ij}|\overline{D}|{\it kl}\}
=
(ij|D|k^\prime l^\prime)\tilde{b}_{k^\prime k}
\tilde{b}_{l^\prime l}
-
(ij|\overline{D}|k^\prime l^\prime)
\tilde{a}_{k^\prime k}\tilde{a}_{l^\prime l}
+
(ij|d|k^\prime l^\prime)\tilde{a}_{k^\prime k}
\tilde{b}_{l^\prime l} ,\\
\\[-6pt]
({\it ij}|F|{\it kl}\}
=
(ij|F|k^\prime l^\prime)\tilde{a}_{k^\prime k}^\star
\tilde{a}_{l^\prime l}^\star
-
(ij|\overline{F}|k^\prime l^\prime)
\tilde{b}_{k^\prime k}^\star\tilde{b}_{l^\prime l}^\star
+
(ij|f|k^\prime l^\prime)
\tilde{b}_{k^\prime k}^\star\tilde{a}_{l^\prime l}^\star ,\\
\\[-6pt]
-({\it ij}|\overline{F}|{\it kl}\}
=
(ij|F|k^\prime l^\prime)
\tilde{b}_{k^\prime k}\tilde{b}_{l^\prime l}
-
(ij|\overline{F}|k^\prime l^\prime)
\tilde{a}_{k^\prime k}\tilde{a}_{l^\prime l}
+
(ij|f|k^\prime l^\prime)\tilde{a}_{k^\prime k}
\tilde{b}_{l^\prime l} ,
\end{array}
\right \}
\label{explicit bold D,F}
\end{eqnarray}\\[-30pt]
\begin{eqnarray}
\begin{array}{r}
~
({\it ij}|D|{\it kl})
\!=
-
[a_ia_k^\star|a_ja_l^\star]
-
[b_k^\star b_i|b_l^\star b_j]
+
[a_ib_j\!-\!a_jb_i|b_k^\star a_l^\star\!-\!b_l^\star a_k^\star] ,
\mbox{etc}. 
\end{array}
\label{explicit bold D,F2}
\end{eqnarray}\\[-18pt]
We have used the notation
$[a_ia_k^\star|a_ja_l^\star]
=
[\alpha\beta|\gamma\delta]
a_{0\alpha i}a_{0\beta k}^\star a_{0\gamma j}a_{0\delta l}^\star
$.
Under the minimization of the residual curvature, 
putting $\widetilde{\cal F}$=$\widetilde{\cal F}_c$,
we get\\[-4pt] 
\begin{equation}
\partial_{\Lambda_\n}\widetilde{\cal F}_c
=
i \!
\left[ \!\!
\begin{array}{cc}
\!-\!
({\bf F}\}\psi_{o-\n}
\!-\!
(\overline{\bf F} \}
\varphi_{o-\n} 
&-(\overline{\bf D}\}^\star\psi_{o-\n}-({\bf D}\}^\star
\varphi_{o-\n}\\ 
\\[-6pt]
({\bf D}\}\psi_{o-\n}
\!+\!
(\overline{\bf D}\}
\varphi_{o-\n} 
&^{\mbox{\scriptsize T}}({\bf F}\}^\star
\psi_{o-\n}
\!+\!
^{\mbox{\scriptsize T}}(\overline{\bf F} \}^\star
\varphi_{o-\n}
\end{array} \!\!
\right].
\label{grad_Lambda tilde F}
\end{equation}\\[-4pt]
Derive a new equation %formally
analogous to the $\!U(N)\!$ RPA equation.
To do this, we decompose the fluctuating
pair mode amplitude 
$
\tilde{g}
\!\!\rightarrow\!\!
\tilde{g}\tilde{g}
(\Hat{\varepsilon}_{\!0},\!\Check{\varepsilon}_{\!0})
$.
We finally get the following 
matrix-valued equations:\\[-12pt]
\begin{eqnarray}
\!\!\!\!\!\!
\begin{array}{c}
\tilde{g}^\dag
(\Hat{\varepsilon}_0,\Check{\varepsilon}_0)
\!\cdot\!\!
\left[ \!\!\!
\begin{array}{ll}
i\hbar\partial_t\xi_{o-\n}
\!+\!
[\Hat{\varepsilon}_0,\xi_{o-\n}]
&
i\hbar\partial_t\varphi_{o-\n}
\!+\!
[\Hat{\varepsilon}_0,\varphi_{o-\n}]_+\\
~\!-\!
\{{\bf F}\}\psi_{o-\n}
-
\{\overline{\bf F}\}\varphi_{o-\n}
&
~\!-\!
\{\overline{\bf D}\}^\star\psi_{o-\n}
-
\{{\bf D}\}^\star\varphi_{o-\n}\\
\\[-8pt]
i\hbar\partial_t\psi_{o-\n}
\!-\!
[\Check{\varepsilon}_0,\psi_{o-\n}]_+
&
\!-\!
i\hbar\partial_t\xi_{o-\n}^{\mbox{\scriptsize T}}
\!+\!
[\Check{\varepsilon}_0,\xi_{o-\n}^{\mbox{\scriptsize T}}]\\
~\!+\!
\{{\bf D}\}\psi_{o-\n}
+
\{\overline{\bf D}\}\varphi_{o-\n}
&
~
\!+\!
^{\mbox{\scriptsize T}} \!
\{{\bf F}\}^\star
\psi_{o-\n}
\!+\!
^{\mbox{\scriptsize T}} \!
\{\overline{\bf F}\}^\star
\varphi_{o-\n}
\end{array} \!\!\!
\right]
\!\!\cdot\!
\tilde{g}
(\Hat{\varepsilon}_0,\Check{\varepsilon}_0)
\!=\!
0,
\end{array}
\label{matrix form of formal RPA}
\end{eqnarray}\\[-6pt]
with the modified new matrices $\{{\bf F}\}$ etc.\/
defined through\\[-18pt]
\begin{eqnarray}
\begin{array}{c}
\{{\bf F}\}
\!=\!\|
\{ij|F|kl\} \|, ~
\{\overline{\bf F}\}
\!=\!\|
 \{ij|\overline{F}|kl\}\| ,~
\{{\bf D}\}
\!=\!\|
\{ij|D|kl\}\| ,~
\{\overline{\bf D}\}
\!=\!\|
\{ij|\overline{D}|kl\}\| ,
\end{array}
\label{definition of bold D, F}
\end{eqnarray}\\[-18pt]
whose matrix elements are given by\\[-14pt]
\begin{eqnarray}
\!\!\!\!\!\!\!\!
\left.
\begin{array}{r}
~\!
-\{ij|F|kl\}
\|=\|
\tilde{b}^\star_{i^\prime i}\tilde{a}_{j^\prime j}
(i^\prime j^\prime|D|kl\}
\|-\|
\tilde{a}^\star_{i^\prime i}\tilde{b}_{j^\prime j}
(i^\prime j^\prime|\overline{D}|kl\}^\star
~~~~~~~~~~~~~~~~~~~~~~~~
~~~~~\\[4pt]
-\tilde{a}^\star_{i^\prime i}\tilde{a}_{j^\prime j}
(i^\prime j^\prime|F|kl\}
\|+\|
\tilde{b}^\star_{i^\prime i}\tilde{b}_{j^\prime j}
(i^\prime j^\prime|\overline{F}|kl\}^\star,\\
\\[-6pt]
\{ij|\overline{F}|kl\}
\|=\|
\tilde{a}^\star_{i^\prime i}\tilde{b}_{j^\prime j}
(i^\prime j^\prime|D|kl\}^\star
\|-\|
\tilde{b}^\star_{i^\prime i}\tilde{a}_{j^\prime~ j}
(i^\prime j^\prime|\overline{D}|kl\}
~~~~~~~~~~~~~~~~~~~~~~~~~~~~
~\\[4pt]
-\tilde{b}^\star_{i^\prime i}\tilde{b}_{j^\prime j}
(i^\prime j^\prime|F|kl\}^\star
\|+\|
\tilde{a}^\star_{i^\prime i}\tilde{a}_{j^\prime j}
(i^\prime j^\prime|\overline{F}|kl\}, 
\end{array} \!\!
\right\}
\label{explicit expres of bold D, F}
\end{eqnarray}\\[-24pt]
\begin{eqnarray}
\!\!\!\!\!\!\!\!
\left.
\begin{array}{r}
\{ij|D|kl\}
\|=\|
\tilde{a}_{i^\prime i}\tilde{a}_{j^\prime j}
(i^\prime j^\prime|D|kl\}
\|-\|
\tilde{b}_{i^\prime i}\tilde{b}_{j^\prime j}
(i^\prime j^\prime|\overline{D}|kl\}^\star
~~~~~~~~~~~~~~~~~~~~~~~~
~~~~~\\[4pt]
-\tilde{b}_{i^\prime i}\tilde{a}_{j^\prime j}
(i^\prime j^\prime|F|kl\}
\|+\|
\tilde{a}_{i^\prime i}\tilde{b}_{j^\prime j}
(i^\prime j^\prime|\overline{F}|kl\}^\star,\\
\\[-6pt]
-\{ij|\overline{D}|kl\}
\|=\|
\tilde{b}_{i^\prime i}\tilde{b}_{j^\prime j}
(i^\prime j^\prime|D|kl\}^\star
\|-\|
\tilde{a}_{i^\prime i}\tilde{a}_{j^\prime j}
(i^\prime j^\prime|\overline{D}|kl\}
~~~~~~~~~~~~~~~~~~~~~~~~~~~~~
~
\\[4pt]
-\tilde{a}_{i^\prime i}\tilde{b}_{j^\prime j}
(i^\prime j^\prime|F|kl\}^\star
\|+\|
\tilde{b}_{i^\prime i}\tilde{a}_{j^\prime j}
(i^\prime j^\prime|\overline{F}|kl\}.
\end{array}
\right\}
\label{explicit expres of bold D, F2}
\end{eqnarray}\\[-10pt]
By making the block off-diagonal matrices 
of the LHS of the equation vanish, 
we get\\[-18pt]
\begin{eqnarray}
\left.
\begin{array}{l}
i\hbar\partial_t \psi_{o-\n}
\!=\!
~~~\!
[\Check{\varepsilon}_0,\psi_{o-\n}]_+
\!\!-\!\!
\{{\bf D} \}\psi_{o-\n}
-\!
\{\overline{{\bf D}}\}
~
\varphi_{o-\n} ,\\
\\[-6pt]
i\hbar\partial_t \varphi_{o-\n}
\!=\!
-
[\Hat{\varepsilon}_0,\varphi_{o-\n}]_+
\!\!+\!\!
\{\overline{{\bf D}} \}^\star\psi_{o-\n}
\!\!+\!\!
\{{\bf D}\}^\star\varphi_{o-\n} .
\end{array}
\right \}
\label{off-diagonal part of formal RPA}
\end{eqnarray}\\[-10pt] 
Due to the vanishing of
the block diagonal matrices in the LHS of 
(\ref{matrix form of formal RPA}) , 
we obtain\\[-8pt]
\begin{equation}
i\hbar\partial_t \xi_{o-\n}
\!=\!
-
[\Hat{\varepsilon}_0,\xi_{o-\n}]
\!+\!
\{{\bf F}\}\psi_{o-\n}
\!+\!
\{\overline{{\bf F}}\}\varphi_{o-\n} .
\label{diagonal part of formal RPA} 
\end{equation}\\[-14pt]
Eqs. 
(\ref{off-diagonal part of formal RPA}) 
and 
(\ref{diagonal part of formal RPA}) 
are very similar to TD equation
\cite{FN.84}
and have 
matrices $\{{\bf D}\}$ etc.
with ($\Lambda,\! \Lambda^\star,\! t$)-dependence.$\!$
Starting from the lowest solution of 
$\psi_{o-\n},~\varphi_{o-\n}$ and $\xi_{o-\n}$,
we proceed to the next leading solution of 
$\psi_{o-\n}$ and $\varphi_{o-\n}$.$\!$
Using these, $\xi_{o-\n}$ 
with the same power is obtained.$\!$
Then, we determine  time dependence of the amplitudes
corresponding to each power iteratively.$\!$
It works well {\em over a wide range} of physics 
beyond the $SO(2N)$ RPA.

Finally,
following Rajeev
\cite{Rajeev.94},
we show the existence of 
symplectic 2-form $\omega$.
Using $\frac{SO(2N)}{U(N)}$ coset variable 
$q( \!=\! ba^{\!-1}) \!\!=\!\! - q^{\mbox{\scriptsize T}}$,
$SO(\!2N\!)\!$ (HB) density matrix ${\cal R}(G)$
is expressed as\\[-18pt]
\begin{eqnarray}
\!\!\!\!\!\!\!\!\!\!\!\!
\left.
\begin{array}{c}
{\cal R}(G)
\!=\!
G \!
\left[ \!\!\!
\begin{array}{cc}
-1_{\!N} & \!\!\!\! 0 \\
\\[-10pt]
0 & \!\!\!\! 1_{\!N}
\end{array} \!\!\!
\right] \!
G^\dagger
\!\!=\!\!
\left[ \!\!
\begin{array}{cc}
2R(G) \!-\!1_N &  \!\!\!\! -2K^\star(G)\\
\\[-10pt]
2K(G) &  \!\!\!\! -2R^\star(G) \!+\! 1_N
\end{array} \!\!
\right] ,\!\!
\begin{array}{c}
R(G)
\!=\!
q^\dag q \!
\left(1_N \!+\! q^\dag q \right)^{\!-1} ,\\
\\[-10pt]
K(G)
\!=\!
- q \!
\left(1_N \!+\! q^\dag q \right)^{\!-1} .
\end{array} \!\!
\end{array} \!\!
\right\} 
\label{densityRg2}
\end{eqnarray}\\[-12pt]
The two-form $\omega$ is given as\\[-8pt]
\beq
\BA{c}
\omega
\!=\!
-
\frac{i}{8}
\mbox{Tr} \!
\left\{\left(d{\cal R}(G)\right)^3 \! \right\}
\!=\!
-
\frac{i}{8}
\mbox{Tr}
\left\{\left(d{\cal R}(G)\right)^3 \!
{\cal R}(G)^2\right\} ,~
d\omega
\!=\!
- d\omega
\!=\!
0 ,
\EA
(\mbox{closed form}) .
\label{twoform}
\eeq\\[-20pt]
\beq
\BA{c}
\omega(U,V)
\!=\!
-
\frac{i}{8}
\mbox{Tr} \!
\left\{ \!
\left[ \!\!\!
\begin{array}{cc}
1_{\!N} & \!\!\!\!\! 0 \\[2pt]
0 & \!\!\!\!\! -1_{\!N}
\end{array} \!\!\!
\right] \!
[U,V]  \!
\right\}
\!=\!
\frac{i}{4}
\mbox{Tr} \!
\left\{ \!
u^\dag v \!-\! v^\dag u \!
\right\} ,
U 
\!\!\equiv\!\!
\left[ \!\!
\begin{array}{cc}
0 & \!\!\! u \\[2pt]
u^\dag & \!\!\! 0
\end{array} \!\!
\right]
\mbox{and~}
V
\!\!\equiv\!\!
\left[ \!\!
\begin{array}{cc}
0 & \!\!\! v \\[2pt]
v^\dag & \!\!\! 0
\end{array} \!\!
\right] ,
\EA
\label{symplectictwoform}
\eeq\\[-4pt]
which is a symplectic form and
makes it possible to practice a geometric quantization on
the FD- and ID- Grassmannians
\cite{RajeevTurgut.98,ToprakTugurt.02}.\\[-12pt]

\newpage

%%%%%%%%%%%%%%%%%%%%%%
%                                                                   %
%   7      Geometric equation with                %
%                                                                   %
%  U_{m+n}/S(U_{m}xU_{n}) structure   %
%                                                                   %
%%%%%%%%%%%%%%%%%%%%%%

\setcounter{equation}{0}
\renewcommand{\theequation}{
\arabic{section}.
\arabic{equation}}
\section{Geometric equation with 
$\frac{SU_{m+n}}{S(U_{m} \times U_{n})}$
structure}

Here we construct alternative geometric equation,
noticing the special structure of the coset space
$\frac{SU_{m+n}}{S(U_{m} \times U_{n})}$.
Composing linearly of the same kind as
the infinitesimal generators
(\ref{Hc Tr-express})
and
(\ref{Oc Tr-express}),
we adopt the following one-form $\Omega_o$
and the integrability condition:\\[-20pt]
\begin{eqnarray}
\begin{array}{c}
\Omega_o
\!=\!
\frac{1}{\hbar} 
{\cal F}_o \!\cdot\! dt
\!+\!
\theta_{o-\n}^\dagger \!\cdot\! d\Lambda_\n
\!+\!
\theta_{o-\n} \!\cdot\! d\Lambda_\n^\star ,~~
d\Omega_o - \Omega_o \!\wedge\! \Omega_o
\!=\!
0 ,
\end{array}
\label{one-form_o eq}
\end{eqnarray}\\[-18pt]
where the expressions for 
${\cal F}_{o},~\theta_{o-\n}^\dagger$
and
$\theta_{o-\n}$
are given as\\[-18pt]
\begin{eqnarray}
\begin{array}{c}
{\cal F}_{o}
\!=\! 
\left[  \!\!
\begin{array}{cc}
\Hat{F}_{o}  & \acute{F}_{o} \\
\grave{F}_{o}       &  \Check{F}_{o}
\end{array}  \!\!
\right] \! , ~
\theta_{o-\n}^\dagger
\!=\!
\left[ \!\!\!
\begin{array}{rr}
\xi_o & \varphi_o\\[1pt]
\psi_o & 
\zeta_o 
\end{array} \!\!\!
\right]_\n,~
\theta_{o-\n}
\!=\!
\left[ \!\!\!
\begin{array}{rr}
\xi_o^\dagger & 
\psi_o^\dagger\\[1pt]
\varphi_o^\dagger & 
\zeta_o^\dagger 
\end{array} \!\!\!
\right]_\n .
\end{array}
\label{one-form-Omega}
\end{eqnarray}\\[-10pt]
Under the conditions
$
\acute{F}_{o}
\!\!=\!\!
- \grave{F}_{o}^\dag ,
\varphi_{o-\n} \!\!=\!\! - \psi_{o-\n}^\dag
$
and
$ n \!\!=\!\! N \!-\! m$,
$\Omega_o$
is divided into as follows:\\[-16pt]
\begin{eqnarray}
\begin{array}{c}
\Omega_o
\!=\!
\left[  \!\!
\begin{array}{cc}
\Hat{\Omega}_{o-m}  & 0 \\
0       &  \Check{\Omega}_{o-n}
\end{array} \!\!
\right]
\!+\!
\left[  \!\!
\begin{array}{cc}
0  & -\grave{\Omega}_{o}^\dag \\
\grave{\Omega}_{o}       &  0
\end{array}  \!\!
\right] ,~
\mbox{Tr} 
{\Omega}_{o}
\!=\!
0 , ~
\left(
\BA{c}
\mbox{using~indices}\\[-4pt]
{\it m}~\mbox{and}~{\it n}~\mbox{instead}\\[-4pt]
~\mbox{of}~
\m~\mbox{and}~\n 
\EA
\right) .
\end{array}
\label{devided-Omega}
\end{eqnarray}\\[-12pt]
where the 
$\Hat{\Omega}_{o-m}$
and
$\Check{\Omega}_{o-n}$
are expressed as\\[-18pt]
\begin{eqnarray}
\!\!
\begin{array}{c}
\Hat{\Omega}_{o-m}
\!=\!
\frac{1}{\hbar}
\Hat{F}_{o} \!\cdot\! dt
\!+\!
\xi_{o-\n} \!\cdot\! d\Lambda_\n
\!+\!
\xi_{o-\n}^\dagger \!\cdot\! d\Lambda_\n^\star ,~
\Check{\Omega}_{o-n}
\!=\!
\frac{1}{\hbar}
\Check{F}_{o} \!\cdot\! dt
\!+\!
\zeta_{o-\n} \!\cdot\! d\Lambda_\n
\!+\!
\zeta_{o-\n}^\dagger \!\cdot\! d\Lambda_\n^\star . 
\end{array} \!\!
\label{devided-Omega-n}
\end{eqnarray}\\[-16pt]
and
according to
Caseller--Megna-Sciuto (CMS)
\cite{CMS.81},
we adopt the $\lambda$-modified 
$\grave{\Omega}_{o}$
given as\\[-18pt]
\begin{eqnarray}
\begin{array}{c}
\grave{\Omega}_{o}
\!=\!
\lambda
\frac{1}{\hbar}
\grave{F}_{o} \!\cdot\! dt
\!+\!
\frac{1}{\lambda} 
\psi_{o-\n} \!\cdot\! d\Lambda_\n
\!+\!
\frac{1}{\lambda} 
\varphi_{o-\n}^\dagger \!\cdot\! d\Lambda_\n^\star .
\end{array} \!\!
\label{Omega-o}
\end{eqnarray}\\[-18pt]
Further,
the integrability condition
is calculated as\\[-14pt]
\begin{eqnarray}
\begin{array}{ll}
\left[  \!\!
\begin{array}{cc}
d\Hat{\Omega}_{o-m}  &\!\! -d\grave{\Omega}_{o}^\dag \\[1pt]
d\grave{\Omega}_{o}  &\!\!  d\Check{\Omega}_{o-n}
\end{array} \!\!
\right]
&
\!\!\!=\!
\left[  \!\!
\begin{array}{cc}
\Hat{\Omega}_{o-m}  &\!\! -\grave{\Omega}_{o}^\dag \\[1pt]
\grave{\Omega}_{o}  &\!\!  \Check{\Omega}_{o-n}
\end{array} \!\!
\right]
\!\wedge\!
\left[  \!\!
\begin{array}{cc}
\Hat{\Omega}_{o-m}  &\!\! - \grave{\Omega}_{o}^\dag \\[1pt]
\grave{\Omega}_{o}  &\!\!  \Check{\Omega}_{o-n}
\end{array} \!\!
\right] \\
\\[-10pt]
&
\!\!\!=\!
\left[  \!\!
\begin{array}{cc}
\Hat{\Omega}_{o-m} \!\wedge\! \Hat{\Omega}_{o-m}  
\!-\!
\grave{\Omega}_{o}^\dag \!\wedge\! \grave{\Omega}_{o}
& -\Hat{\Omega}_{o-m}  \!\wedge\! \grave{\Omega}_{o}^\dag
\!-\!
\grave{\Omega}_{o}^\dag \!\wedge\!  \Check{\Omega}_{o-n}\\[1pt]
\grave{\Omega}_{o} \!\wedge\! \Hat{\Omega}_{o-m}
\!+\!
\Check{\Omega}_{o-n} \!\wedge\! \grave{\Omega}_{o}       
& \Check{\Omega}_{o-n} \!\wedge\! \Check{\Omega}_{o-n}
\!-\!
\grave{\Omega}_{o} \!\wedge\! \grave{\Omega}_{o}^\dag
\end{array} \!\!
\right] ,
\end{array}
\label{integ-condi}
\end{eqnarray}\\[-12pt]
in which
$d\grave{\Omega}_{o}$
is given as\\[-12pt]
\begin{eqnarray}
\!\!\!\!\!\!\!\!
\begin{array}{ll}
&
d\grave{\Omega}_{o}
\!=\!
- \partial_t \grave{\Omega}_{o} \!\wedge\! dt
\!-\! 
\partial_\Lambda{}_\n \grave{\Omega}_{o} 
\!\wedge\! 
d\Lambda_\n
\!-\! 
\partial_\Lambda{}^{\!\!\star}_\n \grave{\Omega}_{o} 
\!\wedge\! 
d\Lambda^\star_\n \\
\\[-12pt]
&
\!=\!
-
\left(
\lambda
\frac{1}{\hbar}
\partial_t \grave{F}_{o} \!\cdot\! dt
\!+\!
\frac{1}{\lambda} 
\partial_t \psi_{o-\n} \!\cdot\! d\Lambda_\n
\!+\!
\frac{1}{\lambda} 
\partial_t \varphi_{o-\n}^\dagger \!\cdot\! d\Lambda_\n^\star
\right) \!
\!\wedge\! dt \\
\\[-10pt]
&
~~\!
-
\left(
\lambda
\frac{1}{\hbar}
\partial_\Lambda{}_\n \grave{F}_{o} \!\cdot\! dt
\!+\!
\frac{1}{\lambda} 
\partial_\Lambda{}_\n 
\psi_{o-\n'} \!\cdot\! d\Lambda_{\n'}
\!+\!
\frac{1}{\lambda} 
\partial_\Lambda{}_\n 
\varphi_{o-\n'}^\dagger \!\cdot\! d\Lambda_{\n'}^\star
\right) \!
\!\wedge\! d\Lambda_\n \\
\\[-10pt]
&
~~\!
-
\left(
\lambda
\frac{1}{\hbar}
\partial_\Lambda{}^{\!\!\star}_\n \grave{F}_{o} \!\cdot\! dt
\!+\!
\frac{1}{\lambda} 
\partial_\Lambda{}^{\!\!\star}_\n 
\psi_{o-\n'} \!\cdot\! d\Lambda_{\n'}
\!+\!
\frac{1}{\lambda} 
\partial_\Lambda{}^{\!\!\star}_\n 
\varphi_{o-\n'}^\dagger \!\cdot\! d\Lambda_{\n'}^\star
\right) \!
\!\wedge\! d\Lambda^\star_\n \\
\\[-10pt]
&
\!=\!
-
\lambda \!
\left( \!
\frac{1}{\hbar} 
\partial_\Lambda{}_\n 
\grave{F}_{o}  
dt \!\wedge\! d\Lambda_\n
\!\!+\!\!
\frac{1}{\hbar}
\partial_\Lambda{}^{\!\!\star}_\n 
\grave{F}_{o}  
dt \!\wedge\! d\Lambda^\star_\n \!
\right) 
\!+\!
\frac{1}{\lambda} \!
\left( \!
\partial_t \psi_{o-\n}  
dt \!\wedge\! d\Lambda_\n
\!\!+\!\!
\partial_t 
\varphi_{o-\n}^\dagger  
dt \!\wedge\! d\Lambda_\n^\star
\right. \\
\\[-14pt]
&
\left.
\!\!+\!
\partial_\Lambda{}_\n 
\psi_{o-\n'} 
d\Lambda_\n \!\wedge\! d\Lambda_{\n'}
\!\!+\!\!
\partial_\Lambda{}^{\!\!\star}_\n 
\varphi_{o-\n'}^\dagger 
d\Lambda_{\n}^\star \!\wedge\! d\Lambda_{\n'}^\star
\!\!+\!\!
\partial_\Lambda{}_\n 
\varphi_{o-\n'}^\dagger 
d\Lambda_\n 
\!\wedge\! d\Lambda_{\n'}^\star
\!\!+\!\!
\partial_\Lambda{}^{\!\!\star}_\n 
\psi_{o-\n'}  
d\Lambda_{\n}^\star \!\wedge\! d\Lambda_{\n'}^\star \!
\right) \! .
\end{array}
\label{d-Omega}
\end{eqnarray}\\[-10pt]
On the other hand,
the term to equalize with 
$d\grave{\Omega}_{o}$
is computed as\\[-14pt]
\begin{eqnarray}
\!\!\!\!
\begin{array}{ll}
&
\grave{\Omega}_{o} \!\wedge\! \Hat{\Omega}_{o-m}
\!+\!
\Check{\Omega}_{o-n} \!\wedge\! \grave{\Omega}_{o} \\
\\[-12pt]
&   
\!=\!
\left(
\lambda
\frac{1}{\hbar}
\grave{F}_{o} \!\cdot\! dt
\!+\!
\frac{1}{\lambda} 
\psi_{o-\n} \!\cdot\! d\Lambda_\n
\!+\!
\frac{1}{\lambda} 
\varphi_{o-\n}^\dagger \!\cdot\! d\Lambda_\n^\star
\right) 
\!\wedge\!
\left(
\frac{1}{\hbar}
\Hat{F}_{o} \!\cdot\! dt
\!+\!
\xi_{o-\n'} \!\cdot\! d\Lambda_{\n'}
\!+\!
 \xi_{o-\n'}^\dagger \!\cdot\! d\Lambda_{\n'}^\star 
\right)\\
\\[-12pt]
&   
\!+\!
\left(
\frac{1}{\hbar}
\Check{F}_{o} \!\cdot\! dt
\!+\!
\zeta_{o-\n} \!\cdot\! d\Lambda_\n
\!+\!
 \zeta_{o-\n}^\dagger \!\cdot\! d\Lambda_\n^\star
\right)
\!\wedge\!
\left(
\lambda
\frac{1}{\hbar}
\grave{F}_{o} \!\cdot\! dt
\!+\!
\frac{1}{\lambda} 
\psi_{o-\n'} \!\cdot\! d\Lambda_{\n'}
\!+\!
\frac{1}{\lambda} 
\varphi_{o-\n'}^\dagger \!\cdot\! d\Lambda_{\n'}^\star
\right) \\
\\[-10pt]
&   
\!=\!
\lambda \!
\left\{ \!
\left( \!
\frac{1}{\hbar}
\grave{F}_{o} \xi_{o-\n}
\!-\! 
\zeta_{o-\n} 
\frac{1}{\hbar}
\grave{F}_{o} \!
\right) \!
dt \!\wedge\! d\Lambda_\n
\!\!+\!\!
\left( \!
\frac{1}{\hbar}
\grave{F}_{o} \xi_{o-\n}^\dag
\!-\! 
\zeta_{o-\n}^\dag 
\frac{1}{\hbar}
\grave{F}_{o} \!
\right) \!
dt \!\wedge\! d\Lambda_\n^\star \!
\right\} \\
\\[-10pt]
&
\!+\!
\frac{1}{\lambda} \!
\left\{ \!
\left( \!
\frac{1}{\hbar}
\Check{F}_{o} \psi_{o-\n}
\!-\! 
\psi_{o-\n} 
\frac{1}{\hbar}
\Hat{F}_{o} \!
\right) \!  
dt \!\wedge\! d\Lambda_\n
\!+\!
\left( \!
\frac{1}{\hbar}
\Check{F}_{o} \varphi_{o-\n}^\dag
\!-\! 
\varphi_{o-\n}^\dag 
\frac{1}{\hbar}
\Hat{F}_{o} \!
\right)  \! 
dt \!\wedge\! d\Lambda_\n^\star
\right. \\
\\[-12pt]
&
\left.
\!\!+
\left( \! 
\psi_{o-\n} \xi_{o-\n'}
\!+\!
\zeta_{o-\n} \psi_{o-\n'} \!
\right) \! 
d\Lambda_\n \!\wedge\! d\Lambda_{\n'}
\!+\!
\left( \! 
\varphi_{o-\n'}^\dagger \xi_{o-\n'}^\dag
\!+\!
\zeta_{o-\n}^\dag \varphi_{o-\n'}^\dagger \!
\right) \! 
d\Lambda_{\n}^\star \!\wedge\! d\Lambda_{\n'}^\star \right.\\
\\[-12pt]
&
\left.
\!\!+
\left( \! 
\psi_{o-\n} \xi_{o-\n'}^\dag
\!+\!
\zeta_{o-\n} \varphi_{o-\n'}^\dagger \!
\right)  \! 
d\Lambda_\n 
\!\wedge\! d\Lambda_{\n'}^\star
\!+\!
\left( \! 
\varphi_{o-\n'}^\dagger \xi_{o-\n'}
\!+\!
\zeta_{o-\n}^\dag \varphi_{o-\n'} \!
\right) \! 
d\Lambda_{\n}^\star \!\wedge\! d\Lambda_{\n'}^\star \!
\right\} \! .
\end{array}
\label{dOmega}
\end{eqnarray}\\[-12pt]
Equating the $\lambda$ term and 
the $\frac{1}{\lambda}$ term
in both sides of
(\!\ref{d-Omega})
and
(\!\ref{dOmega}),
we get the relations\\[-22pt]
\begin{eqnarray}
\begin{array}{c}
\lambda\mbox{-term}:~
\partial_\Lambda{}_\n
\grave{F}_{o} 
\!=\!
\zeta_{o-\n} \grave{F}_{o}
\!-\! 
 \grave{F}_{o} \xi_{o-\n},~
\partial_\Lambda{}^{\!\!\star}_\n 
\grave{F}_{o}
\!=\!
\zeta_{o-\n}^\dag \grave{F}_{o}
\!-\! 
\grave{F}_{o} \xi_{o-\n}^\dag ,
\end{array}
\label{dOmegaEqL}
\end{eqnarray}
\vspace{-1.0cm}
\begin{eqnarray}
\!\!\!\!\!\!\!\!
\left.
\begin{array}{ll}
\frac{1}{\lambda}\mbox{-term}:
&
\hbar \partial_t \psi_{o-\n} 
\!=\!
\Check{F}_{o}\psi_{o-\n}
\!-\!
\psi_{o-\n} \Hat{F}_{o} ,~
\hbar \partial_t 
\varphi_{o-\n}^\dagger
\!=\!
\Check{F}_{o}\varphi_{o-\n}^\dag
\!-\!
\varphi_{o-\n}^\dag \Hat{F}_{o} ,\\
\\[-12pt]
&
\partial_\Lambda{}_\n 
\psi_{o-\n'}
\!=\!
\psi_{o-\n} \xi_{o-\n'}
\!+\!
\zeta_{o-\n} \psi_{o-\n'} \! ,~
\partial_\Lambda{}^{\!\!\star}_\n 
\varphi_{o-\n'}^\dagger
\!=\!
\varphi_{o-\n'}^\dagger \xi_{o-\n'}^\dag
\!+\!
\zeta_{o-\n}^\dag \varphi_{o-\n'}^\dagger \! ,\\
\\[-12pt]
&
\partial_\Lambda{}_\n 
\varphi_{o-\n'}^\dagger
\!=\!
\psi_{o-\n} \xi_{o-\n'}^\dag
\!+\!
\zeta_{o-\n} \varphi_{o-\n'}^\dagger \!,~
\partial_\Lambda{}^{\!\!\star}_\n \!
\psi_{o-\n'}
\!=\!
\varphi_{o-\n'}^\dagger \xi_{o-\n'}
\!+\!
\zeta_{o-\n}^\dag \varphi_{o-\n'} \! ,
\end{array} \!\!
\right\}
\label{dOmegaEq1/L}
\end{eqnarray}\\[-12pt]
using
$\varphi_{\!o-n} \!\!=\!\! - \psi_{\!o-n}^\dag$,
from Eqs. of second and third lines in
(\!\ref{dOmegaEq1/L}),
we acquire   
$
\xi_{o-\n}^\dag
\!=\!
- \xi_{o-\n}
$.

In
(\!\ref{integ-condi}),
we have\\[-18pt]
\begin{eqnarray}
\!\!\!\!\!\!\!\!
\begin{array}{ll}
&
d\Check{\Omega}_{o-n}
\!=\!
- \partial_t \Check{\Omega}_{o-n} \!\wedge\! dt
\!-\! 
\partial_\Lambda{}_\n \Check{\Omega}_{o-n} 
\!\wedge\! 
d\Lambda_\n
\!-\! 
\partial_\Lambda{}^{\!\!\star}_\n \Check{\Omega}_{o-n} 
\!\wedge\! 
d\Lambda^\star_\n \\
\\[-10pt]
&
\!=\!
-
\left(
\partial_t \frac{1}{\hbar} 
\Check{F}_{o} \!\cdot\! dt
\!+\!
\partial_t \zeta_{o-\n} \!\cdot\! d\Lambda_\n
\!+\!
\partial_t \zeta_{o-\n}^\dagger \!\cdot\! d\Lambda_\n^\star
\right) \!
\!\wedge\! dt \\
\\[-10pt]
&
~~\!
-
\left(
\partial_\Lambda{}_\n  \frac{1}{\hbar} 
\Check{F}_{o} \!\cdot\! dt
\!+\!
\partial_\Lambda{}_\n 
\zeta_{o-\n'} \!\cdot\! d\Lambda_{\n'}
\!+\!
\partial_\Lambda{}_\n 
\zeta_{o-\n'}^\dagger \!\cdot\! d\Lambda_{\n'}^\star
\right) \!
\!\wedge\! d\Lambda_\n \\
\\[-10pt]
&
~~\!
-
\left(
\partial_\Lambda{}^{\!\!\star}_\n  \frac{1}{\hbar}
\Check{F}_{o} \!\cdot\! dt
\!+\!
\partial_\Lambda{}^{\!\!\star}_\n 
\zeta_{o-\n'} \!\cdot\! d\Lambda_{\n'}
\!+\!
\partial_\Lambda{}^{\!\!\star}_\n 
\zeta_{o-\n'}^\dagger \!\cdot\! d\Lambda_{\n'}^\star
\right) \!
\!\wedge\! d\Lambda^\star_\n \\
\\[-10pt]
&
\!=\!
-
\partial_\Lambda{}_\n \frac{1}{\hbar}
 \Check{F}_{o} 
dt \!\wedge\! d\Lambda_\n
\!\!-\!\!
\partial_\Lambda{}^{\!\!\star}_\n \frac{1}{\hbar}
 \Check{F}_{o}  
dt \!\wedge\! d\Lambda^\star_\n
\!+\!
\partial_t  \zeta_{o-\n} \! 
dt \!\wedge\! d\Lambda_\n
\!\!+\!\!
\partial_t 
\zeta_{o-\n}^\dagger \!  
dt \!\wedge\! d\Lambda_\n^\star \\
\\[-8pt]
&
\!\!+
\partial_\Lambda{}_\n 
\zeta_{o-\n'} 
d\Lambda_\n \!\wedge\! d\Lambda_{\n'}
\!+\!
\partial_\Lambda{}^{\!\!\star}_\n 
\zeta_{o-\n'}^\dagger  
d\Lambda_{\n}^\star \!\wedge\! d\Lambda_{\n'}^\star
\!+\!
\partial_\Lambda{}_\n 
\zeta_{o-\n'}^\dagger  
d\Lambda_\n 
\!\wedge\! d\Lambda_{\n'}^\star
\!+\!
\partial_\Lambda{}^{\!\!\star}_\n 
\zeta_{o-\n'} 
d\Lambda_{\n}^\star \!\wedge\! d\Lambda_{\n'} \! .
\end{array}
\label{d-Omega-n}
\end{eqnarray}\\[-10pt]
The terms 
to equalize with 
$d\grave{\Omega}_{o-n}\!$
are computed
as\\[-16pt]
\begin{eqnarray}
\!\!\!\!
\begin{array}{ll}
&
\Check{\Omega}_{o-n} \!\wedge\! \Check{\Omega}_{o-n}
\!-\!
\grave{\Omega}_{o} \!\wedge\! \grave{\Omega}_{o}^\dag \\
\\[-10pt]
&   
\!=\!
\left( \!
\frac{1}{\hbar}
\Check{F}_{o} \!\cdot\! dt
\!+\!
\zeta_{o-\n} \!\cdot\! d\Lambda_{\n}
\!+\!
 \zeta_{o-\n}^\dagger \!\cdot\! d\Lambda_{\n}^\star \!
\right) 
\!\wedge\!
\left( \!
\frac{1}{\hbar}
\Check{F}_{o} \!\cdot\! dt
\!+\!
\zeta_{o-\n'} \!\cdot\! d\Lambda_{\n'}
\!+\!
 \zeta_{o-\n'}^\dagger \!\cdot\! d\Lambda_{\n'}^\star \!
\right)\\
\\[-8pt]
&   
\!-\!
\left( \!
\lambda
\frac{1}{\hbar}
\grave{F}_{o} \!\cdot\! dt
\!+\!
\frac{1}{\lambda} 
\psi_{o-\n} \!\cdot\! d\Lambda_\n
\!+\!
\frac{1}{\lambda} 
\varphi_{o-\n}^\dagger \!\cdot\! d\Lambda_\n^\star \!
\right)
\!\wedge\!
\left( \!
\lambda
\frac{1}{\hbar}
\grave{F}_{o}^\dag \!\cdot\! dt
\!+\!
\frac{1}{\lambda} 
\psi_{o-\n'}^\dag \!\cdot\! d\Lambda_{\n'}
\!+\!
\frac{1}{\lambda} 
\varphi_{o-\n'} \!\cdot\! d\Lambda_{\n'}^\star \!
\right) \\
\\[-8pt]
&   
\!=\!
\!-\!
\left( \!
\frac{1}{\hbar}
\grave{F}_{o} \psi_{o-\n}^\dag
\!-\! 
\psi_{o-\n} 
\frac{1}{\hbar}
\grave{F}_{o}^\dag \!
\right) \! 
dt \!\wedge\! d\Lambda_\n
\!-\!
\left( \!
\frac{1}{\hbar}
\grave{F}_{o} \varphi_{o-\n}
\!-\! 
\varphi_{o-\n}^\dag 
\frac{1}{\hbar}
\grave{F}_{o}^\dag \!
\right) \!  
dt \!\wedge\! d\Lambda_\n^\star \! \\
\\[-8pt]
&
\!+\!
\left( \!
\frac{1}{\hbar}
\Check{F}_{o} \zeta_{o-\n}
\!-\! 
\zeta_{o-\n} 
\frac{1}{\hbar}
\Check{F}_{o} \!
\right) \!
dt \!\wedge\! d\Lambda_\n
\!\!+\!\!
\left( \!
\frac{1}{\hbar}
\Check{F}_{o} \zeta_{o-\n}^\dag
\!-\! 
\zeta_{o-\n}^\dag 
\frac{1}{\hbar}
\Check{F}_{o} \!
\right) \!
dt \!\wedge\! d\Lambda_\n^\star \\
\\[-10pt]
&
\!+\!
\zeta_{o-\n} \zeta_{o-\n'}
d\Lambda_\n \!\wedge\! d\Lambda_{\n'}
\!+\!
\zeta_{o-\n}^\dagger \zeta_{o-\n'}^\dag
d\Lambda_{\n}^\star \!\wedge\! d\Lambda_{\n'}^\star
\!+\!
\zeta_{o-\n} \zeta_{o-\n'}^\dag
d\Lambda_\n 
\!\wedge\! d\Lambda_{\n'}^\star
\!+\!
\zeta_{o-\n}^\dagger \zeta_{o-\n'} 
d\Lambda_{\n}^\star \!\wedge\! d\Lambda_{\n'} .
\end{array}
\label{dOmega_n}
\end{eqnarray}\\[-10pt]
Here
we should discard the $\frac{1}{\lambda^2}$ term. 
Equating both sides of
(\!\ref{d-Omega-n})
and
(\!\ref{dOmega_n}),
we get \\[-20pt]
\begin{eqnarray}
\!\!\!\!\!\!\!\!
\begin{array}{ll}
&
\partial_\Lambda{}_\n 
\Check{F}_{\!o} 
\!=\!
\grave{F}_{o} \psi_{o-\n}^\dag
\!-\! 
\psi_{o-\n} 
\grave{F}_{o}^\dag ,~~
\partial_\Lambda{}^{\!\!\star}_\n 
\Check{F}_{\!o}
\!=\!
\grave{F}_{o} \varphi_{o-\n}
\!-\! 
\varphi_{o-\n}^\dag 
\grave{F}_{o}^\dag ,
\end{array}
\label{dOmegaEqL2}
\end{eqnarray}
\vspace{-0.9cm}
\begin{eqnarray}
\!\!\!\!\!\!\!\!\!\!\!\!\!
\left.
\begin{array}{ll}
&
\hbar \partial_t \zeta_{o-\n} 
\!=\!
\Check{F}_{o} \zeta_{o-\n}
\!-\! 
\zeta_{o-\n} 
\Check{F}_{o} ,~
\hbar \partial_t \zeta_{o-\n}^\dag
\!=\!
\Check{F}_{o} \zeta_{o-\n}^\dag
\!-\! 
\zeta_{o-\n}^\dag 
\Check{F}_{o} ,\\
\\[-8pt]
&
\partial_\Lambda{}_\n 
\zeta_{o-\n'}
\!\!=\!\!
\zeta_{o-\n} \zeta_{o-\n'} ,~
\partial_\Lambda{}^{\!\!\star}_\n 
\zeta_{o-\n'}^\dagger
\!\!=\!\!
\zeta_{o-\n}^\dagger \zeta_{o-\n'}^\dag ,~
\partial_\Lambda{}_\n 
\zeta_{o-\n'}^\dag
\!\!=\!\!
\zeta_{o-\n} \zeta_{o-\n'}^\dag ,~
\partial_\Lambda{}^{\!\!\star}_\n 
\zeta_{o-\n'}
\!\!=\!\!
\zeta_{o-\n}^\dag \zeta_{o-\n'} .
\end{array} \!\!\!
\right\}
\label{dOmegaEq1/L2}
\end{eqnarray}\\[-24pt]

Following
CMS
\cite{CMS.81},
under
$
h_m, k_m
\!\in\!
U_m
$
and
$
h_n, k_n
\!\in\!
U_n
$,
we assume\\[-20pt]
\begin{eqnarray}
\begin{array}{ll}
&
\grave{F}_{o}
\!=\!
h_n
\grave{\mathbf{F}}_{0}
h_m,~
\grave{\mathbf{F}}_{0}
\!=\!
\left[ \!\!
\BA{c}
\cdot\cdot\cdot 
\grave{\mathbf{F}}_{0}^\prime 
\cdot\cdot\cdot \\[2pt]
\underbrace{
\cdot\cdot\cdot 0 \cdot\cdot\cdot 
}_{m}
\EA \!\!
\right]
\!\!\!\!
\BA{l}
\left.{}\right\}m\\
\left.{}\right\}n-m 
\EA,~
\psi_{o-\n}
\!=\!
k_n
\psi_{o-\n}^0
k_m,~\\
\\[-8pt]
&
(\grave{\mathbf{F}}_{0}^\prime)_{ij}
\!\!=\!\!
\delta_{ij}
\grave{\mathbf{f}}_{~\!\!0,i} ,~
\partial_\Lambda{}_\n
\grave{\mathbf{f}}_{~\!\!0,i}
\!\!=\!\!
0 ,~~
({\psi}_{o-\n}^0)_{ij}
\!\!=\!\!
\delta_{ij}
{\mathfrak{f}}_{~\!\!i}^0,~ 
\hbar \partial_t 
{\mathfrak{f}}_{~\!\!i}^0
\!\!=\!\!
0 .
\end{array} \!\!
\label{assumption}
\end{eqnarray}\\[-10pt]
As also suggested by
CMS
\cite{CMS.81},
Eqs.
(\!\ref{dOmegaEqL})
and
(\!\ref{dOmegaEq1/L})
imply\\[-20pt]
\begin{eqnarray}
\left.
\begin{array}{c}
\partial_\Lambda{}_\n \!
\mbox{Tr} \!
\left( \!
\grave{F}_{o}^\dag
\grave{F}_{o}
\right)^{\!\!\mbox{\scriptsize T}} \!
\!\!=\!\!
0 ,~
\grave{F}_{o}^\dag
\grave{F}_{o}
\!=\!
c_F \I_m,~
\hbar \partial_t \!
\mbox{Tr} \!
\left( \!
\psi_{o-\n}^\dag
\psi_{o-\n}
\right)^{\!\!\mbox{\scriptsize T}} \!
\!\!=\!\!
0 ,
\BA{c}
\psi_{o-\n}^\dag
\psi_{o-\n}\\[2pt]
\psi_{o-\n}
\psi_{o-\n}^\dag
\EA
\!\!\!\!=\!\!\!\!
\BA{c}
-c_\psi \I_m\\[2pt]
c_\psi \I_m
\EA \! ,
\end{array} \!
\right\}
\label{dTrF-F}
\end{eqnarray}\\[-14pt]
which is proved as\\[-20pt]
\begin{eqnarray}
\!\!\!\!\!\!\!\!
\begin{array}{ll}
&
0
\!=\!
\partial_\Lambda{}_\n \!
\mbox{Tr} \!
\left( \!
\grave{F}_{o}^\dag
\grave{F}_{o}
\right)^{\!\!\mbox{\scriptsize T}} \!
\!\!=\!\!
\mbox{Tr} \!
\left( \!
\partial_\Lambda{}_\n 
\grave{F}_{o}^{\mbox{\scriptsize T}}
\grave{F}_{o}^\star
\!\!+\!\!
\grave{F}_{o}^{\mbox{\scriptsize T}} 
\partial_\Lambda{}_\n 
\grave{F}_{o}^\star
\right) \\
\\[-12pt]
&
\!\!=\!\! 
\mbox{Tr} \!
\left\{ \!\!
\left( \!
\grave{F}_{o}^{\mbox{\scriptsize T}} 
\zeta_{o-\n}^{\mbox{\scriptsize T}} 
\!\!-\!\! 
\xi_{o-\n}^{\mbox{\scriptsize T}} 
\grave{F}_{o}^{\mbox{\scriptsize T}} \!
\right) \!\!
\grave{F}_{o}^\star
\!\!+\!\!
\grave{F}_{o}^{\mbox{\scriptsize T}} \!\!
\left( \!
\zeta_{o-\n}^\star \grave{F}_{o}^\star
\!\!-\!\! 
\grave{F}_{o}^\star \xi_{o-\n}^\star \!
\right) \!\!
\right\}
\!\!=\!\! 
\mbox{Tr} \!
\left\{ \!\!
\left( \!
\grave{F}_{o}
\grave{F}_{o}^{\dag} 
\!\!-\!\!
\grave{F}_{o}^{\dag} 
\grave{F}_{o}  \!
\right) \!\!
\left(  \!
\zeta_{o-\n}
\!\!+\!\! 
\xi_{o-\n}^\dag \!
\right) \!\!
\right\} ,\\
\\[-10pt]
&
0
\!=\!
\hbar \partial_t \!
\mbox{Tr} \!
\left( \!
\psi_{o-\n}^\dag \!\!
\psi_{o-\n}
\right)^{\!\!\mbox{\scriptsize T}} \! \!
\!=\!
\mbox{Tr} \!
\left( \!
\hbar \partial_t 
\psi_{o-\n}^{\mbox{\scriptsize T}}
\psi_{o-\n}^\star
\!\!+\!\!
\psi_{o-\n}^{\mbox{\scriptsize T}} 
\hbar \partial_t 
\psi_{o-\n}^\star \!
\right) \\
\\[-12pt]
&
\!\!=\!\! 
\mbox{Tr} \!
\left\{ \!
\psi_{o-\n}^{\mbox{\scriptsize T}} \!\!
\left( \!
\Hat{F}_{o}^\star \psi_{o-\n}^\star
\!\!-\!\!
\psi_{o-\n}^\star \Check{F}_{o}^\star \!
\right)
\!-\!
\left( \!
\Check{F}_{o}^{\mbox{\scriptsize T}} 
\psi_{o-\n}^{\mbox{\scriptsize T}} 
\!\!-\!\! 
\psi_{o-\n}^{\mbox{\scriptsize T}} 
\Hat{F}_{o}^{\mbox{\scriptsize T}} \!
\right) \!\!
\psi_{o-\n}^\star \!
\right\} \\
\\[-10pt]
&
\!\!=\!\! 
\mbox{Tr} \!
\left\{ \!
\psi_{o-\n}
\psi_{o-\n}^{\dag} \!
\left(  \!
\Hat{F}_{o}
\!\!+\!\! 
\Hat{F}_{o}^\dag \!
\right) \!
\right\}
\!\!-\!\!
\mbox{Tr} \!
\left\{ \!
\psi_{o-\n}^{\dag} 
\psi_{o-\n}  \!
\left(  \!
\Check{F}_{o}
\!\!+\!\! 
\Check{F}_{o}^\dag \!
\right) \!
\right\} ,~
(\mbox{Due~to~Tr}
(\Hat{F}_{o}\!\!+\!\!\Check{F}_{o})\!\!=\!\!0) .
\end{array} \!\! 
\label{proofdTrF-F}
\end{eqnarray}\\[-12pt]
Further
putting 
$\zeta_{o-\n} \!=\! 0$
and
substituting it into
(\!\ref{dOmegaEqL})
and
(\!\ref{dOmegaEq1/L}),
we have the relations\\[-22pt]
\begin{eqnarray}
\begin{array}{c}
\partial_\Lambda{}_\n
\grave{F}_{o} 
\!=\!
- \grave{F}_{o} \xi_{o-\n},~
\mbox{i.e.},
\xi_{o-\n}
\!=\!
-
h_m^{\!-1}\partial_\Lambda{}_\n
h_m~~
\mbox{and}~~
\partial_\Lambda{}_\n 
\psi_{o-\n'}
\!\!=\!\!
\psi_{o-\n} \xi_{o-\n'} .
\end{array}
\label{dOmegaEqL1}
\end{eqnarray}\\[-34pt]

Lastly, 
in
(\!\ref{integ-condi}),
we have\\[-20pt]
\begin{eqnarray}
\!\!\!\!\!\!\!\!
\begin{array}{ll}
&
d\Hat{\Omega}_{o-m}
\!=\!
- \partial_t \Hat{\Omega}_{o-m} \!\wedge\! dt
\!-\! 
\partial_\Lambda{}_\n \Hat{\Omega}_{o-m} 
\!\wedge\! 
d\Lambda_\n
\!-\! 
\partial_\Lambda{}^{\!\!\star}_\n \Hat{\Omega}_{o-m} 
\!\wedge\! 
d\Lambda^\star_\n \\
\\[-14pt]
&
\!=\!
-
\left(
\partial_t \frac{1}{\hbar} 
\Hat{F}_{o} \!\cdot\! dt
\!+\!
\partial_t \xi_{o-\n} \!\cdot\! d\Lambda_\n
\!+\!
\partial_t \xi_{o-\n}^\dagger \!\cdot\! d\Lambda_\n^\star
\right) \!
\!\wedge\! dt \\
\\[-10pt]
&
~~\!
-
\left(
\partial_\Lambda{}_\n  \frac{1}{\hbar} 
\Hat{F}_{o} \!\cdot\! dt
\!+\!
\partial_\Lambda{}_\n 
\xi_{o-\n'} \!\cdot\! d\Lambda_{\n'}
\!+\!
\partial_\Lambda{}_\n 
\xi_{o-\n'}^\dagger \!\cdot\! d\Lambda_{\n'}^\star
\right) \!
\!\wedge\! d\Lambda_\n \\
\\[-10pt]
&
~~\!
-
\left(
\partial_\Lambda{}^{\!\!\star}_\n  \frac{1}{\hbar}
\Hat{F}_{o} \!\cdot\! dt
\!+\!
\partial_\Lambda{}^{\!\!\star}_\n 
\xi_{o-\n'} \!\cdot\! d\Lambda_{\n'}
\!+\!
\partial_\Lambda{}^{\!\!\star}_\n 
\xi_{o-\n'}^\dagger \!\cdot\! d\Lambda_{\n'}^\star
\right) \!
\!\wedge\! d\Lambda^\star_\n \\
\\[-12pt]
&
\!=\!
-
\partial_\Lambda{}_\n \frac{1}{\hbar}
 \Hat{F}_{o}  
dt \!\wedge\! d\Lambda_\n
\!\!-\!\!
\partial_\Lambda{}^{\!\!\star}_\n \frac{1}{\hbar}
 \Hat{F}_{o} 
dt \!\wedge\! d\Lambda^\star_\n
\!+\!
\partial_t  
\xi_{o-\n}  
dt \!\wedge\! d\Lambda_\n
\!\!+\!\!
\partial_t 
\xi_{o-\n}^\dagger  
dt \!\wedge\! d\Lambda_\n^\star \\
\\[-10pt]
&
\!\!+
\partial_\Lambda{}_\n 
\xi_{o-\n'}  
d\Lambda_\n \!\wedge\! d\Lambda_{\n'}
\!+\!
\partial_\Lambda{}^{\!\!\star}_\n 
\xi_{o-\n'}^\dagger 
d\Lambda_{\n}^\star \!\wedge\! d\Lambda_{\n'}^\star
\!+\!
\partial_\Lambda{}_\n 
\xi_{o-\n'}^\dagger 
d\Lambda_\n 
\!\wedge\! d\Lambda_{\n'}^\star
\!+\!
\partial_\Lambda{}^{\!\!\star}_\n 
\xi_{o-\n'} 
d\Lambda_{\n}^\star \!\wedge\! d\Lambda_{\n'} \! .
\end{array}
\label{d-Omega-n2}
\end{eqnarray}\\[-10pt]
While,
the term to equalize with 
$d\Hat{\Omega}_{o-m}$
is computed 
as\\[-18pt]
\begin{eqnarray}
\!\!\!\!\!\!\!\!
\begin{array}{ll}
&
\Hat{\Omega}_{o-m} \!\wedge\! \Hat{\Omega}_{o-m}
\!-\!
\grave{\Omega}_{o}^\dag
\!\wedge\!
\grave{\Omega}_{o}   \\
\\[-12pt]
&   
\!=\!
\left( \!
\frac{1}{\hbar}
\Hat{F}_{o} \!\cdot\! dt
\!+\!
\xi_{o-\n} \!\cdot\! d\Lambda_{\n}
\!+\!
 \xi_{o-\n}^\dagger \!\cdot\! d\Lambda_{\n}^\star \!
\right) 
\!\wedge\!
\left( \!
\frac{1}{\hbar}
\Hat{F}_{o} \!\cdot\! dt
\!+\!
\xi_{o-\n'} \!\cdot\! d\Lambda_{\n'}
\!+\!
\xi_{o-\n'}^\dagger \!\cdot\! d\Lambda_{\n'}^\star \!
\right)\\
\\[-12pt]
&   
\!-\!
\left( \!
\lambda
\frac{1}{\hbar}
\grave{F}_{o}^\dag \!\cdot\! dt
\!+\!
\frac{1}{\lambda} 
\psi_{o-\n'}^\dag \!\cdot\! d\Lambda_{\n'}
\!+\!
\frac{1}{\lambda} 
\varphi_{o-\n'} \!\cdot\! d\Lambda_{\n'}^\star \!
\right)
\!\wedge\!
 \left( \!
\lambda
\frac{1}{\hbar}
\grave{F}_{o} \!\cdot\! dt
\!+\!
\frac{1}{\lambda} 
\psi_{o-\n} \!\cdot\! d\Lambda_\n
\!+\!
\frac{1}{\lambda} 
\varphi_{o-\n}^\dagger \!\cdot\! d\Lambda_\n^\star \!
\right) \\
\\[-12pt]
&   
\!=\!
\left( \!
\frac{1}{\hbar}
\Hat{F}_{o} \xi_{o-\n}
\!-\! 
\xi_{o-\n} 
\frac{1}{\hbar}
\Hat{F}_{o} \!
\right) \!
dt \!\wedge\! d\Lambda_\n
\!\!+\!\!
\left( \!
\frac{1}{\hbar}
\Hat{F}_{o} \xi_{o-\n}^\dag
\!-\! 
\xi_{o-\n}^\dag 
\frac{1}{\hbar}
\Hat{F}_{o} \!
\right) \!
dt \!\wedge\! d\Lambda_\n^\star \! \\
\\[-12pt]
&
\!-\!
\left( \!
\frac{1}{\hbar}
\grave{F}_{o}^\dag \psi_{o-\n}
\!-\! 
\psi_{o-\n}^\dag 
\frac{1}{\hbar}
\grave{F}_{o} \!
\right) \! 
dt \!\wedge\! d\Lambda_\n
\!-\!
\left( \!
\frac{1}{\hbar}
\grave{F}_{o}^\dag \varphi_{o-\n}^\dag
\!-\! 
\varphi_{o-\n} 
\frac{1}{\hbar}
\grave{F}_{o} \!
\right) \! 
dt \!\wedge\! d\Lambda_\n^\star \\
\\[-12pt]
&
\!+\!
\xi_{o-\n} \xi_{o-\n'}  
d\Lambda_\n \!\wedge\! d\Lambda_{\n'}
\!+\!
\xi_{o-\n}^\dagger \xi_{o-\n'}^\dag  
d\Lambda_{\n}^\star \!\wedge\! d\Lambda_{\n'}^\star 
\!\!+\!\!
\xi_{o-\n} \xi_{o-\n'}^\dag
d\Lambda_\n 
\!\wedge\! d\Lambda_{\n'}^\star
\!\!+\!\!
\xi_{o-\n}^\dagger \xi_{o-\n'}
d\Lambda_{\n}^\star \!\wedge\! d\Lambda_{\n'} .
\end{array}
\label{dOmega_n2}
\end{eqnarray}\\[-12pt]
We should discard the $\frac{1}{\lambda^2}$ term. 
Equating both sides of
(\!\ref{d-Omega-n2})
and
(\!\ref{dOmega_n2}),
we get \\[-18pt]
\begin{eqnarray}
\left.
\begin{array}{ll}
&
\partial_\Lambda{}_\n 
\Hat{F}_{\!o} 
\!-\!
\hbar \partial_t \xi_{o-\n}
\!=\!
[ \xi_{o-\n},\!\Hat{F}_{\!o}]
\!+\!
\grave{F}_{o}^\dag \psi_{o-\n}
\!-\! 
\psi_{o-\n}^\dag 
\grave{F}_{o}, \\
\\[-10pt]
&
\partial_\Lambda{}^{\!\!\star}_\n 
\Hat{F}_{\!o}
\!-\!
\hbar \partial_t \xi_{o-\n}^\dag
\!=\!
[\xi_{o-\n}^\dag,\! \Hat{F}_{\!o}]
\!+\!
\grave{F}_{o}^\dag \psi_{o-\n}
\!-\! 
 \psi_{o-\n}^\dag 
\grave{F}_{o} ,
\end{array}
\right\}
\label{dOmegaEqL3}
\end{eqnarray}
\vspace{-0.6cm}
\begin{eqnarray}
\!\!\!\!\!\!\!\!\!\!\!\!\!\!
\begin{array}{ll}
&
\partial_\Lambda{}_\n 
\xi_{o-\n'}
\!\!=\!\!
\xi_{o-\n} \xi_{o-\n'} ,~
\partial_\Lambda{}^{\!\!\star}_\n 
\xi_{o-\n'}^\dagger
\!\!=\!\!
\xi_{o-\n}^\dagger \xi_{o-\n'}^\dag ,~
\partial_\Lambda{}_\n 
\xi_{o-\n'}^\dag
\!\!=\!\!
\xi_{o-\n} \xi_{o-\n'}^\dag ,~
\partial_\Lambda{}^{\!\!\star}_\n 
\xi_{o-\n'}
\!\!=\!\!
\xi_{o-\n}^\dag \xi_{o-\n'} .
\end{array} \!\!\!
\label{dOmegaEq1/L3}
\end{eqnarray}\\[-32pt]

By exploiting the invariance of 
the integrability condition 
(\!\ref{integ-condi}),
especially by noticing Eqs.
(\!\ref{dOmegaEqL2})
and
(\!\ref{dOmegaEq1/L2}),
under the transformation
$
U
\!=\!
U(\Lambda,\!\Lambda^\star,\!t)
\!\in\!
S(U_m \!\times\! U_n)
$,\\[-18pt]
\begin{eqnarray}
\begin{array}{c}
\left[  \!\!
\begin{array}{cc}
\Hat{\Omega}_{o-m}  & 0 \\
0       &  \Check{\Omega}_{o-n}
\end{array} \!\!
\right]
\Rightarrow
U^{-1} 
\left[  \!\!
\begin{array}{cc}
\Hat{\Omega}_{o-m}  & 0 \\
0       &  \Check{\Omega}_{o-n}
\end{array} \!\!
\right]
U
\!-\!
U dU ,~
\grave{\Omega}_{o}
\Rightarrow
U^{-1} 
\grave{\Omega}_{o}
U , 
\end{array} \!\!
\label{InvarianceEq.Mat-psiCondi}
\end{eqnarray}\\[-12pt]
we have reasonably put\\[-24pt]
\begin{eqnarray}
\begin{array}{c}
\grave{F}_{o}
\!=\!
\grave{\mathbf{F}}_{0}
h_m,~
\zeta_{o-\n} \!=\! 0 ,~
\psi_{o-\n}
\!=\!
k_n
{\psi}_{o-\n}^0 .
\end{array} \!\!
\label{InvarianceEq.Mat-psiCondi2}
\end{eqnarray}\\[-20pt]
According to
CMS
\cite{CMS.81},
we take\\[-18pt]
\begin{eqnarray}
\begin{array}{c}
\psi_{o-\n}
\!\equiv\!
\left[
\BA{c}
\psi_{o-\n,1} \\[2pt]
\underbrace{
\psi_{o-\n,2}
}_{m} 
\EA
\right]
\!\!\!\!
\BA{l}
\left.{}\right\}m\\[6pt]
\left.{}\right\}n-m 
\EA,~
\Hat{F}_{o}
\!\equiv\!
i \alpha 
\!\cdot\!
\I_m,~ 
\Check{F}_{o} 
\!\equiv\! 
\left[ \!\!\!
\begin{array}{cc}
S 
&\!\!\!\! -T^\dag\\
\\[-14pt]
\underbrace{
T
 }_{~m~}
&\!\!\!\!
\underbrace{
\!\! 0 
}_{~n-m~}
\end{array} \!\!\!\!
\right]
\!\!\!\!
\BA{l}
\left.{}\right\}m\\[6pt]
\left.{}\right\}n-m 
\EA,
\end{array} \!\!
\label{Matdtzeta-psi}
\end{eqnarray}\\[-10pt]
where
$\psi_{o-\n,1}$
is an invertible
$m \!\times\! m$
matrix.
Then,
Eq. of the first line in
(\!\ref{dOmegaEq1/L})
becomes\\[-16pt]
\begin{eqnarray}
\!\!\!\!
\begin{array}{c}
\hbar \partial_t \! 
\left[ \!\!\!
\BA{c}
\psi_{o-\n,1} \\[2pt]
\psi_{o-\n,2} 
\EA \!\!\!
\right]
\!=\! 
\left[ \!\!\!
\begin{array}{cc}
S &\!\! -T^\dag\\[2pt]
T &\!\! 0 
\end{array} \!\!\!
\right] \!\!
\left[ \!\!\!
\BA{c}
\psi_{o-\n,1} \\[2pt]
\psi_{o-\n,2} 
\EA \!\!\!
\right] 
\!-\!
\left[ \!\!
\BA{c}
\psi_{o-\n,1} \\[2pt]
\psi_{o-\n,2} 
\EA \!\!
\right] \!
i \alpha 
\!\cdot\!
\I_n
\!=\!
\left[ \!\!\!
\BA{c}
S \psi_{o-\n,1}
\!\!-\!\!
T^\dag
\psi_{o-\n,2}
\!\!-\!\!
i \alpha
\psi_{o-\n,1}\\[2pt]
T \psi_{o-\n,1}
\!\!-\!\!
i \alpha
\psi_{o-\n,2} 
\EA \!\!
\right] ,
\end{array} \!\!
\label{Eq.Mat-psi}
\end{eqnarray}\\[-10pt]
from which we have
the explicit expressions for
$T$
and
$S$
as\\[-18pt]
\begin{eqnarray}
\left.
\begin{array}{ll}
T
\!=\!
&
\!\!\!
\left( \!
\hbar \partial_t 
\!\!+\!\!
i \alpha \!
\right) \!
\psi_{o-\n,2}
\!\cdot\!
\psi_{o-\n,1}^{-1} ,\\
\\[-8pt]
S
&
\!\!\!\!\!\!\!
\!=\!
\left\{ \!
\left( \!
\hbar \partial_t 
\!\!+\!\!
i \alpha \!
\right) \!
\psi_{o-\n,1}
\!\!+\!\!
T^\dag \!
\psi_{o-\n,2} \!
\right\} \!\!
\psi_{o-\n,1}^{\!-1} \\
\\[-10pt]
&
\!\!\!\!\!\!\!
\!=\!
(\!\psi_{\!o-\n,1}^\dag\!){}^{\!-1} \!\!
\left\{ \!
i \alpha \!\!
\left( \!
\psi_{\!o-\n,1}^\dag
\!\!-\!\!
\psi_{\!o-\n,2}^\dag
\psi_{\!o-\n,2} \!
\right) \!\!
\psi_{\!o-\n,1}
\!\!+\!\!
\psi_{\!o-\n,1}^\dag 
\hbar \partial_t \!
\psi_{\!o-\n,1}
\!\!+\!\!
\hbar \partial_t \!
\psi_{\!o-\n,2}^\dag
\!\cdot\!
\psi_{\!o-\n,2} \!
\right\} \!\!
\psi_{\!o-\n,1}^{\!-1} .
\end{array} \!\!
\right\}
\label{Eq.Mat-psi}
\end{eqnarray}\\[-10pt]
The unknown field $\alpha$ can be determined through Eq.
(\!\ref{Eq.Mat-psi})
by imposing the condition
$
\mbox{Tr}
{\Omega}_{o}
\!=\!
\mbox{Tr}
(
\Hat{\Omega}_{o-m}
\!+\!
\Check{\Omega}_{o-n}
)
=
0
$
given in 
(\!\ref{devided-Omega})
which reads\\[-16pt]
\begin{eqnarray}
\left.
\!\!\!\!\!\!\!\!\!\!\!\!
\begin{array}{ll}
&
\mbox{Tr}
{\Omega}_{o}
\!\!=\!\!
i \alpha m
\!+\!
\mbox{Tr} S
\!=\!
0 ,\\
\\[-10pt]
&
\mbox{Tr} S
\!\!=\!\!
\mbox{Tr} \!\!
\left[ \!
i \alpha \!
\left\{ \!
\I_m
\!\!-\!\!
(\!\psi_{\!o-\!\n,1}^\dag \! \psi_{\!o-\!\n,1}\!){}^{\!-\!1} \!
\psi_{\!o-\!\n,2}^\dag \!
\psi_{\!o-\!\n,2} \!
\right\} 
\!\!+\!\!
\psi_{\!o-\!\n,1}^{\!-\!1} \!
\hbar \partial_t \!
\psi_{\!o-\!\n,1}
\!\!-\!\!
(\!\psi_{\!o-\!\n,1}^\dag \! \psi_{\!o-\!\n,1}\!){}^{\!-\!1} \!
\hbar \partial_t \!
\psi_{\!o-\!\n,2}^\dag
\!\cdot\!
\psi_{\!o-\!\n,2} \!
\right] \! ,
\end{array} \!\!
\right\}
\label{Eq.Mat-psiCondiAlpha}
\end{eqnarray}\\[-10pt]
and one can determine
$\alpha$
as\\[-18pt]
\begin{eqnarray}
\begin{array}{c}
\alpha
\!=\!
i
{\displaystyle
\frac{
\mbox{Tr} \!
\left( 
\psi_{o-\n,1}^{\!-1} 
\hbar \partial_t 
\psi_{o-\n,1} 
\right)
\!-\!
\mbox{Tr} \!
\left[ 
(\!\psi_{o-\n,1}^\dag \psi_{o-\n,1}\!){}^{\!-1} 
\hbar \partial_t 
\psi_{o-\n,2}^\dag
\cdot
\psi_{o-\n,2} 
\right]}
{
2m
\!-\!
\mbox{Tr} \!
\left[ 
(\!\psi_{o-\n,1}^\dag \psi_{o-\n,1}\!){}^{\!-1} 
\psi_{o-\n,2}^\dag 
\psi_{o-\n,2} 
\right]} .
}
\end{array} \!\!
\label{Eq.Mat-psiCondiAlpha}
\end{eqnarray}\\[-10pt]
Using
$
\xi_{\!o-\n}
\!\!=\!\!
-
h_m^{\!-1} \!
\partial_\Lambda{}_\n
h_m,
\grave{F}_{\!o}
\!\!=\!\!
\grave{\mathbf{F}}_{\!0}
h_m
\!\!\equiv\!\!
\grave{\mathbf{F}}_{\!0,1}
$,
$
\psi_{\!o-\n}
\!\!=\!\!
k_n
{\psi}_{\!o-\n}^{0}
\!$
and
$\!
\Hat{F}_{\!o}
\!=\!
i \alpha 
\!\cdot\!
\I_m
$,
(\!\ref{dOmegaEqL3})
is changed as\\[-20pt]
\begin{eqnarray}
\!\!\!\!\!\!\!\!
\begin{array}{ll}
&
i \alpha
\!\cdot\! 
\I_m
\!+\!
\hbar \partial_t \!
\left(
h_n^{\!-1}\partial_\Lambda{}_\n h_n
\right) 
\!=\!
h_m^\dag 
\grave{\mathbf{F}}_0^{ \dag}
k_n
{\psi}_{o-\n}^{0}
\!-\! 
{\psi}^{0 \star}_{o-\n} 
k_n^\dag
\grave{\mathbf{F}}_{0}
h_m 
\!=\!
\grave{\mathbf{F}}_{0,1}^\dag
\psi_{o-\n,1}
\!-\!
\psi_{o-\n,1}^\dag
\grave{\mathbf{F}}_{0,1}.
\end{array} \!\!\!\!
\label{dOmegaEq1/L4}
\end{eqnarray}\\[-20pt]
With the aid of the relation
(\!\ref{dOmegaEqL2})
and
the expression for
$\Check{F}_{o}$ 
(\!\ref{Matdtzeta-psi}),
we obtain the differential operation
$\partial_\Lambda{}_\n$
on $\Check{F}_{o}$
in the following forms:\\[-16pt]
\begin{eqnarray}
%\left.
\!\!\!\!\!\!\!\!
\begin{array}{ll}
&
\left[ \!\!\!
\begin{array}{cc}
\partial_\Lambda{}_\n S &
\!\! -\partial_\Lambda{}_\n T^\dag\\[6pt]
\partial_\Lambda{}_\n T &\!\! 0 
\end{array} \!\!\!
\right]
\!\!=\!
\grave{F}_{o} \psi_{o-\n}^\dag
\!-\! 
\psi_{o-\n} 
\grave{F}_{o}^\dag
\!=\!
\left[ \!\!\!
\begin{array}{c}
\grave{\mathbf{F}}^\prime_{\!0,1}\\[2pt]
0
\end{array} \!\!\!
\right] \!
[\psi_{o-\n,1}^\dag, \psi_{o-\n,2}^\dag ]
\!-\!
\left[ \!\!\!
\BA{c}
\psi_{o-\n,1} \\[2pt]
\psi_{o-\n,2} 
\EA \!\!\!
\right] \!
[\grave{\mathbf{F}}^\prime{}^\dag_{\!0,1}, 0 ] \\
\\[-6pt]
&
\!=\!
\left[ \!\!\!
\begin{array}{cc}
\grave{\mathbf{F}}^\prime_{\!0,1} \psi_{o-\n,1}^\dag &\!\! 
\grave{\mathbf{F}}^\prime_{\!0,1} \psi_{o-\n,2}^\dag \\[6pt]
0 &\!\! 0 
\end{array} \!\!\!
\right] 
\!-\!
\left[ \!\!\!
\begin{array}{cc}
\psi_{o-\n,1} \grave{\mathbf{F}}^\prime{}^\dag_{\!0,1} &\!\! 0\\[6pt]
\psi_{o-\n,2} \grave{\mathbf{F}}^\prime{}^\dag_{\!0,1} &\!\! 0 
\end{array} \!\!\!
\right]
\!=\!
\left[ \!\!\!
\begin{array}{cc}
\grave{\mathbf{F}}^\prime_{\!0,1} \psi_{o-\n,1}^\dag
\!-\! 
\psi_{o-\n,1} \grave{\mathbf{F}}^\prime{}^\dag_{\!0,1} &\!\! 
\grave{\mathbf{F}}^\prime_{\!0,1} \psi_{o-\n,2}^\dag \\[6pt]
\!-\! 
\psi_{o-\n,2} \grave{\mathbf{F}}^\prime{}^\dag_{\!0,1} &\!\! 0 
\end{array} \!\!\!
\right] \! ,
\end{array} \!
%\right\}
\label{differential-operation on S&T}
\end{eqnarray}\\[-10pt]
from which,
thus we have\\[-20pt]
\begin{eqnarray}
\begin{array}{c}
\partial_\Lambda{}_\n S
\!=\!
\grave{\mathbf{F}}^\prime_{\!0,1} \psi_{o-\n,1}^\dag
\!-\! 
\psi_{o-\n,1} \grave{\mathbf{F}}^\prime{}^\dag_{\!0,1} ,~
\partial_\Lambda{}_\n T
\!=\!
-
\psi_{o-\n,2} \grave{\mathbf{F}}^\prime{}^\dag_{\!0,1} .
\end{array}
\label{differential-operation on S&T}
\end{eqnarray}\\[-34pt]

$\!\!\!\!\!\!\!\!\!$Taking the idea of
construction of
(\!1\!+\!1\!)D model with 
$\frac{SU_{m+n}}{S(\!U_{m} \!\times\! U_{n}\!)}$
structure $\!$by$\!$
CMS$\!$
\cite{CMS.81},
we have extended the model to the
$(\!1\!\!+\!\!2\m\!)$D one.$\!$
We have successfully derived the
geometric equation with 
$\frac{SU_{m+n}}{S(U_{m} \!\times\! U_{n})}$
structure to get the classical equation of motion
for integrable  system
an application of which
to the Lipkin-Meshkov-Glick model$\!$
\cite{LMG.65}$\!$
will be published eksewhere. 
The present particle-hole formalism 
for the geometric equation is also applicable
to the superconducting formalism
on the coset space
$\frac{SO(2N)}{U(N)}$
for the paired mode
\cite{Nishi.81}
and on the coset space its extension
$\frac{SO(2N+2)}{U(N+1)}$
for both the paired and unpaired modes
\cite{FYN.77,Fu.81,Nishi.98}.
This work will be possible in the near future.

\newpage

%%%%%%%%%%%%%%%%%%%%%%%%%%%%
%                                                                                       %
%    8   Bilinear differential equation in SCF method      %
%                                                                                       %
%%%%%%%%%%%%%%%%%%%%%%%%%%%%

\setcounter{equation}{0}
\section{Bilinear differential equation in SCF method}

\vspace{-0.1cm}

The fermion operators 
${E}^\alpha_{~\beta}
\!=\!
c_\alpha^\dagger  c_\beta$ 
span the $U(N)$ Lie algebra
$
[{E}^\alpha_{~\beta},{E}^\gamma_{~\delta}]
\!=\!
\delta_{\gamma \beta}{E}^\alpha_{~\delta}
\!-\!
\delta_{\alpha \delta}{E}^\gamma_{~\beta}
$
and generate a canonical transformation 
$
U(g)( \!=\! e^{\gamma _{\alpha\beta}c_\alpha^\dagger c_\beta };
\gamma^\dagger 
\!=\!
-\gamma )
$ 
specified 
by a $U(N)$ matrix 
$g 
(
\!=\!
e^\gamma 
)$ 
as\\[-20pt]
\begin{eqnarray}
\begin{array}{rl}
&U(g)c_\alpha^\dagger U^{-1} (g)
\!=\!
c_\beta^\dagger 
g_{\beta\alpha},~
U(g) c_\alpha U^{-1} (g)
\!=\!
c_\beta 
g_{\beta\alpha}^* , ~
g^\dagger g\!=\!gg^\dagger \!=\!1_{N} 
~\mbox{and}~
c_\alpha |0\rangle \!\!=\!\! 0 .
\end{array}
\label{Cano.Tra.}
\end{eqnarray}\\[-18pt]
An $m$ particle S-det 
$\!|\phi_m \rangle \!\!=\!\! c_m^\dagger \!\cdots\! c_1^\dagger 
|0 {\rangle}$
is 
exterior product of $m$ 
single-particle state ({\em simple} state). 
Due to the Thouless theorem
\cite{Th.60},
different S-det
is produced as \\[-18pt]
\begin{eqnarray}
U(g) |\phi_m \rangle 
=
(c^\dagger g)_m 
\cdots (c^\dagger g)_1 
|0 \rangle \stackrel{d}{=}  |g {\rangle} ,~
U(g) |0 {\rangle} 
=
|0 {\rangle}.
\label{ThoulessTra.}
\end{eqnarray}\\[-18pt] 
The set of all the simple states of unit modulas 
together with the equivalence relation 
identifying distinct states only 
in phase with the same state, 
constitute a manifold ${\bf Gr}_{\!m}$
(group orbit). 
Any simple state $| \phi_m \rangle$  
defines a decomposition 
of single-particle Hilbert spaces into 
sub-Hilbert spaces of occupied and unoccupied states 
\cite{RR.81}. 
Thus the ${\bf Gr}_m$ corresponds to a coset space\\[-16pt] 
\begin{eqnarray}
\BA{c}
{\bf Gr}_m \!\sim\! \frac{U(m\!+\!n)}{U(m) \!\times\! U(n)},
~ N\!=\!m\!+\!n .
\EA
\end{eqnarray}\\[-16pt]
Following Fukutome$\!$ 
\cite{Fu.81,Fu.Int.J.Quantum Chem.81},  
for the coset variable $p$ defined later,$\!$
we express it as
$U(g)|\phi_m \rangle 
\!\!=\!\!
\langle \phi_m|U\!(\!g_\xi g_\upsilon\!)|\phi_m \rangle
e^{p_{ia} c_i^\dagger c_a}|\phi_m {\rangle}$,
using the relations 
$g\!\!=\!\!g_\xi g_\upsilon $,
$
\langle \phi_m | U\!(\!g_\xi g_\upsilon\!)
|\phi_m \rangle
\!\!=\!\! 
[\det (1\!\!+\!\!p^\dagger p)
]^{\!-\frac{1}{2}} \! \det \upsilon 
$
and \\[-16pt]
\begin{eqnarray}
\!\!\!\!\!\!\!\!
\begin{array}{rl}
&\sum_{\rho =0}^{M_{\mbox{\scriptsize max}}} \!\!
\sum_{\!1 \!\leq\! a_1 <\cdots <a_\rho \!\leq\! m,~
m\!+\!1 \!\leq\! i_1 <\cdots <i_\rho \!\leq\! N} \!
{\cal A}(\!p_{{i_1}{a_1}} \!\cdots\! 
p_{{i_\rho}{a_\rho}}\!)
c_{i_1}^\dagger \! c_{a_1} \!\cdots\! 
c_{i_\rho}^\dagger \! c_{a_\rho}
\!\!=\!
e^{p_{ia} c_i^\dagger c_a}, 
M_{\mbox{\scriptsize max}}
\!\!=\!\!\mbox{min}(m,\!n) .
\end{array} 
\label{Slaterdet1}
\end{eqnarray}\\[-14pt]
An anti-symmetrizer 
$
{\cal A}(p_{{i_1}{a_1}} \cdots p_{{i_\rho}{a_\rho}})
$
is defined as  \\[-14pt]  
\begin{eqnarray}
{\cal A}(p_{{i_1}{a_1}} \cdots p_{{i_\rho}{a_\rho}}) 
\stackrel{d}{=} 
\det \!
\left[ \!\!\!
\begin{array}{ccc}
p_{{i_1}{a_1}}&\cdots&p_{{i_1}{a_\rho}} \\[-4pt]
\vdots&&\vdots \\[-6pt]
p_{i_\rho a_1}& \cdots & 
p_{{i_\rho}{a_\rho}} 
\end{array} \!\!\!
\right] ,~
(\det \upsilon \mbox{: determinant~of}~\upsilon). 
\label{Slaterdet2}
\end{eqnarray}\\[-18pt] 

We give the 
Pl\"{u}cker coordinate studied in the textbook
of Miller and Sturmfels
\cite{EMBS.05},
which has played 
important roles of algebraic construction 
of soliton theory by Sato
\cite{Sa.81},\\[-16pt]
\begin{eqnarray}
\left.
\begin{array}{c}
U(g)|\phi_m \rangle 
\!=\!
\sum_{n\geq \alpha_m >\dots >\alpha_1 \geq 1} 
v_{\alpha_m ,\dots ,\alpha_1 }^ {m,\cdots ,1}
c_{\alpha_m}^\dagger \cdots 
c_{\alpha_1}^\dagger |0 \rangle ,~
v_{\alpha_m ,\cdots ,\alpha_1}^{m,\cdots,1} 
\!=\!
\mbox{det} \!
\left[ \!\!\!
\begin{array}{rrl}
g_{\alpha_1 ,1} & \cdots & g_{\alpha_1 ,m} \\[-4pt]
\vdots & & \vdots \\[-6pt]
g_{\alpha_m ,1} & \cdots & g_{\alpha_m ,m}
\end{array} \!\!\!
\right] , 
\end{array} \!\!
\right \}
\label{Pluecker coordinates}
\end{eqnarray}\\[-12pt] 
We easily find that
the Pl\"{u}cker coordinate 
$v_{\alpha_m ,\cdots ,\alpha_1}^{m,\cdots,1}$
has a relation\\[-16pt]  
\begin{eqnarray}
\begin{array}{rl}
\sum_{i=1}^{m+1} (-1)^{i-1}v_{\beta_i ,
\alpha_{m-1}, \cdots ,\alpha_1}^
{m, \cdots ,1}
\!\cdot\! 
v _{\beta_{m+1} ,\cdots ,\beta_{i+1} ,
\beta_{i-1} ,\cdots ,\beta_1 }^
{m ,\cdots ,1} 
=
0,~
\mbox{(Pl\"{u}cker~relation)} .
\end{array}
\end{eqnarray}\\[-16pt]
The indices denote 
the distinct sets $\!1 \!\!\leq\!\! \alpha_1 \!,
\!\cdots\!,\!\alpha_{m\!-\!1} \!\!\leq\!\! N\!$ 
and 
$\!1 \!\!\leq\!\! \beta_1 \!,\!\cdots\! ,\!\beta_{m\!+\!1} \!\!\leq\!\! N\!$.$\!\!$
The Pl\"{u}cker relation 
is equivalent to a bilinear identity equation\\[-18pt] 
\begin{eqnarray}
\begin{array}{rl}
\sum_{\alpha =1}^N c_\alpha^\dagger U(g) | 
\phi_m \rangle \otimes c_\alpha U(g) 
| \phi_m \rangle 
=
\sum_{\alpha =1}^N U(g) c_\alpha^\dagger |
\phi_m \rangle \otimes U(g) c_\alpha 
| \phi_m \rangle 
\!=\!
0 ,
\end{array}
\label{bilinear eq.a}
\end{eqnarray}\\[-16pt] 
%The ${\bf Gr}_m$ is an $SU(N)$ 
%group manifold
%since the phase equivalence theorem does hold.
%The Pl\"{u}cker relation 
and analogous to the Hirota form.$\!$
Following the regular rep 
of $\!SO(2N\!\!+\!\!1)\!$ group by Fukutome$\!$  
\cite{Fu.77}
and
introducing a phase variable 
$\tau \!=\! i\ln \det \upsilon$,
it is shown that
the Lie commutation relation
is satisfied by the 
differential operators for particle-hole pairs
given below\\[-16pt]  
\begin{eqnarray}
\left. 
\begin{array}{rl}
&
e_{ia}^* 
\stackrel{d}{=}
-(p_{ib}p_{ja}
\frac{\partial}{\partial p_{jb}}
+\frac{\partial}{\partial p_{ia}^*}
+\frac{i}{2}p_{ia} \frac{\partial}
{\partial \tau}) ,~
e_{ai}^* \stackrel{d}{=} p_{ja}^* p_{ib}^* 
\frac{\partial}{\partial p_{jb}^*} 
+\frac{\partial}{\partial p_{ia}} 
-\frac{i}{2}p_{ia}^*\frac{\partial}
{\partial \tau}, \\
\\[-14pt] 
&
e_{ab}^* 
\stackrel{d}{=} 
p_{ia}^* 
\frac{\partial}{\partial p_{ib}^*} 
-p_{ib}\frac{\partial}{\partial p_{ia}}
-i\delta_{ab} \frac{\partial}{\partial \tau} ,~
e_{ij}^* \stackrel{d}{=} p_{ia} 
\frac{\partial}{\partial p_{ja}} 
-p_{ja}^* \frac{\partial}{\partial p_{ia}^*},
\end{array} \!\!
\right\}
\label{p-h differential operators} 
\end{eqnarray}\\[-10pt] 
which obey\\[-20pt]  
\begin{eqnarray}
\left. 
\begin{array}{rl}
&e_{ia}^* \Phi_{m,m}(p,p^*,\tau)=p_{ia}
\Phi_{m,m}(p,p^*,\tau) ,~~
e_{ai}^* \Phi_{m,m}(p,p^*,\tau)=0 , \\
\\[-8pt]
&e_{ij}^* \Phi_{m,m}(p,p^*,\tau)=0 ,~~
e_{ab}^* \Phi_{m,m}(p,p^*,\tau)=\delta_{ab}
\Phi_{m,m}(p,p^*,\tau),
\end{array} 
\right\}
\label{p-h differential operators onto vacuum}  
\end{eqnarray}\\[-8pt] 
and
$[e_{ia}^*~,~p_{jb}]=-p_{ib}p_{ja}$. 
The vacuum function 
$\Phi_{m,m}(p,p^*,\tau)$  
is defined as\\[-16pt] 
\begin{eqnarray}
\Phi_{m,m}(p,p^*,\tau) \stackrel{d}{=} 
v_{m\cdots 1}^{m \cdots 1}(g_\xi g_\upsilon )
=
[\det (1+p^\dagger p)]^{-\frac{1}{2}}e^{i\tau}.
\end{eqnarray}\\[-14pt] 
The present framework becomes 
the dual of regular rep by Fukutome.
In both the SCF theory and  
soliton theory on a group, 
we  find common features that 
${\bf Gr}$ is identical 
with solution space of 
bilinear differential equation. 
The solution space of differential equation 
becomes 
an integral surface, 
subspace in the ${\bf Gr}_m$ on which 
the residual coset variables 
are remained to be constant.

\newpage

%%%%%%%%%%%%%%%%%%%%%%%
%                                                                      %    
%   9   Summary and further perspective        %
%                                                                      %
%%%%%%%%%%%%%%%%%%%%%%%

\section{Summary and further perspective}

By $\!$KN$\!$ 
\cite{WigSym99,KN.00,NishiProviKoma2.07}$\!$ 
we have clarified the relation of concept 
of particle-hole and collective motion 
in the TDHF to that of soliton theory 
from the $loop$ group.$\!$
The subgroup orbits made of {\em loop-group} paths 
exist innumerably in ${\bf Gr}_m$.$\!$
To develop a perturbative method
with the use of the collective variables, 
using the ID fermion, 
we have aimed at constructing 
the SCF theory
on affine KM algebra
along the soliton theory.$\!$ 
The ID fermion operator has been introduced 
through the Laurent expansion of 
finite-dimensional (FD) fermion operators 
with respect to the degrees of freedom of 
fermions related to the mean-field (MF) potential.$\!$ 
${\bf Gr}_m$ 
is identified with ${\bf Gr}_{\!\infty}$
which is affiliated 
with the manifold obtained 
by reduction of $gl_{\!\infty}$ to $su_{\!N}$.$\!$ 
The extraction of the subgroup orbits 
from ${\bf Gr}_m$ is equivalent to the construction of 
differential equation (Hirota equation)
for the $su_{\!N}$ reduced KP hierarchy.$\!$ 
The SCF theory on $F_\infty$ results in
the {\em gauge theory of fermion}.$\!$ 
The {\em collective motion} 
due to quantal fluctuations of 
the SC MF potential
is attributed to {\em the motion of the gauge of fermion}.$\!$
A {\em common factor} explains interference
among fermions.$\!$
The {\em concept of particle-hole and collective motions} is
regarded as the {\em compatible condition for particle-hole
and collective modes}.$\!$
The SCF theory on $F_\infty$
gives us a {\em new algebraic method} 
for the understanding of the fermions.$\!$ 
Prescribing the fermion to form a pair 
by absorbing the {\em change of gauge}, 
the SCF Hamiltonian made of only $H_{F_\infty;HF}$ 
is induced.$\!$
Through the compatible condition for particle 
and collective modes and 
a special choice of the Laurent expansion,
the fermion gauge arises.$\!$
We have the expressions 
for pair operators of the ID fermions 
in terms of the Laurent spectral numbers.$\!$
Though TDHF theory on $F_\infty$
describes the dynamics on the real fermion-harmonic oscillators, 
soliton theory does on the complex 
fermion-harmonic oscillators.$\!$
This gives a problem on the relation
of the present theory to the resonating (Res) MF one$\!$
\cite{FukuNishi.88.91,Nishi.94}
and a task
how to construct 
the {\em ID} boson variable$\!$
from the group parameter 
\cite{Nishi.99}.

We start with $su_{\!N}$ algebra consisting of
particle-hole (p-h) component and $u_{\!N}$ algebra including
particle-particle (p-p) and hole-hole (h-h) ones
for the state $|\phi\rangle$
\cite{Fu.Int.J.Quantum Chem.81}.$\!$
Then we have $U(N)$ group orbit.
We must notice the equivalence relation which
identifies the states different from each other in phases
dependent only on diagonal components in h-h types,
with the same state
\cite{RR.81}.$\!$
Under the equivalence relation, we ought to
treat $SU\!(\!N\!)$ group orbit.$\!$
However the HF Hamiltonian has value
on $\!u_{\!N}\!$ but not on $\!su_{\!N}$.$\!$
To describe the dynamics on $SU\!(\!N\!)$ group orbit,
we must remove extra components not satisfying
$su_{\!N}$ from the fully parametrized HF Hamiltonian.$\!$
With the help of the equivalence relation
we ought to take diagonal components in p-p and h-h types
into account to assign them.$\!$
These quantities can play a role of 
fermion gauge phase
as the canonical transformation describes.$\!$
The gauge-phase is separated
into a term which comes from single pair of fermions
and a term coming from the particle-number operator,
on the Lie algebra through which the fermion pairs are governed.$\!$
The former relates to the particle mode
and the latter to the collective one.
Removing the above superfluous components of
the HF Hamiltonian,
assignment of them to both the modes turns out to
bring the concept of particle-hole and collective motions.
The usual TDHF theory has not a complete scheme
to treat separately both the motions.$\!$
The TDHF equation on $S^1$,
however, 
has such a scheme.$\!$
It provides not only manifest and algebraic understanding
of the motions but also a scheme to describe
a large amplitude collective motion.
As for the particle-hole and collective motions,
assuming a time-periodicity of motions,
we can derive a new unified equation for both the motions
beyond the HF and RPA ones
\cite{KN.01,NK.02}.$\!$
The new TDHF theory on $S^1$ is constructed on a collection of
various subgroup orbits consisting of {\em loop paths}
in the ${\bf Gr}_{\!m}\!$
of the FD fermion Fock space
and is shown
to be built on the ID
${\bf Gr}$ of the ID fermion Fock space $F_\infty$.$\!$
If we choose a broken-symmetry vacuum on $S^1$,
the vacuum state is able to deform by through
the shift operators associated with collective modes
so that $\tau$-function in soliton theory is also deformed.
We find an algebraic mechanism for appearance
of the collective motion induced by the TD MF potential.$\!$
This gives an algebraic understanding of
physical concepts of the symmetry breaking and
the successive occurrence
of the collective motion due to the recovery of the symmetry.

$\!\!\!$A multi-circle TDHF theory is exciting
but has a problem
how the Pl\"{u}cker relation on multi circle is constructed.$\!$
It is related to a multi-soliton theory$\!$
\cite{DJKM.81,KBK.97}.$\!$
The $\!$ID$\!$ algebraic approaches
have been proposed using the Bethe ansatz ($\!$BA$\!$) $\!$WF
\cite{BA.31},$\!$
Lipkin-Meshkov-Glick
(LMG)-model
\cite{MOPN.06,LMG.65}
and 
pairing theory
\cite{Rowe.91,Ri.65}.$\!$ 
Rowe has showed that
a number-projected WF satisfies 
some recursion relations
and
expressed it in a determinantal form
\cite{Rowe.91}
with completely anti-symmetric Schur function
in the theory of group character by
Littlewood
\cite{Littlewood.58}
and
MacDonald
\cite{MacDonald.79}.$\!$
The WF is described
by an ID Lie algebra
\cite{PD.98.99,PZDD.17}.$\!$
The algebra is constructed by
the power series expansion of the FD Lie algebra
with respect to the parameters involved in Hamiltonian.$\!$
See $\!$
\cite{PD.98.99}(b)$\!$
which has
given an ID affine Lie algebra $\widehat{su(2)}$ 
and an exactly solvable pairing Hamiltonian.$\!$
It has also shown the conditions 
for solving the eigenvalue problem 
using the BA method.$\!$
It is a very exciting problem to compare 
the ID affine Lie algebra $\widehat{su(2)}$ 
with 
the ID affine Lie algebra $\widehat{sl(2,\!\C)}$ 
of p-h LMG-model.$\!$
They have the $\widehat{su(2)}$ and $\widehat{sl(2,\!\C)}$ 
invariant Casimir operators
which provide almost the pairing and the LMG Hamiltonians.$\!$ 
This will be solved and presented elsewhere
in the near future.
Further one has
possible generalization and 
initial value problem for solution of 
Sine-Gordon equation,
Mansfield
\cite{Mansfield.85}.
This suggests an infinitesimal form of
the corresponding nonlinear algebra
generated by the building blocks of the BA WF
\cite{Gau76}.$\!$
We obtain
a collective sub-manifold decided by the SCF Hamiltonian
of LMG-model.$\!$
Then we acquire a clue to build the relation between
the methods using Gaudin model
\cite{Skl99,Gau76},
Oritz exactly-solvable model
\cite{OSDR.05}
and further using
the minimal matrix product state and the geminal WF
\cite{LJ-HC.20}.$\!$
Finally
we give some remarks:
Daboul has shown that
the dynamical symmetry of hydrogen atom leads in a natural way to
the ID algebra, twisted affine KM algebras of 
$\widehat{so(4)}$ and $\widehat{so(3,\!1)}$
\cite{Daboul.93}
based on a suggestive paper by
Goddard and Oliv
\cite{GO.86}.

We have studied
the {\em curvature equation} along the idea of Lax
%for integrable system,$\!$
unfamiliar to
the conventional treatments of
fermion system
and 
adopted the geometric equation  
to truncate the collective sub-manifolds out of the 
TDHF manifold.$\!$
We show below the following subjects to be studied:\\
(I) The expectation values of the zero curvatures  
for the state function  become the set 
of  equations of motion in the analytical mechanics, 
imposing  
the weak orthogonal conditions among the infinitesimal generators,
i.e., the equation for 
the {\em tangent vector fields} on the group sub-manifold
consisting of the collective variables.$\!$
The expectation values of the non-zero curvatures
become the gradients of potential, which arises from
the existence of the residual Hamiltonian,
along the collective variables.
These quantities can be expected to give a
criterion how the collective sub-manifolds is effectively 
truncated,
if we might understand such the treatments.\\
(II) The expression for the zero curvature conditions
is nothing but the formal RPA equation imposed by
the weak orthogonal conditions.
The formal RPA equation has a simple geometrical interpretation:
relative vector fields made of the SCF Hamiltonian
around each point on an integral curve
also constitute the solutions for the formal RPA equation around
the same point which is in turn a fixed point.$\!$
It means that the formal RPA equation
is a natural extension of the usual RPA
equation for small-amplitude quantal fluctuations
around the ground state to that at any point
on the collective sub-manifold which should be studied.
Moreover, the enveloping curve, made of the solutions of 
the formal RPA at each point
on an integral curve, becomes another integral curve.$\!$
The integrability condition
 is just the infinitesimal condition
to transfer a solution to another one for the evolution equation
under consideration.$\!$
Then the usual treatment of
RPA equation for small amplitude around the ground state 
becomes nothing but
a method of determining an infinitesimal transformation of symmetry
under the assumption that 
the fluctuating fields are composed of only normal-modes.$\!$
See $\!$Klein-Walet-Dang$\!$
\cite{KleinWaletDang.91}.$\!$
At the beginning, 
their descriptions of dynamical fermion systems 
in both the methods had looked
very different manners at first glance.$\!$
In the abstract fermion Fock spaces, 
each solution space 
in both the methods
belongs to the corresponding ${\bf Gr}$.$\!$
There is 
a difference of the FD and ID fermion systems.$\!$  
Overcoming the difference, 
we have aimed at  clearing a close connection between
the concepts of MF potential and gauge of fermion
making a role of the loop group.\\
(1) The {\em Pl\"{u}cker relations} 
on the coset variables becomes 
analogous io 
the {\em Hirota's bilinear form}
\cite{Hi.76}.
The SCF method has been devoted to
a construction of boson-coordinate systems
rather than 
soliton solution by the $\tau$-FM. 
Both the methods are equivalent to each other,
if we stand on the viewpoint of
the Pl\"{u}cker relation or the bilinear identity equation
defining the ${\bf Gr}$.\\
(2) The ID fermion operators are introduced
through the Laurent expansion of 
the FD fermions with respect to 
the degrees of freedom of fermion related to 
the MF potential.$\!$
Inversely, the collectivity of the MF potential
is attributed to the gauge of interacting ID fermion.$\!$
The construction of the fermion operator is contained in that of
Clifford algebra.$\!$
This fact permits us to introduce the {\it affine} KM algebra.$\!$
It means that the usual perturbative method in terms of
the collective variables with time periodicity
has implicitly stood on the reduction to $su_N$.
The TDHF theory 
becomes  {\em the gauge theory of fermion} and 
the collective motion
appears as the motion of fermion gauge 
with a common factor.$\!$
The physical concept 
of {\em particle-hole and vacuum} in the
SCF method dependent on $S^1$
connects to {\em the Pl\"{u}cker relation}.$\!$ 
The algebraic treatment of
extracting sub-group orbit consisting of loop paths
out of ${\bf Gr}_{\!m}$ is just the formation of 
the Hirota equation
for the $su_{\!N}\!$ reduced KP hierarchy
in the soliton theory.$\!$ 
The present framework gives a manifest structure of
the gauge theory of fermion inherent in the SCF method
and
provides a {\em new algebraic tool
for the microscopic understandings} of the fermion systems.\\
%It means that
%the finite-dimensional fermion systems
%have the algebraic structure embedded into 
%that of the {\em infinite ones}.
(3) Through the investigation of physical meanings for 
the ID shift operators and the conditions of reduction
to $sl_{\!N}\!$ from the $loop$ group viewpoint
\cite{Ottesen.95}, 
it is induced that there is 
the close connection between the {\em collective variables}
and the {\em spectral parameter} in soliton theory and that
the algebraic mechanism bringing
the physical concept of particle and collective motions
arises from the reduction from $u_N$ to $su_N$ 
for the ID HF Hamiltonian.\\
(4)  Though the TDHF $\!$theory$\!$ 
describes a$\!$ dynamics 
on$\!$ {\em the real fermion-harmonic oscillator}, 
the soliton theory does 
on {\em the complex fermion-harmonic oscillator}.$\!$
This remark gives important tasks 
to extend the$\!$ TDHF $\!$theory$\!$ on 
the real space {\it affine} $\widehat{su_{\!N}}\!$ 
to the complex space {\it affine} $\widehat{sl_{\!N}}$
and to understand
the concept of 
particle energy and boson energy.
They are itemized as follows:\\[2pt] 
(i) The algebraic mechanism can be derived from 
the three important elements: first,$su_N$ 
condition for HF Hamiltonian;
second, the vacuum state %(highest weight vector) 
according to the idea of Dirac;
and last, the phase of 
fermion gauge  can be separated
into the {\em particle and collective modes}.\\
(ii) From this mechanism,
a new theory for unified description of
both the motions can be presented. 
Introducing values with a time periodicity, 
a unified theory of
both the motions 
beyond the static HF equation and RPA 
one can be obtained.
With the help of the {\it affine} KM algebra,
the theory  provides the algebraic mechanism 
to elucidate the physical concepts clearly:
collective motion induced by a TD MF potential
and the symmetry breaking of fermion systems
and the successive occurrence of the collective motion
due to the recovery of the symmetry
in which
circle $S^1$ takes an active role in causing 
the resonance
(interference) between fermions.\\
(iii) The theory gives a toy model
to clear algebraic structure among the original fermion field,
vacuum field defined in the SCF potential
and bosonic field associated with Laurent spectra.\\
(iv) To solve the new equation,
we must study how to extract the various subgroup orbits
satisfying the Pl\"{u}cker relation and
how to determine
the solution
in the $su_N( \!\in\! sl_N)$ hierarchy.

We attempt to construct an optimal coordinate system
on group manifold.$\!$
Then
the relation of
boson expansion method for FD fermion system
to $\tau$-FM for ID one
is investigated to
clarify the algebro-geometric structure of integrable system.$\!$
Some physical concepts and mathematical methods
work well in the ID system.$\!$
The SCF method based on the global symmetry
is much improved,
if we notice a local symmetry of
the systems.$\!$
The generator coordinate (GC) method gives 
the superposition principle
on nonlinear space
\cite{Boyd90,Fu.81}.$\!$
Standing on the local symmetry of
the system behind the global symmetry in the FD system,
we have reconstructed the GC and 
the nonlinear superposition methods.$\!$
We have many problems related to further perspective. 
They are itemized as follows
\cite{KN.00}:\\[-20pt]
%%%%%%%%%
\begin{enumerate}
\item To study path integral method ($\!$PIM$\!$)$\!$
\cite{Nishi.81}-formal RPA from the viewpoint of symmetry:\\%[-2pt]
In PIM, the RPA equation is 
described as the fluctuational
mode with time periodicity of  the Jacobi field around a classical
path on the phase space. 
Then we intend to illustrate the relation between both the methods
from the viewpoint of symmetry.\\[-24pt]

\item To extend the
particle-hole formalism for
($1+2\m$)-dimensional 
$\frac{SU_{m+n}}{S(U_{m} \times U_{n})}$
model,
Caseller--Megna-Sciuto$\!$ 
\cite{CMS.81}
to a superconducting paired formalism
on the
$\!\frac{SO(2N)}{U(N)}\!$
coset space:\\[-4pt]
We have successfully derived the
$\frac{SU_{m+n}}{S(U_{m} \times U_{n})}$
geometric equation 
to get the classical equation of motion
for an integrable  system. 
The particle-hole formalism 
is applicable
to the superconducting paired formalism on the
$\!\frac{SO(2N)}{U(N)}\!$
and 
$\!\frac{SO(2N\!+\!2)}{U(N\!+\!1)}\!$
coset spaces
\cite{Nishi.81,Nishi.98}.
We will study the relation of the present model
with the $\sigma$-model on the ${\bf Gr}$
\cite{NPB.08}.\\[-22pt]

\item To study the hiddenness behind 
the gauge of state-function 
$\!$and$\!$ formation of fermion pair:\\%[-2pt]
For forming a pair we give
{\em classifications} of Laurent spectra
in the ID fermion.$\!$
The Laurent coefficient of $n$-soliton solution
and 
$\tau$-function 
for {\it affine} $\widehat{su}(\!N\!)$
have been given in
\cite{NishiProviKoma2.07}.\\[-24pt]

\item To clarify the relation
between the spectral parameter and the collective variables:\\%[-2pt]
A spectral parameter of the iso-spectral equation$\!$
\cite{AdlerMoerbeke.94}
in soliton theory and collective variable in the SCF method,
though being seen different aspect,
work as scaling parameters on $S^1\!$.$\!$
The former relates to the scaling parameters in description by
analytical continuation.$\!$
The latter makes a role of the deformation parameter
of {\em loop paths} in ${\bf Gr}_m$.\\[-24pt]

\item To study the relation between 
collective motions in SCF theory and Far Fields 
\cite{TW.98}:\\%[-2pt]
In the SCF method, 
we have not known 
approaches from the viewpoint of Far Fields.
How do we bring in the idea of 
the {\em Far Fields} to relate it to
the {\em adiabatic} one
which is introduced to separate a collective motion 
from the others?
The theory bases on that 
the speed of collective motion is much slower
than that of any other non-collective motions.\\[-22pt]

\item To study why soliton solution for
classical wave equation shows fermion-like behavior in
quantum dynamics and what symmetry is hidden 
in soliton equation
\cite {TW.97}:\\%[-2pt]
The nonlinear Schr\"{o}dinger equation as a classical image
of corresponding bose field 
\cite{YK.87}
has multi-soliton solutions and appears as 
solutions for
non-scattering potential
\cite{NW.78}.$\!$
%The potential is derived with the use of $\tau$-function.$\!$
Nogami
used 
a multi-component nonlinear Schr\"{o}dinger equation 
instead of TDHF equation
\cite{NW.78}.\\[-22pt]

\item To study another expression of the TDHB equation
using the K\"{a}hler coset space:\\%[-2pt]
To describe the classical motion on the coset manifold,
we start from the local equation of motion
which becomes a Riccati-type equation.
We get a general solution of the TD Riccati-HB equation
for coset variables.
We obtain the Harish-Chandra decomposition
for the $SO(2N)$ matrix
based on the nonlinear M\"{o}bius transformation
\cite{SeiyaJoao.15}.\\[-22pt]

 \item To study the relation between 
the nonlinear superposition principle in soliton theory
and the Res MF theory in SCF method
and the algebraic (Alg) MF one:\\%[-2pt]
What relation does exist between 
the construction of exact solutions based on the idea of
imbricate series in soliton equation 
\cite{TW.97}
and Res- and Alg- MF theories
\cite{NishiProvi.19,FukuNishi.88.91}?
Such the idea bases on
the group integration on the solution space of soliton equation.$\!$
The GC method in SCF method
stands on the group integration
\cite{NMO.04,NishiProvi.16}.$\!$
Further 
we discuss on the ordinary MF theory related to
the Alg MF 
theory based on 
the coadjoint orbit leading to
the non-degenerate symplectic form
\cite{Kirillov.76}.
\end{enumerate}

\newpage

%%%%%%%%%%%%%
%                                    %
%      Appendix              %
%                                    %
%%%%%%%%%%%%%

\leftline{\large{\bf Appendix}}

\vspace{-0.5cm}
 
\appendix
\def\thesection{\Alph{section}}
\setcounter{equation}{0}
\renewcommand{\theequation}{\Alph{section}.
\arabic{equation}}

\setcounter{equation}{0}
\section{Derivation of (\ref{Hc def}) and (\ref{Oc def})
and density matrix}

%%%%%%%%%%%%%%%%%%%%%%%%%%%%%%%%%%%%

We give a decomposition of the generator
$U(g)$,$\!$ i.e.,$\!$
$U(N)$ canonical transformation
(\ref{SO(2N) Canonical Trans}),
$
U^{\!-1}(g (\Lambda,\!\Lambda^\star\!,\!t))
\!=\!
e^{\widehat{\Xi} (\Lambda,\!\Lambda^\star,\!t)}
e^{\widehat{\Upsilon} (\Lambda,\!\Lambda^\star,\!t)} 
\left( \!
U^{\!-1}(g)
\!\!=\!\!
U^\dag(g) ,
U^\dag(g)
\!\!=\!\!
U(g^\dag) \!
\right)
$.~$\!$
We express 
$g (\Lambda,\!\Lambda^\star\!,\!t)$
as $g$.$\!$ 
For our aim,
we set up 
variable $\xi$,
$(N \!-\! m) \!\times\! m$ matrix
$(\xi_{ia})$,
$a \!\!=\!\! 1,2 \cdots m\mbox{~(occupied~state)}$
and
$i \!=\! m + 1, \!\cdots\! N~\mbox{(unoccupied~state)}$ 
and
variable 
$\upsilon 
~(=\!-\upsilon^\dag),
~m \times m$ 
matrix
$(\upsilon_{ab}), a, b \!=\! 1,2 \cdots m$
and 
variable 
$\upsilon^\star 
(\!=\!-\upsilon^{\mbox{\scriptsize T}}),
(N \!-\! m) \!\times\! (N \!-\! m)\!$
matrix
$\!(\upsilon^\star_{ij}), i, j \!\!=\!\! m  \!+\! 1, \!\cdots\! N $,
respectively.~$\!$
Further,
we decompose
the creation operator
${[c^\dag}_{\!\alpha}]$
as
$
{[c^\dag}_{\!\alpha}]
\!=\!
{[\Hat{c}^\dag_a, \Check{c}^\dag_i}]
$
and
the annihilation operator
${[c_{\!\alpha}}]$
as
$
{[c_{\!\alpha}]
\!=\!
{[\Hat{c}_a, \Check{c}_i}}]
$
and prepare the following two operators
$\widehat{\Xi}$
and
$\widehat{\Upsilon}$:\\[-16pt]
\begin{eqnarray}
\!\!\!\!
\left.
\begin{array}{c}
%\\
%\\[-10pt]
\widehat{\Xi} 
\!\equiv\! 
[\Hat{c}^\dag, \Check{c}^\dag]
\Xi \!
\left[ \!
\begin{array}{l}
\Hat{c} \\
\Check{c}
\end{array} \! 
\right], 
\Xi 
\!=\!
\left[ \!\!\!
\begin{array}{cc}
0 &\!\! -\xi^\dag \\
 \xi       &\!\!  0
\end{array} \!\!\!
\right] ,~
g_\xi
\!=\!
e^{\Xi}
\!=\!
\left[ \!\!\!
\BA{cc} C(\xi ) &\!\!\! 
-S^\dag (\xi ) \\ S(\xi )  &\!\!\! 
\tilde{C}(\xi ) 
\EA \!\!\!
\right],\\
\\[-6pt]
\widehat{\Upsilon} 
\!\equiv\! 
[\Hat{c}^\dag, \Check{c}^\dag] 
\Upsilon \!
\left[ \!
\begin{array}{l}
\Hat{c} \\
\Check{c}
\end{array} \! 
\right], 
\Upsilon
\!\equiv\!
\left[ \!\!\!
\begin{array}{cc}
\upsilon &\!\!\! 0 \\
 0      &\!\!\!  \upsilon^\star
\end{array} \!\!\!
\right],
g_\upsilon
\!=\!
e^\Upsilon
\!=\! 
\left[ \!\!\!
\begin{array}{cc}
e^\upsilon &\!\!\! 0 \\
 0      &\!\!\!  e^\upsilon{}^\star
\end{array} \!\!\!
\right] \! ,
\end{array}
\right\}
\label{generators}
\end{eqnarray}\\[-10pt]
where
the triangular matrix functions
$(S( \xi), C( \xi ), \tilde{C}( \xi ))$
are given by
Fukutome$\!$
\cite{Fu.81,Fu.Int.J.Quantum Chem.81}
as, \\[-18pt]
\begin{eqnarray}
\left.
\begin{array}{rl}
&S( \xi )
\!=\!
\sum^\infty_{k=0}(-1)^k 
{\displaystyle \frac{1}{(2k+1)!}}\, 
\xi (\xi ^{\dagger } \xi)^k
\!=\!
{
\displaystyle 
\frac{\sin \sqrt{ \xi ^{\dagger } \xi})}
{\sqrt{ \xi ^{\dagger } \xi}}
}
\xi, \\
\\[-10pt]
&C( \xi )
\!=\!
1_m
\!+\!
\sum^\infty_{k=1}(-1)^k 
{\displaystyle \frac{1}{(2k)!}}\,
( \xi ^{\dagger } \xi)^k
\!=\!
\cos \sqrt{\xi ^{\dagger } \xi}
=
C^{\dagger }( \xi ), \\
\\[-10pt]
&\tilde C( \xi ) 
= 
1_{n}
\!+\!
\sum^\infty_{k=1}(-1)^k 
{\displaystyle \frac{1}{(2k)!}}
( \xi^{\dagger })^k
=
\cos \sqrt{ \xi \xi ^{\dagger }}
=
\tilde C^{\star }( \xi ), 
\end{array} 
\right\}
\label{SinCosMat}
\end{eqnarray}\\[-10pt]
which hold the properties analogous to 
the usual triangular functions\\[-16pt]
\beqa
C^2( \xi )+S^{\dagger }( \xi )S(\xi ) \!=\! 1_m,~
\tilde C^2( \xi )
+
S( \xi )S^{\dagger }( \xi ) \!=\! 1_{n} (\!\!=\!\! 1_{N\!-\!m}),~
S( \xi )C( \xi )
\!=\!
\tilde{C}( \xi )S( \xi ).
\label{SinCosRel}
\eeqa\\[-18pt]
Then using
the matrices 
$g_{ \xi }$ and $g_{ \upsilon }$,
the matrix $g$ 
is given as\\[-12pt]
\begin{eqnarray}
\begin{array}{c}
g 
\!=\! 
g_{\xi } g_{\upsilon}
\!=\!
e^{\Xi}e^{\Upsilon }
\!=\!
\left[ \!\!\!
\BA{cc} C(\xi ) &\!\!\! 
-S^\dag (\xi ) \\ S(\xi )  &\!\!\! 
\tilde{C}(\xi ) 
\EA \!\!\!
\right] \!
\left[ \!\!\!
\BA{cc} 
e^{ \upsilon } &\!\!\! 0 \\ 
0  &\!\!\! e^{ \upsilon^\star } 
\EA \!\!\!
\right] 
\!=\!
\left[ \!\!\!
\BA{cc} C(\xi ) e^{ \upsilon } &\!\!\! 
-S^\dag (\xi ) e^{ \upsilon^\star } \\ 
S(\xi )e^{ \upsilon }  &\!\!\! 
\tilde{C}(\xi ) e^{ \upsilon^\star }
\EA \!\!\!
\right] 
\!=\!
\left[ \!\!
\BA{cc} 
\Hat{a} & \acute{b} \\ 
\grave{b} & \Check{a} 
\EA \!\!
\right] \! .
\end{array} 
\label{matrix-g}
\end{eqnarray}\\[-10pt]
From
(\ref{generators}),
we get,\\[-16pt]
\begin{eqnarray}
\!\!\!\!\!\!
\left.
\begin{array}{c}
U(g_{ \xi})
\Hat{c}_a^\dagger 
U^{-1}(g_{ \xi})
\!\!=\!\!
\Hat{c}_b^\dagger \!
\left[C(\xi )\right]_{ba}
\!+\!
\Check{c}_i^\dagger \!
\left[S(\xi )\right]_{ia},~
U(g_{ \xi})
\Check{c}_i^\dagger 
U^{-1}(g_{ \xi})
\!\!=\!\!
\Check{c}_j^\dagger \!
\left[\tilde{C}(\xi )\right]_{ji}
\!-\!
\Hat{c}_a^\dagger \!
\left[S^\dag(\xi )\right]_{ai},\\
\\[-12pt]
U(g_{ \xi})
[\Hat{c}^\dagger,\Check{c}^\dagger]
U^{-1}(g_{ \xi})
\!=\!
[\Hat{c}^\dagger,\Check{c}^\dagger]g_{ \xi} ,
\left[ \!
\begin{array}{c}
\Hat{d} \\
\Check{d} 
\end{array} \!\!
\right] 
\!=\!
g_{ \xi}^\dag \!
\left[ \!
\begin{array}{c}
\Hat{c} \\
\Check{c} 
\end{array} \!
\right] \!
g_{ \xi} ,~
g^\dag_{ \xi}g_{ \xi}
\!=\!
g_{ \xi}g^\dag_{ \xi}
\!=\!
1_N ,\\
\\[-2pt]
U(g_{ \upsilon })
\hat{c}_a^\dagger
U^{-1}(g_{\upsilon})
\!=\!
\Hat{c}_b^\dagger (e^\upsilon)_{ba},~
U(g_{ \upsilon })
\Check{c}_i^\dagger
U^{-1}(g_{\upsilon})
\!=\!
\Check{c}_j^\dagger (e^\upsilon{}^\star)_{ji} ,\\
\\[-6pt]
U(g_{ \upsilon })
[\Hat{c}^\dagger,\Check{c}^\dagger]
U^{-1}(g_{\upsilon})
\!=\!
\left[\hat{c}^\dagger, 
\Check{c}^\dagger  \right] \!
\left[ \!\!\!
\BA{cc} 
e^{ \upsilon } &\!\!\! 0 \\ 
0  &\!\!\! e^{ \upsilon{}^\star } 
\EA \!\!\!
\right] .
\end{array} \!\!
\right \}
\label{SO(2N) Canonical Trans2}
\end{eqnarray}
Finally, 
from
(\ref{SO(2N) Canonical Trans2}),
we obtain the following formula:\\[-16pt]
\begin{eqnarray}
\left.
\begin{array}{c}
\\[-12pt]
[\Hat{d}^\dagger,\Check{d}^\dagger]
\!\!=\!\!
U(g_{ \xi})U(g_{ \upsilon })
[\Hat{c}^\dagger,\Check{c}^\dagger]
U^{-1}(g_{\upsilon})
U^{-1}(g_{ \xi})
\!\!=\!\!
U(g_{ \xi })
[\Hat{c}^\dagger,\Check{c}^\dagger]
U^{-1}(g_{\xi})
g_{\upsilon }
\!\!=\!\!
[\Hat{c}^\dagger,\Check{c}^\dagger]
g ,\\
\\[-2pt]
\left[ \!
\begin{array}{c}
\Hat{d} \\
\Check{d}
\end{array} \!
\right] \!
[\Hat{d}^\dagger,\Check{d}^\dagger]
\!=\!
g^\dag\!
\left[ \!
\begin{array}{c}
\Hat{c} \\
\Check{c}
\end{array} \!
\right] \!
[\Hat{c}^\dagger,\Check{c}^\dagger]
g .
\end{array}
\right \}
\label{SO(2N) Canonical Trans3}
\end{eqnarray}\\[-8pt]
The differential 
$\partial_t e^{\widehat{\Xi}}$
is calculated as\\[-16pt]
\begin{eqnarray}
\BA{rl}
\partial_t e^{\widehat{\Xi}}
&\!\!\!\!=\!
{\displaystyle \int_0^t }\! d\tau 
e^{\tau \widehat{\Xi}(t)}
\partial_t \widehat{\Xi}(t)
e^{-\tau \widehat{\Xi}(t) }
e^{\widehat{\Xi}(t)}
\!=\!
{\displaystyle \int_0^t } \! d\tau 
e^{\tau \Delta_{\widehat{\Xi}(t) }}
\partial_t \widehat{\Xi}(t)
e^{\widehat{\Xi}(t)}
\!=\!
{
\displaystyle 
\frac{e^{\Delta_{\widehat{\Xi}}}
\!-\!
{\bf 1}}{\Delta_{\widehat{\Xi}}}
}
\partial_t \widehat{\Xi}(t)
e^{\widehat{\Xi}(t)} \\
\\[-14pt]
&\!\!\!\!=\!
\left[
{\displaystyle  \frac{1}{1!}}
\partial_t \widehat{\Xi}
+
{\displaystyle  \frac{1}{2!}}
[\widehat{\Xi},~\! \partial_t \widehat{\Xi}]
+
\cdots
+
{\displaystyle  \frac{1}{n!}}
[\widehat{\Xi}, \cdots
[\widehat{\Xi},~\! \partial_t \widehat{\Xi}
\underbrace{] \cdots]}_{n-1}
+
\cdots
\right] \!
e^{\widehat{\Xi}}  \\
\\[-16pt]
&\!\!\!\!=\!
[\Hat{c}^\dag, \Check{c}^\dag]~\!
\partial_t e^{\Xi}
e^{-\Xi} \!
\left[ \!\!
\begin{array}{l}
\Hat{c} \\
\Check{c}
\end{array} \!\! 
\right] \!
e^{\widehat{\Xi}}
\!=\! 
\partial_t \widehat{\Xi}_L
e^{\widehat{\Xi}} ,~
(
\mbox{due~to}~
[\widehat{\Xi},~\! \partial_t \widehat{\Xi}]
\!=\!
0
) ,
\EA
\label{Xi_generator}
\end{eqnarray}\\[-10pt]
Similarly,
the differential 
$\partial_t e^{\widehat{\Upsilon}}$
is also computed as\\[-16pt]
\begin{eqnarray}
\BA{rl}
\partial_t e^{\widehat{\Upsilon}}
&
\!\!\!\!=\!
[\Hat{c}^\dag, \Check{c}^\dag] 
\partial_t e^{\Upsilon}
e^{-\Upsilon} \!
\left[ \!\!
\begin{array}{l}
\Hat{c} \\
\Check{c}
\end{array} \!\! 
\right] \!
e^{\widehat{\Upsilon}}
\!=\! 
\partial_t \widehat{\Upsilon}_L
e^{\widehat{\Upsilon}} ,~
(
\mbox{due~to}~
[\widehat{\Upsilon},~\! \partial_t \widehat{\Upsilon}]
\!=\!
0
) .
\EA
\label{Gamma_generator}
\end{eqnarray}\\[-12pt]
The proof
of
$
[\widehat{\Xi},~\! \partial_t \widehat{\Xi}]
\!=\!
0
$
is given as follows:\\[-16pt]
\begin{eqnarray}
\BA{ll}
&\!\!\!\!
[\widehat{\Xi},~\! \partial_t \widehat{\Xi}]
\!=\!
[\Hat{c}^\dag, \Check{c}^\dag] 
\Xi \!
\left[ \!\!
\begin{array}{l}
\Hat{c} \\
\Check{c}
\end{array} \!\! 
\right] \!
[\Hat{c}^\dag, \Check{c}^\dag] 
\partial_t \Xi \!
\left[ \!\!
\begin{array}{l}
\Hat{c} \\
\Check{c}
\end{array} \!\! 
\right] \!
-
[\Hat{c}^\dag, \Check{c}^\dag] 
\partial_t \Xi \!
\left[ \!\!
\begin{array}{l}
\Hat{c} \\
\Check{c}
\end{array} \!\! 
\right] \!
[\Hat{c}^\dag, \Check{c}^\dag] 
 \Xi \!
\left[ \!\!
\begin{array}{l}
\Hat{c} \\
\Check{c}
\end{array} \!\! 
\right] \\
\\[-10pt]
&\!\!\!\!=\!
-
\mbox{Tr} \!
\left[
\left\{ 
\Xi \!
\left[ \!\!
\begin{array}{l}
\Hat{c} \\
\Check{c}
\end{array} \!\! 
\right] \!
[\Hat{c}^\dag, \Check{c}^\dag] ~\!
\partial_t \Xi 
-
\partial_t \Xi \!
\left[ \!\!
\begin{array}{l}
\Hat{c} \\
\Check{c}
\end{array} \!\! 
\right] \!
[\Hat{c}^\dag, \Check{c}^\dag] ~\!
\Xi
\right\} \!
\left[ \!\!
\begin{array}{l}
\Hat{c} \\
\Check{c}
\end{array} \!\! 
\right] \!
[\Hat{c}^\dag, \Check{c}^\dag] 
\right] \\
\\[-8pt]
&\!\!\!
\!=\!
-
\mbox{Tr} \!\!
\left[ \!
\left\{ \!
\Xi \!
\left[ \!\!
\begin{array}{cc}
E^\bullet_{~\bullet} &\!\!\!\! E^{\bullet\bullet} \\
E_{\bullet\bullet} 
&\!\!\!\! -E^{\bullet \dagger}_{~\bullet}
\end{array} \!\!
\right] \!
\partial_t \Xi \!
\left[ \!\!
\begin{array}{cc}
E^\bullet_{~\bullet} &\!\!\!\! E^{\bullet\bullet} \\
E_{\bullet\bullet} 
&\!\!\!\! -E^{\bullet \dagger}_{~\bullet}
\end{array} \!\!
\right] 
\!\!-\!\!
\partial_t \Xi \!
\left[ \!\!
\begin{array}{cc}
E^\bullet_{~\bullet} &\!\!\!\! E^{\bullet\bullet} \\
E_{\bullet\bullet} 
&\!\!\!\! -E^{\bullet \dagger}_{~\bullet}
\end{array} \!\!
\right]  \!
\Xi \!
\left[ \!\!
\begin{array}{cc}
E^\bullet_{~\bullet} &\!\!\!\! E^{\bullet\bullet} \\
E_{\bullet\bullet} 
&\!\!\!\! -E^{\bullet \dagger}_{~\bullet}
\end{array} \!\!
\right] \!
\right\} \!
\right] 
\!=\!
0 .
\EA
\label{proof}
\end{eqnarray}\\[-6pt]
Using
(\ref{Hc def})
and
(\ref{Xi_generator}),
collective Hamiltonian $H_c$ is rewritten as\\[-14pt]
\begin{eqnarray}
\BA{ll}
H_c
&\!\!\!\!=\!
i\hbar \partial_t
U(g) U^{-1}(g)
\!=\!
\left\{
i\hbar \partial_t \!
\left(e^{\widehat{\Xi}}e^{\widehat{\Upsilon}}\right) \!
\right\}
e^{-\widehat{\Upsilon}}
e^{-\widehat{\Xi}}\\
\\[-10pt]
&\!\!\!\!=\!
\left\{
i\hbar \partial_t e^{\widehat{\Xi}} e^{\widehat{\Upsilon}}
+
e^{\widehat{\Xi}} i\hbar \partial_t e^{\widehat{\Upsilon}}
\right\}
e^{-\widehat{\Upsilon}}
e^{-\widehat{\Xi}}
\!=\!
\left\{
i\hbar \partial_t \widehat{\Xi}_L
e^{\widehat{\Xi}} e^{\widehat{\Upsilon}}
+
e^{\widehat{\Xi}} 
i\hbar \partial_t \widehat{\Upsilon}_L
e^{\widehat{\Upsilon}}
\right\}
e^{-\widehat{\Upsilon}}
e^{-\widehat{\Xi}}\\
\\[-8pt]
&\!\!\!\!=\!
i\hbar \partial_t \widehat{\Xi}_L
+
e^{\widehat{\Xi}} 
i\hbar \partial_t \widehat{\Upsilon}_L
e^{\widehat{\Upsilon}}
U^{-1}(g)
\!=\!
i\hbar \partial_t \widehat{\Xi}_L
+
U(g)
e^{-\widehat{\Upsilon}} 
i\hbar \partial_t \widehat{\Upsilon}_L
e^{\widehat{\Upsilon}}
U^{-1}(g) .
\EA
\label{H-c}
\end{eqnarray}\\[-6pt]
We demand that
the canonical transformation 
$U(g_\upsilon), g_\upsilon \!\!\in\!\! U(N)\!$
$(c_\alpha |0 \rangle \!\!=\!\! 0)$
leaves the free-particle vacuum $|0 \rangle$
invariant;
$U(g_\upsilon) |0 \rangle \!\!=\!\! |0 \rangle$.
Due to this demand,
we set up the condition,
$i\hbar \partial_t \! \widehat{\Upsilon}_{\!L} 
\!\!=\!\!
i\hbar \partial_t e^{\!\Upsilon} \!\!=\! 0$.
From this condition and
(\ref{matrix-g})
we get the final expression for
$H_{c}$
(\ref{Hc Tr-express})
as\\[-16pt]
\begin{eqnarray}
\!\!\!\!\!\!
H_c
\!\!=\!\!
[\Hat{c}^\dag, \Check{c}^\dag]
i\hbar \partial_t e^{\Xi}
e^{-\Xi} \!\!
\left[ \!\!
\begin{array}{l}
\Hat{c} \\
\Check{c}
\end{array} \!\! 
\right] \!\!
e^{\widehat{\Xi}}
\!\!=\!\!
[\Hat{c}^\dag, \Check{c}^\dag]
i\hbar \partial_t \!
\left( \! e^{\Xi}e^{\Upsilon } \! \right) \!
e^{-\Upsilon } \! e^{-\Xi} \!\!
\left[ \!\!
\begin{array}{l}
\Hat{c} \\
\Check{c}
\end{array} \!\! 
\right] \!\!
e^{\widehat{\Xi}}
\!\!=\!\!
[\Hat{c}^\dag, \Check{c}^\dag]
i\hbar \partial_t
g \!\cdot\! g^\dag \!\!
\left[ \!\!
\begin{array}{l}
\Hat{c} \\
\Check{c}
\end{array} \!\! 
\right] \!\!
e^{\widehat{\Xi}} .
\label{finalH-c}
\end{eqnarray}\\[-12pt]
For the Lie-algebra-valued
coordinates $(O_\n^\dagger,O_\n)$
(\ref{Oc Tr-express}), 
we also get the final expressions as\\[-16pt]
\begin{eqnarray}
O_\n^\dagger
\!\!=\!\!
[\Hat{c}^\dag, \Check{c}^\dag]
i\hbar \partial_{\Lambda_\n}
g \!\cdot\! g^\dag \!\!
\left[ \!\!
\begin{array}{l}
\Hat{c} \\
\Check{c}
\end{array} \!\! 
\right] \!
e^{\widehat{\Xi}} ,~\!
O_\n
\!\!=\!\!
[\Hat{c}^\dag, \Check{c}^\dag]
i\hbar \partial_{\Lambda^\star_\n}
g \!\cdot\! g^\dag \!\!
\left[ \!\!
\begin{array}{l}
\Hat{c} \\
\Check{c}
\end{array} \!\! 
\right] \!
e^{\widehat{\Xi}} ,~\!
e^{\widehat{\Xi}}
\!\!=\!\!
\exp \!
\left\{ \!
[\Hat{c}^\dag, \Check{c}^\dag]~\!
\Xi \!
\left[ \!\!
\begin{array}{l}
\Hat{c} \\
\Check{c}
\end{array} \!\! 
\right] \!
\right\} \! .
\label{finalO-c}
\end{eqnarray}\\[-10pt]
Following
Fukutome$\!$
\cite{Fu.81,Fu.Int.J.Quantum Chem.81},
let 
$q^L_{\xi,\alpha a}$ 
and
$q^R_{\xi,\alpha a}$
be the 
$N \!\times\! m$ matrix
and the
$N \!\times\! n$ matrix\\[-8pt]
\beq
(q^L_{\xi,\alpha a})
\!=\!
\left[ \!\!
\begin{array}{l}
\Hat{a} \\
\grave{b}
\end{array} \!\! 
\right] ,~
(q^R_{\xi,\alpha i})
\!=\!
\left[ \!\!
\begin{array}{l}
\acute{b} \\
\Check{a}
\end{array} \!\! 
\right] ,
\left(
g 
\!=\! 
\left[  \!\!
\begin{array}{cc}
\Hat{a}  & \acute{b} \\
 \grave{b}       &  \Check{a}
\end{array}  \!\!
\right] 
\stackrel{\mbox{mode}}{\Longrightarrow} 
\left[  \!\!
\begin{array}{cc}
(hh)  & (hp) \\
 (ph)       &  (pp)
\end{array}  \!\!
\right]
\right) .
\label{column_q}
\eeq
The HF density matrix 
$
{\cal Q}
\!=\!
({\cal Q}_{\alpha \beta})
$
is expressed as\\[-8pt]
\beq
{\cal Q}(g)
\!\!=\!\!
q^L_{\xi}~\!\!
q^{L\dag}_{\xi}
\!\!=\!\!
\left[ \!\!
\BA{cc} C^2(\xi ) &\!\! 
C(\xi ) S^\dag (\xi ) \\[2pt] 
S(\xi ) C(\xi )  &\!\! 
S(\xi ) S^\dag (\xi )  
\EA \!\!
\right] 
\!\!=\!\!
\left[ \!\!
\BA{cc} C^2(\xi ) &\!\! 
C(\xi ) S^\dag (\xi ) \\[2pt]  
S(\xi ) C(\xi ) &\!\! 
1_{N\!-\!m} \!-\! \tilde C^2( \xi )  
\EA \!\!
\right] \! ,~
{\cal Q}
\!\!=\!\!
{\cal Q}^\dag \!,
{\cal Q}^2
\!\!=\!\!
{\cal Q}.
\label{DM_Q}
\eeq
We also have
$
q^R_{\xi}~\!\!
q^{R\dag}_{\xi}
\!=\!
1_N
-
{\cal Q}(g) 
$.
The HF energy functional  is given in terms of 
${\cal Q}$
as\\[-12pt]
\beqa
\BA{ll}
E_{HF}
&\!\!
\!\stackrel{d}{=}\!
\langle U(g)|
H
|U^{-1}(g) \rangle
\!=\!
\langle \phi (g)|
H
|\phi (g) \rangle
\!=\!
h_{\beta\alpha}{\cal Q}_{\alpha \beta}
\!+\!
\frac{1}{2}
[\gamma\alpha|\delta\beta]
{\cal Q}_{\alpha \gamma}
{\cal Q}_{\beta \delta}\\
\\[-4pt]
&\!\!
\!=\!
\mbox{Tr} 
\{h{\cal Q}\}
\!+\!
\frac{1}{2}
\left[
\mbox{Tr}\{{\cal Q}\bullet\bullet\}|
\mbox{Tr}\{\bullet\bullet{\cal Q}\}
\right] .
\EA
\label{HFenergyfunc}
\eeqa

\newpage

%%%%%%%%%%%%%%
%                                        %
%   Acknowledgements     %
%                                        %
%%%%%%%%%%%%%%

\begin{center}
{\bf Acknowledgements}
\end{center}
One of the authors (S. N.)
expresses his sincere thanks to
Professor Constan\c{c}a Provid\^{e}ncia for kind and
warm hospitality extended to him at
CFisUC,
Departamento de F\'{i}sica,
Universidade de Coimbra, Portugal.
This work was supported by FCT (Portugal) under the Project
UID/FIS/04564/2016 and UID/FIS/04564/2019.

\vspace{0.5cm}

\begin{center}
{\bf Note}
\end{center}
Using the particle--hole representation
this work is made, basing on a part of the paper:
S.Nishiyama,
J.daProvid\^{e}ncia,$\!$
C.Proid\^{e}ncia, $\!$
F.Cordeiro and T.Komatsu,$\!$
{\it Self-Consistent-Field Method and $\tau\!$-$\!$Functional Method
on Group Manifold in Soliton Theory:$\!$ a Review and  New Results},
SIGMA {\bf 5} (2009) 009-1-76.

%%%%%%%%%%%%%%%%%%%%%%%%%%%%%%

%%%%%%%%%%
%                           %
%   References      %
%                           %
%%%%%%%%%%


\newpage



\begin{thebibliography}{999}
\bibitem{YK.87}
%\vspace{-0.3cm}
M. Yamamura and A. Kuriyama, 
Time-Dependent Hartree-Fock Method and Its Extension,
{\it Prog. Theor. Phys. Suppl.} {\bf 93} (1987).
\bibitem{BM.53}
\vspace{-0.1cm}
A. Bohr and B. Mottelson,
Collective and Individual-Particle Aspects of Nuclear Structure,
{\it Mat. Fys. Medd. Dan. Vid. Selsk.} {\bf 27} (1953)  No.16
\bibitem{AM.60}
\vspace{-0.1cm}
R. Arvieu and M. Verononi,
Quasi-particles and collective states 
of spherical nuclei,
{\it Compt. Rend.} {\bf 250} (1960) 992-994; 2155-2157,\\
M. Baranger,
Extension of the Shell Model for Heavy Spherical Nuclei,
{\it Phys. Rev.} {\bf 120} (1960) 957-968.%\\
T. Marumori,
On the Collective Motion in Even-Even Spherical Nuclei,
{\it Prog. Theor. Phys.} {\bf 24} (1960) 331-356.
\bibitem{Mo.59}
\vspace{-0.1cm}
B.R. Mottelson,
Nucear Structure,
The Many Body Ploblem, Le Probl\`{e}me \`{a} n Corps,
Lectures at Les Houches Summer School, Paris, Dunod (1959)
283-315.
\bibitem{Bogo.59}
\vspace{-0.1cm}
N.N. Bogoliubov,
The Compensation Principle and 
the Self-Consistent Field Method,
{\it Soviet Phys.-Uspekhi} {\bf 67} (1959) 236-254.
%\bibitem{SIGMA.09}
%\vspace{-0.3cm}
%S. Nishiyama., J. da Provid\^{e}ncia, C. Provid\^{e}ncia, 
%F. Cordeiro and T. Komatsu,
%{\it Self-Consistent-Field Method and $\tau$-Functional Method
%on Group Manifold in Soliton Theory :
%a Review and  New Results},
%SIGMA {\bf 5} (2009) 009-1-76.
\bibitem{BZ.62}
\vspace{-0.1cm}
S.T. Belyaev and V.G. Zelevinsky,
Anharmonic Effects of Quadrupole Oscillations of Spherical Nuclei,
{\it Nucl. Phys.} {\bf 39} (1962) 582-604.
\bibitem{MYT.64}
\vspace{-0.1cm}
T$\!$. $\!$Marumori, $\!\!$M. $\!\!$Yamamura $\!$and$\!$ A. $\!\!$Tokunaga,$\!$
On the $\!$Anharmonic $\!$Effects on the Collective Oscillations 
in Spherical Even Nuclei. $\!$I,$\!$
{\it Prog.$\!$Theor.$\!$Phys.} $\!${\bf 31}$\!$ (1964)$\!$1009-1025.
\bibitem{PWM.68}
\vspace{-0.1cm}
%J. da Providencia,
%An extension of the random phase approximation,
%{\it Nucl. Phys. A} {\bf 108} (1968) 589-608,\\
J. da Providencia and J. Weneser,
Nuclear Ground-State Correlations and Boson Expansions,
{\it Phys. Rev. C}{\bf 1} (1970) 825-833,\\
E.R. Marshalek,
On the relation between Beliaev-Zelevinsky and 
Marumori boson expansions, 
{\it Nucl. Phys. A} {\bf 161} (1971) 401-409.
\bibitem{FYN.77}
\vspace{-0.1cm}
H. Fukutome, M. Yamamura and S. Nishiyama,
A New Fermion Many-Body Theory Based on 
the SO(2N+1) Lie Algebra of the Fermion Operators,
{\it Prog. Theor. Phys.} {\bf 57} (1977) 1554-1571.
\bibitem{YN.76}
\vspace{-0.1cm}
M. Yamamura and S. Nishiyama,
An a priori Quantized Time-Dependent 
Hartree-Bogoliubov Theory.-A Generalization of 
the Schwinger Representation of Quasi-Spin to 
the Fermion Pair Algebra-, {\it Prog. Theor. Phys.}
{\bf 56} (1976) 124-134.
\bibitem{Fu.81}
\vspace{-0.1cm}
H. Fukutome,The Group Theoretical Structure of 
Fermion Many-Body Systems 
Arising from the Canonical Anticommutation Relation. I
{\it-Lie Algebras of Fermion Operators and 
Exact Generator Coordinate Representations of State Vectors-},
{\it Prog. Theor. Phys.} {\bf 65} (1981) 809-827.
\bibitem{Nogami.55}
\vspace{-0.1cm}
M. Nogami,
Dynamical Hartree Field and Collective motion,
Soryushiron Kenkyu (Kyoto), in Japanese,
{\bf 10} (1955-1956) 600-608.
\bibitem{MH.72}
\vspace{-0.1cm}
E.R. Marshalek and G. Holzwarth, Boson expansion and 
Hartree-Bogoliubov Theory,
{\it Nucl. Phys. A} {\bf 191} (1972) 438-448.
\bibitem{Inglis.56}
\vspace{-0.1cm}
D. R. Inglis,
Nuclear Moments of Inertia due to Nucleon Motion
in a Rotating
{\it Phys. Rev.}{\bf 103} (1956) 1786-1795.
\bibitem{ST.59}
\vspace{-0.1cm}
Y. Shono and H. Tanaka,
Nuclear Collective Motion and 
the Effective Two-Body Potential,
{\it Prog. Theor. Phys.} {\bf 22} (1959) 17-191.
\bibitem{TV.62}
\vspace{-0.1cm}
D.J. Thouless and J.G. Valatin,
Time-dependent Hartree-Fock equations and rotational states of nuclei,
{\it Nucl. Phys.} {\bf 31} (1962) 211-230.
\bibitem{BV.78}
\vspace{-0.1cm}
M. Baranger and M. V\'{e}n\'{e}roni, 
An adiabatic time-dependent Hartree-Fock theory of 
collective motion in finite systems,
{\it Ann. of Phys.} {\bf 114} (1978) 123-200.
\bibitem{BGV.78}
\vspace{-0.1cm}
D.M. Brink, M.J. Giannoni and M. Veneroni,
Derivation of an Adiabatic Time-Dependent Hartree-Fock 
Formalism from a Variational Principle,
{\it Nucl. Phys. A} {\bf 258} (1976) 237.
\bibitem{GR.78}
\vspace{-0.1cm}
K. G\"{o}ke and P.G. Reinhard,  A consistent microscopic theory 
of collective motion 
in the framework of an ATDHF approach,
{\it Ann. of Phys.} {\bf 112} (1978) 328-355.\\
A.K. Mukherjee and M.K. Pal,
Evaluation of the optimal path in ATDHF theory,
{\it Nucl. Phys. A} {\bf 373} (1982) 289-304.
\bibitem{Villars.77}
\vspace{-0.1cm}
F.H. Villars,
Adiabatic Time-Dependent Hartree-Fock Theory in Nuclear Physics,
{\it Nucl. Phys. A} {\bf 285} (1977) 269-296.
\bibitem{HY.74}
\vspace{-0.1cm}            
G. Holzwarth and T. Yukawa,
Choice of the constraining operator in the constrained 
Hartree-Fock method,
{\it Nucl. Phys. A} {\bf 219} (1974) 125-140.
\bibitem{RB.76}
\vspace{-0.1cm}
D. J. Rowe and R. Bassermann,
Coherent state theory of large amplitude collective motion,
{\it Can. J. Phys.} {\bf 54} (1976) 1941-1968.
\bibitem{Ma.80}
\vspace{-0.1cm}
T. Marumori, T. Maskawa, F. Sakata and A. Kuriyama,
Self-consistent Collective Coordinate Method for 
the Large-Amplitude Nuclear Collective Motion,
{\it Prog. Theor. Phys.} {\bf 64} (1980) 1294.%\\
\bibitem{NishiProvi.14}
\vspace{-0.1cm}
S. Nishiyama and J. da Provid\^{e}ncia,
Exact canonically conjugate momenta to
quadrupole-type collective coordinates and
derivation of nuclear quadrupole-type collective Hamiltonian,
{\it Nucl. Phys. A} {\bf 923} (2014) 51-88.
\bibitem{MYT.55}
\vspace{-0.1cm}
T. Marumori, J. Yukawa and R. Tanaka,
On the Foundation of the Unified Nuclear Model, I,
{\it Prog. Theor. Phys.} {\bf 13} (1955) 442-454.
\bibitem{Tomonaga.55}
\vspace{-0.1cm}
S. Tomonaga,
Elementary Theory of Quantum-Mechanical 
Collective Motion of Particles I,
{\it Prog. Theor. Phys.} {\bf 13} (1955) 467-481, 
ibidem II,
{\bf 13} (1955) 482-496.
\bibitem{NishiProvi.15}
\vspace{-0.1cm}
S. Nishiyama and J. da Provid\^{e}ncia,
Description of collective motion in two-dimensional nuclei;
Tomonaga's method revisited,
{\it Nucl. Phys. A} {\bf 935} (2015) 1-17.
\bibitem{Candlin.56}
\vspace{-0.1cm}
D. Candlin,
On Sums over Trajectories for Systems wth Fermi Statistics,
{\it Il Nuovo Cimento} {\bf 4} (1956) 231-239.
\bibitem{Berezin.66}
\vspace{-0.1cm}
F.A. Berezin,The Method of Second Quantization,Pure and 
Applied Physics,A Series of Monographs and Textbooks
{\bf 24} Academic Press NewYork and London 1966.
\bibitem{Grass variable}
\vspace{-0.1cm}
R. Casalbuoni,
The Classical Mechanics for Bose-Fermi Systems,
{\it Nuovo Cim A} {\bf 33} (1976) 389-431.
\bibitem{Dirac.58}
\vspace{-0.1cm}
P.A M. Dirac,The Principles of Quantum Mechanics,
4th Edition,
Oxford University Press, 1958.
\bibitem{YK.81}
\vspace{-0.1cm}        
M. Yamamura and A. Kuriyama,
A Microscopic Theory of Collective and Independent $\!\!$-Particle Motions,
{\it Prog. Theor. Phys.} {\bf 65} (1981) 550-564. 
\bibitem{Olver}
\vspace{-0.1cm}
P.J. Olver,
{\it Applications of Lie Groups to
Differential Equations},
Second Edition,
Graduates texts in mathematics; 107,
Springer-Verlag, New York, 1993. 
\bibitem{AKH.74}
\vspace{-0.1cm}
J. Ablowitz, J. Kaup, C. Newell and H. Segur,
The Inverse Scattering Transform-Fourier Analysis 
for$\!$ Nonlinear Problems.$\!$
{\it Studies $\!$in$\!$ Applied Mathematics},$\!\!$ 
Vol.LIII,$\!$ No.4 $\!$(1974) 249.
\bibitem{Satt.82}
\vspace{-0.1cm}
D.H. Sattinger,
Gauge Theories For Soliton Problems in
{\em Nonlinear Problems: Present and Future},$\!$
pp.51-64,$\!$
eds. by 
A.R. Bishop, D.K. Cambell and B. Nicolaenko.
North-Holland Publishing company, 1982.
\bibitem{GPM}
\vspace{-0.2cm}
M. O. Stephen,
{\it Geometric Perturbation Theory in Physics},
World Scientific Publishing Co. Pte. Ltd., 1984.
\bibitem{JM.83}
\vspace{-0.2cm}
E. Date, M. Jimbo, M. Kashiwara and T. Miwa,$\!$
(a) $\!$Transformation groups for soliton equations,
Nonlinear 
Integrable Systems-Classical Theory and 
Quantum Theory-,
Ed. by M. Jimbo and T. Miwa
World Scientific Publishing 
Co. Pte. Ltd., 1983), pp.39-119.~%\\
(b) Mathematical science of soliton, in Japanese,
Monographs in
{\em Iwanami Lecture on Applied Mathematics}
{\bf 3} (1993) pp.1-112.
Iwanami Publishing Company.
\bibitem{Dickey}
\vspace{-0.1cm}
L.A. Dickey,
Soliton Equations and Hamiltonian System,
World Scientific Publishing Co. Pte. Ltd., 1991.
\bibitem{AR.85}
\vspace{-0.1cm}
G.M. D'Ariano and M.G. Rasetti,
Soliton equations, $\tau$-functions
and coherent states,
Integrable Systems in Statistical Mechanics,
Ed. by G.M. D'Ariano, A. Montorsi
and M.G. Rasetti
World Scientific Publishing Co. Pte. Ltd., 1985,
pp.143-152.
G.M. D'Ariano and M.G. Rasetti,
Soliton Equations and Coherent States,
{\it Phys. Lett. A} {\bf 107} (1985) 291-294.
\bibitem{Wiegmann.07}
\vspace{-0.1cm}
E. Bettelheim, A.G. Abanov and P.B. Wiegmann,
Orthogonality Catastrophe and Shock Waves in a Nonequilibrium
Fermi Gas,
{\it Phys. Rev. Lett.} {\bf 97} (2006) 246402.
%Nonlinear Dynamics of Quantum Systems and Soliton Theory,
%{\it J. Phys. Math. Gen. A} {\bf 40} (2007) F193-F208
\bibitem{RS.80}
\vspace{-0.1cm}
P. Ring and P. Schuck, 
{\it The nuclear many-body problem}, 
Springer, Berlin, 1980.
\bibitem{Fu.Int.J.Quantum Chem.81}
\vspace{-0.1cm}
H. Fukutome, 
Unrestricted Hartree-Fock Theory and Its Applications 
to Molecules and Chemical Reactions,
{\it Int. J. Quantum Chem.} 
{\bf 20} (1981) 955.
\bibitem{Th.60}
\vspace{-0.1cm}
D. J. Thouless, 
Stability conditions and nuclear rotations
in the Hartree-Fock theory,
{\it Nucl. Phys.} {\bf 21} (1960) 225-232.
\bibitem{Pere.72}
\vspace{-0.1cm}
A.M. Perelomov, 
Coheret States for Arbitrary Lie Group,
{\it Comm. Math. Phys.} {\bf 20} (1972) 222;
Generalized coherent states and some of their applications,
{\it Soviet Phys. -Uspekhi.} {\bf 20} (1977) 703.
\bibitem{PS.86}
\vspace{-0.1cm}
A. Pressley and G. Segal,
{\it Loop Groups}, Clarendon Press, Oxford, 1986. 
\bibitem{Hi.76}
\vspace{-0.1cm}
R. Hirota,
Direct method of finding exact solutions
of nonlinear evolution equation,
Lecture Notes in Mathematics, $\!${\bf 515}
ed. $\!$by 
$\!$Miura $\!$R.M. Springer, $\!$New York, $\!$1976 p.40.
\bibitem{LW.78}
\vspace{-0.1cm}
J. Lepowsky and R.L. Wilson,
Construction of the Affine Lie Algebra $A_1 ^{(1)}$,
{\it Commun. Math. Phys.} {\bf 62} (1978) 43-53.
\bibitem{TW.97}
\vspace{-0.1cm}
M. Tajiri and Y. Watanabe,
Periodic Wave Solutions as Imbricate Series of 
Rational Growing Modes: 
Solutions to the Boussinesq Equation,
{\it J. Phys. Soc. Jpn.} {\bf 66} (1997) 1943-1949;~%\\
Breather solutions to the focusing nonlinear 
Schr\"{o}dinger equation,
{\it Phys. Rev. E} {\bf 57} (1998) 3510-3519.
\bibitem{Nishi.88}
\vspace{-0.1cm}
S.Nishiyama,Microscopic$\!$ Theory $\!$of$\!$ Large-Amplitude 
$\!$Collective$\!$ Motions 
Based $\!$on$\!$
the SO(2N$\!$+$\!$1) Lie Algebra $\!$of$\!$ the Fermion 
Operators,{\it Nuovo Cimento A}{\bf 99}(1988) 239-256.
\bibitem{Nishi.98}
\vspace{-0.1cm}
S. Nishiyama,
Time Dependent Hartree-Bogolibov Equation on 
the Coset Space
SO(2N+2)/U(N+1) and QuasiI AntiI-Commutation 
Relation Approximation,
{\it Int. J. Mod. Phys. E} {\bf 7} (1998) 677-707.
\bibitem{Boyd90}
\vspace{-0.1cm}
J.P. Boyd,
New directions in solitons and nonlinear periodic waves: 
Polycnoidal waves, imbricated solitons, 
weakly non-local solitary waves and 
numerical boundary value algorithms,
in {\em Advances in Applied Mechanics}
edited by J. W. Hutchinson and T. Y. Wu,
Vol. {\bf27} Academic Press, 1989, 1-82.
\bibitem{Nishi.81}
\vspace{-0.1cm}
S. Nishiyama,
Path Integral on the Coset Space of 
the SO(2N) Group and 
the Time-Dependent Hartree-Bogoliubov Equation,
{\it Prog. Theor. Phys.} {\bf 66} (1981) 348-350.
\bibitem{Nishi.82.83}
\vspace{-0.1cm}
S. Nishiyama,
Note on the New Type of the SO(2N+1) Time-Dependent 
Hartree-Bogoliubov Equation,
{\it Prog. Theor. Phys.} {\bf 68} (1982) 680-683.\\
An Equation for the Quasi-Particle RPA 
Based on the SO(2N$\!$+$\!$1) 
Lie Algebra of the Fermion Operators,
ibidem {\bf 69} (1983) 1811-1814.
\bibitem{Lax.68}
\vspace{-0.1cm}
P. D. Lax,
Integrals of nonlinear equations of evolution and
solitary waves,
{\it Comm. Pure Appl. Math.} {\bf 21} (1968) 467-490.
\bibitem{BLMP.88}
\vspace{-0.1cm}
M. Boiti, J.J.-P. L\'{e}on, L. Martina and F. Pempinelli,
Scattering of Localized Solutions in the Plane,
{\it Phys. Lett. A} {\bf 132} (1988) 432-439.
\bibitem{FS.88}
\vspace{-0.1cm}
A.S. Fokas and P.M. Santini,
Coherent Structures in Multidimensions,
{\it Phys. Rev. Lett.} {\bf 63} (1989) 1329-1333;
Dromions and a Boundary Value Problem 
for the Davey-Stewartson Equation, {\it Physica D} 
{\bf 44} (1990) 99-130.
\bibitem{DS.74}
\vspace{-0.1cm}
A. Davey and K. Stewartson,
On three-dimensional packets of surface waves,
{\it Proc. R. Soc. London A} {\bf 338} (1974) 101-110.
\bibitem{HH.90}
\vspace{-0.1cm}
J.Hietrinta $\!$and$\!$ R.Hirota,Mutidromion $\!$Soltions $\!$to$\!$ the 
$\!$Davey-Stewartson$\!$ Equation,{\it Phys.Lett.A} 
$\!${\bf 145}(1990)$\!$ 237-244.M. $\!$Jaulent,M.A.Manna $\!$and$\!$ 
L.$\!$Martinez-Alonso, Fermionic analysis of 
Davey-Stewartson dromions,{\it Phys.Lett.A} {\bf 151}(1990) 303-307.
\bibitem{HMM.91}
\vspace{-0.1cm}
R. Hernandes Heredero, L. Martinez-Alonso $\!$and$\!$ E. Medina Reus,
Fusion and fission $\!$of$\!$ dromions in the Davey-Stewartson Equation,
{\it Phys. Lett. A} {\bf 152} (1991) 37-41.
\bibitem{NK.84.87}
\vspace{-0.1cm}
S. Nishiyama and T. Komatsu,
Integrability Conditions for a Determination of 
Collective Submanifolds.
I. - Group-Theoretical Aspects,
{\it Nuovo Cimento A} {\bf 82} (1984) 429-442;
Integrability Conditions for a Determination of 
Collective Submanifolds.\\
II. - On the Validity $\!$of$\!$ the 
$<$ $\!$Maximally Decoupled$\!$ $>$ Theory,
{\bf 93} (1984) 255-267;
Integrability Conditions for a Determination of 
Collective Submanifolds.\\
III. - An Investigation of the Nonlinear Time Evolution 
Arising from the Zero-Curvature Equation,
{\bf 97} (1987) 513-522:\\
Integrability Conditions for a Determination of 
Collective Submanifolds.
A Solution Procedure,
{\it J. Phys. G: Nucl. Part. Phys.} {\bf 15} (1989) 1265-1274.
\bibitem{Schutz.80}
\vspace{-0.1cm}
B. Schutz,
{\it Geometrical methods of mathematical physics},
Cambridge University Press
Cambridge 1980.
\bibitem{HouHou.97}
\vspace{-0.1cm}
Bo-Yu Hou and Bo-Yuan,
{\it Differential Geometry for Physicists},
World Scientific Publishing Co. Pte. Ltd., 1997.
\bibitem{SMHU.83}
\vspace{-0.1cm}
F. Sakata, T. Marumori, Y. Hashimoto, and T. Une,
Geometry of the Self-Consistent Collective-Coordinate Method 
for the Large-Amplitude Collective Motion:\\ 
Stability Condition of ``Maximally-Decoupled"
Collective Submanifold, 
{\em Prog. Theor. Phys.} {\bf 70} (1983) 424-438.
\bibitem{FN.84} 
\vspace{-0.1cm}         
H. Fukutome and S. Nishiyama,
Time Dependent SO(2N+1) Theory for Unified Description of 
Bose and Fermi Type Collective Excitations,
{\it Prog. Theor. Phys.} {\bf 72} (1984) 239-251. 
\bibitem{Rajeev.94}
\vspace{-0.1cm}
S.G. Rajeev,
Quantum Hydrodynamics in Two Dimensions,
{\it Int. J. Mod. Phys. A} {\bf 9} (1994) 5583-5624.
\bibitem{RajeevTurgut.98}
\vspace{-0.1cm}
S.G. Rajeev and O.T. Turgut,
Geometric Quantization and Two Dimensional QCD,
{\it Commun. Math. Phys.} {\bf 192} (1998) 493-517.
\bibitem{ToprakTugurt.02}
\vspace{-0.1cm}
E. Topra and O.T. Turgut,
Large $N$ limit of SO(2N) scalar gauge theory,
{\it J. Math. Phys.} {\bf 43} (2002) 1340-1352.
Large $N$ 
limit of SO(2$N$) gauge theory of fermions and bosons,
ibidem {\bf 43} (2002) 3074-3096.
Wave Functions of Fermion Many-Body Systems,
\bibitem{CMS.81}
\vspace{-0.1cm}
M. Caseller, R. Megna and S. Sciuto,
Generalizations of the Sine-Gordon Equation
with $SU_{p+q}/S(U_p \!\times\! U_q)$ Structure,
{\it Nuovo Cimento A} {\bf 63} (1981) 339-352.\\
R. D'Auria, T. Regge and S. Sciuto,
A general scheme for bidimensional models with associated linear set,
{\it Phys. Lett.} {\bf 89B} (1980) 363-366.
\bibitem{LMG.65}
\vspace{-0.2cm}
H.J. Lipkin, N. Meshkov. and A.J. Glick,
Validity of Many-Body Approximation Methods 
for a Solvable Model (I). Exact solutions and Perturbation Theory,
{\it Nucl. Phys.} {\bf 62} (1965) 188-198.
\bibitem{RR.81}
\vspace{-0.1cm}
D.J. Rowe, A. Ryman and G. Rosensteel,
Many-body quantum mechanics as a symplectic dynamical system,
{\it Phys. Rev. A} {\bf 22} (1980) 2362-2373.
\bibitem{EMBS.05}
\vspace{-0.1cm}
E. Miller and B. Sturmfels,
Pl\"{u}cker coordinates,
Chapter 14 pp. 273-287 in
Combinatorial Commutative Algebra, 
Graduate Texts in Mathematics, 
Graduate Texts in Mathematics (227),
Springer, 2005.
\bibitem{Sa.81}
\vspace{-0.1cm}
M. Sato,
Soliton equations as dynamical systems on infinite dimensional 
Grassmann manifolds,
{\it RIMS Kokyuroku} {\bf 439} (1981) 30-46.
\bibitem{Fu.77}
\vspace{-0.1cm}
H. Fukutome, 
On the SO(2N+1) Regular Representation of Operators  
and Wave Functions of Fermion Many-Body Systems,
{\it Prog. Theor. Phys.} {\bf 58} (1977) 1692-1708.
\bibitem{WigSym99}
\vspace{-0.1cm}
T. Komatsu  and S. Nishiyama,
Self Consistent Field Method and $\tau$-Functional Method 
on Group Manifold in Soliton Theory,
Proceedings of the 
{\it Sixth International Wigner Symposium},
Bogazici University Press-Istanbul 2002, 381-409.
\bibitem{KN.00}
\vspace{-0.1cm}
T. Komatsu and S. Nishiyama,
Self-consistent field method from a $\tau$-functional viewpoint,
{\it J. Phys. Math. Gen. A} {\bf 33} (2000) 5879-5899;\\
T. Komatsu,
Self Consistent Field Method and
$\tau$-Functional Method 
in Fermion Many-Body Systems,
Doctoral Thesis at Osaka Prefecture University (2000).
\bibitem{NishiProviKoma2.07}
\vspace{-0.1cm}
S. Nishiyama, J. da Provid\^{e}ncia and T. Komatsu,
Self-consistent field-method and $\tau$-functional method 
on group manifold in soliton theory. II. Laurent coefficients 
of solutions for $\widehat{sl}_n$ and for $\widehat{su}_n$,
{\it J. Math. Phys.} {\bf 48} (2007) 053502.
\bibitem{FukuNishi.88.91}
\vspace{-0.1cm}
H. Fukutome,
Theory of Resonating $\!$Quantum$\!$ Fluctuations in a Fermion System,
{\it Prog. Theor. Phys.} {\bf 80} (1988) 417-432.~%\\
S. Nishiyama and H. Fukutome,
Resonating Hartree-Bogoliubov Theory for 
a Superconducting Fermion System 
with Large Quantum Fluctuations,
ibidem {\bf 85} (1991) 1211-1222.
\bibitem{Nishi.94}
\vspace{-0.1cm}
S. Nishiyama,
Application of the resonating Hartree-Fock theory to the Lipkin model,
{\it Nucl. Phys. A} {\bf 576} (1994) 317-350;\\
S. Nishiyama, M. Ido and K. Ishida,
Parity-Projectef Resonating Hartree-Fock 
Approximation to the Lipkin Model,
{\it Int. J. Mod. Phys. E} {\bf 8} (1999) 443-460.
\bibitem{Nishi.99}
\vspace{-0.1cm}
S. Nishiyama,
First-Order Approximation of the Number-Projected 
SO(2N) Tamm-Dancoff Equation and 
its Reduction by the Schur Function,
{\it Int. J. Mod. Phys. E} {\bf 8} (1999) 461-483.
\bibitem{KN.01}
\vspace{-0.1cm}
T. Komatsu and S. Nishiyama,
Toward a unified algebraic understanding of concepts of particle 
and collective motions in fermion many-body systems,
{\it J. Phys. Math. Gen. A} {\bf 34} (2001) 6481-6493.
\bibitem{NK.02}
\vspace{-0.1cm}
S. Nishiyama and T. Komatsu,
RPA Equation Embedded into Infinite-Dimensional Fock Space 
$F_\infty$,
{\it Physics of Atomic Nuclei} {\bf 65} (2002) 1076-1082.\\
S. Nishiyama, J. da Provid\^{e}ncia and T. Komatsu,
The RPA equation embedded into infinite-dimensional Fock space 
$F_\infty$,
{\it J. Phys. Math. Gen. A} {\bf 38} (2005) 6759-6775.
%\bibitem{KacLeur.03}
%\vspace{-0.3cm}
%V.G. Kac and J.W. van de Leur,
%The n-component KP hierarchy and representation theory,
%in {\em  Important developments in soliton theory},
%eds. Fokas A.S. and Zakharov V.E.,
%Springer Series in Nonlinear Dynamics (1993), pp.302-343.
\bibitem{DJKM.81}
\vspace{-0.2cm}
E. Date, M. Jimbo, M. Kashiwara and T. Miwa,
Operator Approach to the Kadomtsev-Petviashvili Equation
- Transformation Groups for Soliton Equations III -,
{\it J. Phys. Soc. Jpn.} {\bf 50} (1981) 3806-3812.
\bibitem{KBK.97}
\vspace{-0.2cm}
E.A. de Kerf,  G.G.A. B\"{a}uerle and A.P.E. Kroode,
{\it Lie Algebras,
Finite and Infinite Dimensional Lie Algebras and Applications in Physics}
Part 2, Elsevier Science V.B., Amsterdam, 1997.
\bibitem{BA.31}
\vspace{-0.2cm}
H. Bethe,
On the Theory of Metals, I. Eigenvalues and 
Eignefunctions of a Linear Chain of Atoms,
{\it Zeits. Physik} {\bf 71} (1931) 205-226.
\bibitem{MOPN.06}
\vspace{-0.2cm}
H.Morita,H.Ohnishi,J.daProvid\^{e}ncia  and S.Nishiyama,
Exact solutions for the LMG model Hamiltonian 
based on the Bethe ansatz,{\it Nucl.Phys.B} {\bf 737} [FS] (2006) 337-350.
\bibitem{Rowe.91}
\vspace{-0.2cm}
D.J. Rowe, T. Song. and H. Chen.,
Unified pair-coupling theory of fermion systems,
{\it Phys. Rev. C} {\bf 44} (1991) R598-R601.~%\\
H. Chen, T. Song and D.J. Rowe,
The pair-coupling model,
{\it Nucl. Phys. A} {\bf 582} (1995) 181-204.
\bibitem{Ri.65}
\vspace{-0.2cm}
R.W. Richardson,
Exact Eigenstates of the Pairing-Force Hamiltonian. II,
{\it J. Math. Phys.} {\bf 6} (1965) 1034-1051.
\bibitem{Littlewood.58}
\vspace{-0.2cm}
D.E. Littlewood,
The theory of group characters and matrix representation of groups,
Clarendon, Oxford, 1958.
\bibitem{MacDonald.79}
\vspace{-0.2cm}
I. G. MacDonald, Symmetric Functions and Hall Polynomials,
Oxford University, 
Oxford,1979. 
\bibitem{PD.98.99}
\vspace{-0.2cm}
F. Pan $\!$and$\!$ J.P. Draayer, $\!$(a) 
New algebraic approach for an exact solution of the nu-\\clear mean-field 
$\!$plus$\!$ orbit-dependent $\!$pairing$\!$ hamiltonian,$\!\!$
{\it Phys.Lett.B}{\bf 442}(1998)7-13.\\
(b) Exact solutions for some nuclear many-body problems, %
Ann. Phys. (NY) {\bf 271} (1999) 120-140.
\bibitem{PZDD.17}
\vspace{-0.2cm}
F. Pan, D. Zhou, L. Dai and J.P. Draayer,
Exact solution of the mean-field plus separable pairing model reexamined,
{\it Phys. Rev. C} {\bf 95} (2017) 034308-1-8.
\bibitem{Mansfield.85}
\vspace{-0.2cm}
P. Mansfield,
Solution of the Initial Value Problem for the sine-Gordon Equation
Using a Kac-Moody Algebra,
{\it Commun. Math. Phys.} {\bf 98} (1985) 525-537.
\bibitem{Gau76}
\vspace{-0.2cm}
M. Gaudin,
Diagonalisatiom D'une Classe D'hamiltoniens de Spin,
{\it J. Physique} {\bf 37} (1976) 1087-1098;
{\it La Fonction d'Onde de Bethe}, Masson, Paris, 1983.
\bibitem{Skl99}
\vspace{-0.1cm}
E.K. Sklyanin,
Generating Function of Correlators in the sl$_2$ Guadin Model,
{\it Lett. Math. Phys.} {\bf 47} (1999) 275-292.
\bibitem{OSDR.05}
\vspace{-0.1cm}
G. Oritz, R. Somma, J. Dukelsky and S. Rombouts,
Exactly-solvable models derived from 
a generalized Gaudin algebra,
{\it Nucl. Phys. B} 
{\bf 707} [FS] (2005) 421-457.
\bibitem{LJ-HC.20}
\vspace{-0.1cm}
H.R. Larsson, C.A. Jim\'{e}nez-Hoyos and G.K-L. Chan,
Minimal matrix product states and generalizations of mean-field and
geminal wavefunctions,
{\it J. Chem. Theory Comput.} (2020).
DOI:10.102/acs. jctc. 0c00463.
\bibitem{Daboul.93}
\vspace{-0.1cm}
J. Daboul, P. Slodowy and C. Daboul,
The hydrogen algebra as centerless twisted Kac-Moody algebra,
{\it Phys. Lett. B} {\bf 317} (1993) 321-328.
\bibitem{GO.86}
\vspace{-0.1cm}
P. Goddard and D. Olive,
Kac-Moody and Virasoro Algebras 
in Relation to Quantum Physics,
{\it Int. J. Mod. Phys. A} {\bf 1} (1986) 303-414.
\bibitem{KleinWaletDang.91}
\vspace{-0.1cm}
A. Klein, N.R. Walet and G.D. Dang,
Classical Theory of Collective Motion
in the Large Amplitude, Small Velocity Regime,
{\it Ann. of Phys.} {\bf 208} (1991) 90-148.
\bibitem{Ottesen.95}
\vspace{-0.1cm}
J.T. Ottesen,
Infinite Dimensional Groups and 
Algebras in Quantum Physics,
Lecture Note in Physics, New Series m:
Monographs; 27,Berlin: Springer, 1995.
\bibitem{NPB.08}
\vspace{-0.1cm}
S. Nishiyama., J. da Provid\^{e}ncia, C. Provid\^{e}ncia, 
and F. Cordeiro,
Extended supersymmetric $\sigma$-model based on the
SO(2N+1) Lie algebra of the fermion operators,
{\it Nucl. Phys. B} {\bf 802} (2008) 121-145.
\bibitem{AdlerMoerbeke.94}
\vspace{-0.1cm}
M. Adler, P. van Moerbeke and S. Birkhoff,
B\"{a}cklund Transformations and 
Regularization of Isospectral Operators,
{\it Adv. in Math.} {\bf 108} (1994) 140-204.
\bibitem{TW.98}
\vspace{-0.1cm}
T. Taniuti,
Part 1. General Theory, Reductive Perturbation Method and
Far Fields of Wave Equations,
{\it Prog. Theor. Phys. Suppl.} {\bf 55} (1974) 1-35.
\bibitem{NW.78}
\vspace{-0.1cm}
M. Nogami and C.S. Warke,
Exactly solvable time-dependent Hartree-Fock equations,
{\it Phys. Rev. C} {\bf 17} (1978) 1905-1913.
\bibitem{SeiyaJoao.15}
\vspace{-0.1cm}
S. Nishiyama and J. da Provid\^{e}ncia,
SO(2N)/U(N) 
Riccati-Hartree-Bogoliubov equation
based on the SO(2N) Lie algebra
of the fermion operators,
{\it Int. J. Geom. Methods Mod. Phys.} {\bf 12} (2015) 1550035.
\bibitem{NishiProvi.19}
\vspace{-0.1cm}
S. Nishiyama and J. da Provid\^{e}ncia,
Remarks on the mean-field theory based on the
$\mbox{SO(2N\!+\!1)}$ Lie algebra of the fermion operators,$\!$
{\it Int. J. Geom. Methods Mod. Phys.} {\bf 12} (2019) 1950184;~%\\
Mean-field theory based on the $\mathfrak{Jacobi~hsp}$
:= semi-direct sum  
 $\mathfrak{h}_N \!\rtimes\! \mathfrak{sp}(2N,\mathbb{R})_\mathbb{C}$
algebra of boson operators,
{\it J. Math. Phys.} {\bf 60} (2019) 081706.
\bibitem{NMO.04}
\vspace{-0.1cm}
S. Nishiyama, H. Morita and H. Ohnishi,
Group-theoretical deduction of a dyadic Tamm-Dancoff 
equation by using a matrix-valued generator coordinate,
{\it J. Phys. Math. Gen. A} {\bf 37} (2004) 10585-10607.
\bibitem{NishiProvi.16}
\vspace{-0.1cm}
S. Nishiyama and J. da Provid\^{e}ncia,
Modified Non-Euclidian Transformation  
on the
SO(2N+2)/U(N+1) Grassmannian and 
SO(2N+1) Random Phase Approximation 
for Unified Description of 
Bose and Fermi Type Collective Excitations,
{\it Int. J. Geom. Methods Mod. Phys.} {\bf 13} (2016) 1650043.
\bibitem{Kirillov.76}
\vspace{-0.1cm}
A.A. Kirillov,
{\it Lectures on the Orbit Method},
Graduate Studies in Mathematics Volume 64:
American Mathematical Society
Providence, Rhode Island, 2004.\\
{\it Elements of the Theory of Representations},
Springer-Verlag, New York, 1976.
%\bibitem{Jimbo.90}
%\vspace{-0.3cm}
%M. Jimbo,
%{\it Quantum Group and Yang-Baxter equation}
%in Japanese,
%Springer-Verlag Tokyo 1990.
\end{thebibliography}
\end{document}